# Kinematically cold and warm planetary nebulae samples, HII regions and supernovae remnants in the disc of the face-on spiral galaxy NGC 628 (M74) - The Planetary Nebulae Spectrograph with the H*α* arm

**Magda Arnaboldi[1*], Ortwin Gerhard[2], Surya Aniyan[3], Kenneth C. Freeman[3], Anastasia Ponomareva[3,4], Lodovico Coccato[1], Johanna Hartke[5,6,7] and Steven P. Bamford[8], Arianna Cortesi[9], Nigel Douglas†, Crescenzo Tortora[10], Michael Merrifield[8], Konrad Kuijken[11], Massimo Capaccioli[12], Nicola R. Napolitano[12], Claudia Pulsoni[2], Aaron J. Romanowsky[13]**

1 ESO, Karl Schwarzschild Str. 2, D-85748 Germany
2 Max-Planck-Institüt fur Extraterrestrisches Physik, Giessenbach Str., D-85741 Garching, Germany
3 Research School of Astronomy and Astrophysics, Australia National University, Canberra ACT 2611, Australia
4 Oxford Astrophysics, University of Oxford, Denys Wilkinson Building, Keble Rd, Oxford OX1 3RH, UK
5 Finnish Centre for Astronomy with ESO, (FINCA), University of Turku, FI-20014 Turku, Finland;
6 Tuorla Observatory, Department of Physics and Astronomy, University of Turku, FI-20014 Turku, Finland;
7 Turku Collegium for Science, Medicine and Technology (TCSMT), University of Turku, FI-20014 Turku, Finland;
8 School of Physics and Astronomy, University of Nottingham, Nottingham, NG7 2RD, UK
9 Universidade Federal do Rio de Janeiro, Observatório do Valongo, CEP 20080-090 Rio de Janeiro, RJ, Brazil
10 Osservatorio Astronomico di Capodimonte, INAF, Salita Moiariello 16, I-80131, Italy
11 Leiden Observatory, Leiden University, the Netherlands
12 Department of Physics "E. Pancini", University of Naples Federico II, Via Cintia, 21, 80126 Naples, Italy
13 Department of Physics and Astronomy, San Jose State University, One Washington Square, San Jose, CA 95192, USA

*** Correspondence:**
Corresponding Author: marnabol@eso.org

## Abstract

We present the results from the observations of the galaxy NGC 628 with the Planetary Nebulae Spectrograph (PN.S) equipped with the Hα arm. With the third PN.S arm, the Hα arm, we measure the Hα fluxes, in addition to fluxes and line-of-sight velocities (LOSV) of monochromatic, spatially unresolved [OIII] 5007Å sources, in the nearly face-on disc of NGC 628. The narrow band color ([OIII] 5007Å – Hα) vs $m_{5007}$ magnitude diagram does separate planetary nebulae (PNe) from single compact ionized HII regions and supernovae remnants (SNRs), which also emit in [OIII]5007 Å. The goals are to detect *bona fide* PNe in the face-on spiral galaxy NGC 628 (M74) so that we can measure the velocity dispersion of the stars perpendicular to the main plane of the disc. By associating the velocity dispersion orthogonal to the disc of the evolved stars whose scale height is measured for edge-on discs in the near-infrared wavelengths, we can measure the surface mass density directly and break the disc-halo degeneracy. This study motivates and validates the empirical selection criteria for PNe with the PN.S in star forming discs, while the modelling of the disc dynamics, based on the selected PN samples, is done in companion papers. We classified 442 PNe and 251 spatially isolated, unresolved HII regions: the PN.S with the Hα arm increased the number of known PNe in NGC 628 by a factor 4. In this study, we find evidence for two kinematically distinct PN populations in the NGC 628 disc. The kinematically 'cold' PN population dominates the PN luminosity function (LF) close to the bright cut-off magnitude, indicating that the PN massive, short-lived progenitors dominate the PNLF bright cut-off in NGC 628. The 'warmer' PN component increasingly dominates at fainter magnitudes. The velocity dispersion orthogonal to the disc plane of these two populations are $\sigma_{z,cold} = 8.8$ kms$^{-1}$ and $\sigma_{z,warm} = 26.1$ kms$^{-1}$ respectively, over a range of radii 80" to 425". The cold and warmer components contribute to the total sample of the PN population in NGC 628 with the ratio 46% (cold) and 54% (warm). Once the velocity dispersion of the old component is matched with the population's scale height, the decomposition of the rotation curve for NGC 628 leads to a maximal disc, with the rotation of the baryonic component accounting for 78% of the total rotational velocity in NGC 628.

## 1. Introduction

The Planetary Nebulae Spectrograph (PN.S; Douglas et al. 2002) mounted on the 4.2 meter William Herschel Telescope on La Palma has been used extensively to map the planetary nebula (PN) populations as kinematical tracers in the low surface brightness regions of nearby ($D \leq 20$ Mpc) early-type galaxies (hereafter ETGs), to trace the their total angular momentum, mass distributions and intrinsic 3D shapes (Coccato et al. 2009; Cortesi et al. 2013; Morganti et al. 2013; Arnaboldi et al. 2017,2020; Pulsoni et al. 2018, 2023). As PNe are the nebular end phases of intermediate to low mass stars, e.g. 8 – 1 $M_{\odot}$, whose strongest optical emission is at [OIII] 5007 Å, once detected they can be used as beacons to trace the motions of the parent stars in any 1 Gyr and older populations. In ETGs, the continuum light is dominated by old stellar populations (Thomas et al. 2005; Greene et al. 2012; Barbosa et al. 2021); hence those stars in the PN phase can be detected and their line-of-sight velocity ($V_{LOS}$) measured with the PN.S. Because of its large field of view (FoV) 10 arcmin × 11 arcmin and a high spectral resolution which allows to reach accuracy of few $kms^{-1}$ for any velocity measurements, the PN.S delivers an observing parameter space which is competitive for mapping discrete velocity fields in nearby galaxies. The PN.S data map the velocity fields out to six $R_e$ on average for massive ETGs (Pulsoni et al. 2018, 2023), where $R_e$ is the effective radius containing half of the stellar light. When measuring the kinematics of discrete tracers, the PN.S instrument is advantageous vs IFU observations, for example with MUSE whose FoV covers the very central high surface brightness regions out to 1 - possibly $2R_e$ (see Roth et al. 2021; Spriggs et al. 2021; Scheuermann et al. 2022; Soemitro et al. 2023; Jacoby et al. 2024). To cover O(100) sq. arcmin area with MUSE, one would need 100 MUSE pointings, while this area coverage is achievable with one single PN.S pointing.

The study of discrete stellar tracers like PNe in star forming discs presents additional challenges with respect to ETGs, since ionized HII regions (Ciardullo et al. 2002) from OB stars and young supernovae remnants (SNRs), caused by ejecta shocks in a cold interstellar medium (ISM, Kreckel et al. 2017), also emits in [OIII] 5007 Å and blur the picture. The morphological constraints applied to the [OIII] 5007 Å emissions to select the PN candidates in ETGs (e.g. selection of spatially unresolved-monochromatic [OIII] 5007 Å emission with no-continuum, see Arnaboldi et al. 2002, Douglas et al. 2002), may not be sufficient to select only *bona-fide* PNe in these systems. The selection of PN candidates in star-forming galaxies requires additional diagnostics that are based on the line-ratios among strong lines, like [OIII] 5007 Å and Hα 6563 Å plus [NII] lines (Baldwin et al. 1981; Ciardullo et al. 2002; Arnaboldi et al. 2003; Kreckel et al. 2017; Roth et al. 2021), by requesting that the [OIII] 5007 Å flux is at least three times that of the Hα 6563 Å plus [NII] lines (Herrmann & Ciardullo 2009a) .

The identification of PNe in nearly face-on disc galaxies is kinematically interesting because their line-of-sight velocity distribution (LOSVD) allows the measurement of the velocity dispersion orthogonal to the disc plane *for an evolved population of stars* which dominates the continuum light at redder wavelengths. Both kinematic and scale length measurements for nearly face-on discs are at the basis of any attempts to determine the surface mass density in discs (van der Kruit & Freeman 1984) and resolve the disc-halo degeneracy (Herrmann & Ciardullo 2009b; Bershady et al. 2010; Aniyan et al. 2016, 2018, 2021). For these efforts, it is essential that the vertical disc scale heights $h_z$ and the vertical velocity dispersion $\sigma_z$ should refer to the same population of stars (Aniyan et al. 2016).

An example of such studies is the measurements of the kinematics for the PN population in the nearly face-on spiral NGC 628 (M74). This spiral galaxy is at ~ 9.0 Mpc distance, and its PN population was studied extensively by Herrmann et al. (2008); Herrmann & Ciardullo (2009a) and Kreckel et al. (2017). NGC 628 has a strong metallicity gradient $12 + \log(O/H) = 8.834 - 0.485 \times R$ dex $R_{25}^{-1}$ as measured using direct abundances based on observations of the temperature sensitive auroral lines (Berg et al. 2015). NGC 628 has been observed at 21 cm wavelength as part of the THINGS HI survey (Walter et al. 2008). Based on previous studies of the HII regions (Leliévre & Roy 2000), SNRs (Kreckel et al. 2017) and PNe (Herrmann & Ciardullo 2009a), we can thus compare, calibrate and assess any systematics affecting the [OIII] 5007 Å discrete emitter sample that is selected from the PN.S left/right arms plus the Hα images. Once validated, these empirical calibrations and selection criteria for the PN candidates become the basis for the surface mass determinations carried out in the companion papers by Aniyan et al. (2018) for NGC 628 and Aniyan et al. (2021) for NGC 6946. After validating the empirical criteria for the selection of PNe in star forming populations, we are then equipped to study the dynamics of discrete stellar tracers in discs with the PN.S CDI + Hα imaging in any star-forming galaxies, thus reaching potentially out to $R_{25}$ and beyond in these systems.

In this paper we present the optical layout of the PN.S instrument, the data acquisition and calibrations of the PN.S equipped with the Hα arm in Section 2. In Section 3 we describe the catalog extraction, selection and validation using a combination of [OIII] and Hα colors. In Section 4, we illustrate the luminosity functions and spatial distributions of PNe and HII region candidates in NGC 628; we present the residual velocity distributions of the different nebulae populations with respect to the HI 21 cm rotation in Section 5. We illustrate the results on the kinematics of the PN population and the properties of the cold and warm components in the disc of NGC 628 in Section 6. In Section 7 we discuss our findings and in Section 8 we present our summary and conclusions. In the Appendix we provide the complete catalogue of PNe and of the isolated, spatially unresolved HII regions in NGC 628. We also further provide the flux calibration for the Hα arm. In the rest of the paper, we adopt a distance to NGC 628 of D = 8.6 ± 0.3 Mpc (Herrmann et al. 2008) hence 1”= 41.7 pc.

## 2. PN.S equipped with the third arm: optical layout, operations and calibrations

The PN.S is a slitless imaging spectrograph designed for the detections of extragalactic PNe (Douglas et al. 2002). It was mounted at the 4.2 meter William Herschel telescope at La Palma. The PN.S has a field of view of 10.4 × 11.3 $arcmin^2$. Figure 1-a shows the optical layout of PN.S with the two [OIII] 5007Å arms (left/right), that are produced by a concave grating (obtained with two plane gratings). Once the light passes through a narrow band filter centered on the red- shifted [OIII] 5007Å emission at the systemic velocity of the galaxy under study, it is dispersed in opposite directions in the two arms, such that the continuum light from stars will appear as streaks, while the nearly monochromatic [OIII] 5007Å emission from PNe associated with the galaxy light will appear as unresolved sources, in space and wavelength. By combining these two counter-dispersed images, the PNe can be identified and their line-of-sight (LOS) velocity measured in a single observation. The addition of the Hα arm includes the insertion of a beam-splitter, i.e. a dichroic, that reflects the red light off the beam, through the Hα camera filter and optics, see Figure 1-b. The beam-splitter has a 97% transmission in the wavelength range 4950 – 5050Å and a front reflection larger than 80% at the wavelength range 6500 – 6640Å.

The third arm is equipped with a narrow band filter centered at the Hα emission, plus a broad band filter that matches the R band. Because the two filters for the Hα arm have different optical thickness, an offset is required to keep the Hα camera on focus, when changing filters from narrow to broad, and vice versa.

The filter specification for the broad R and narrow Hα filters are such that the R band pass band samples the spectra energy distribution around the Hα emission, and the FWHM of the narrow Hα filter covers the entire velocity range of the PN.S [OIII] narrow band filter set. The Hα filter has a diameter 139 mm ± 0.5 mm, and a thickness of 8 mm. The central wavelength is 6595 Å± 5Å and FWHM of 120 Å ± 5Å with a transmission T > 0.8. The broad band filter for the Hα arm has a special R band continuum (top hat) filter with a specification of the Cut-on> 6100Å, and a Cut-off< 6900Å, i.e. Half-power width of < 800Å with a diameter 139 mm ± 0.5 mm and a thickness 8 mm. The Hα arm is operated with the 2k×2k Marconi2 CCD and it has a pixel scale of 0".27 pix$^{-1}$.

The PN.S was designed to measure velocities in addition to detecting nebulae. This is especially valuable in face-on discs: because of scatter by giant molecular clouds in the disc, young and intermediate-old population of stars may have different velocity dispersion $\sigma_z$ orthogonal to the disc plane, see Aniyan et al. (2016) and references therein. Order of magnitude estimates of the velocity dispersion $\sigma_z$ for the young thin disc population is given by the HI 21 cm RMS, of the order of 10 kms$^{-1}$ (Das et al. 2020). To be able to disentangle different populations according to their $\sigma_z$ and measure the velocity dispersion of the cold component in the thin disc, the wavelength calibrations require a velocity accuracy of few kms$^{-1}$. In addition to image processing and implementing color selection, to remove contribution from HII regions and SNRs, the identification of PN population in face-on discs motivates further developments in order to improve the wavelength calibration and geometric distortion correction in the PN.S original pipeline and thus achieve more accurate LOSV measurements.

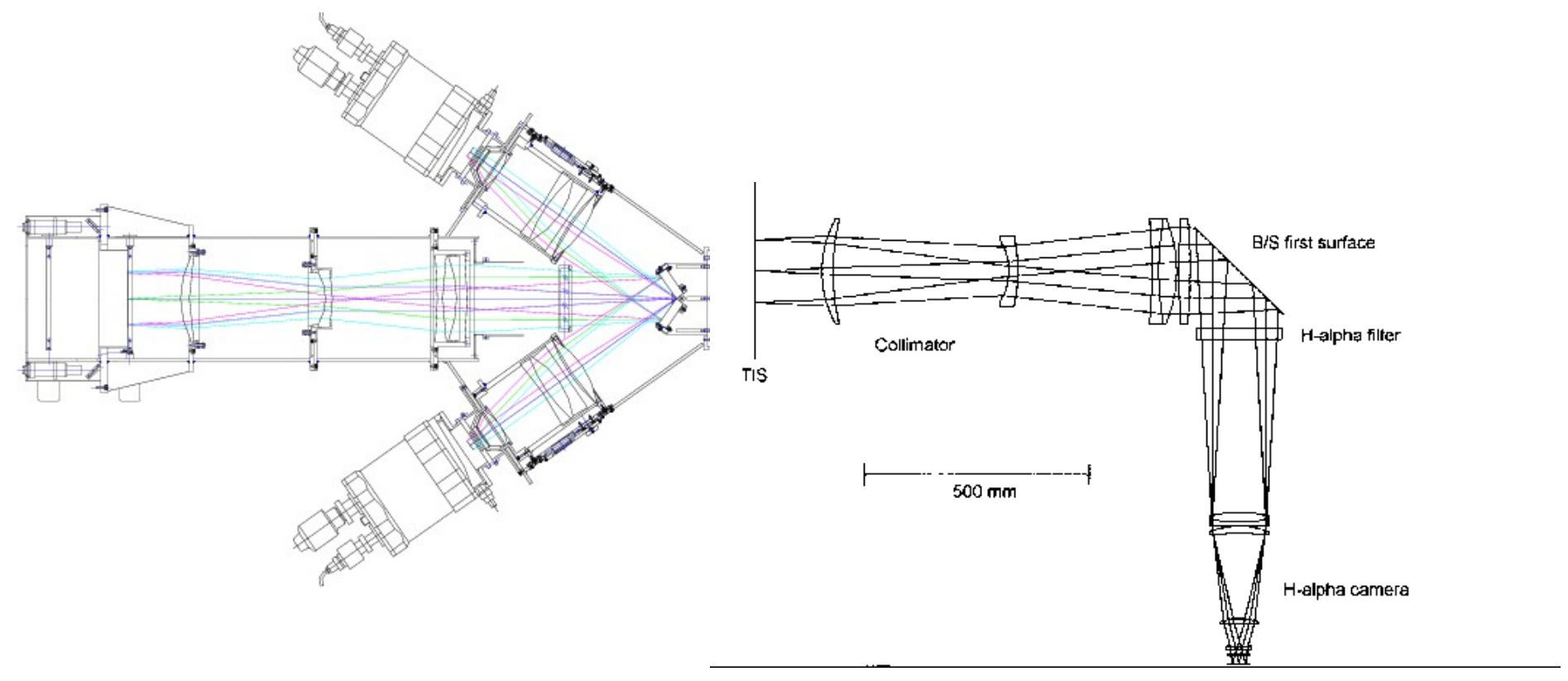

(a) (b)

**Figure 1. (a):** Left panel. The optical layout of the PN.S showing the [OIII] left and right arms. The pair of diffraction gratings on the right split the light beam into two identical spectrograph arms, each with opposite dispersion directions. **(b):** Right panel. The optical arrangement of the PN.S showing the Hα arm. The "B/S first surface" label indicates the beam splitter. The optics to the left of the Beam splitter are the standard PN.S optics. Below the Beam splitter sits the Hα filter and the Hα camera.

## 2.1. Observations and calibration steps for [OIII] 5007 Å CDI and Hα imaging

Once the PN.S left/right arms are operated in standard mode as described in Douglas et al. (2002, 2007), the Hα arm can acquire images of the same pointing of the sky simultaneously. In standard operations, the PN.S is equipped with three CCDs, that are remotely operated from the instrument controller. Exposure times for the two CCDs mounted on the left and right arms are set independently from the exposure time of the third CCD on the Hα arm.

The data for NGC 628 were acquired over two nights in September 2014 at the WHT. The weather during the run was photometric with typical seeing FWHM ~ 1**"**. The observations for the galaxy NGC 628 consisted of 14 left and right exposures of 1800 sec each with the PN.S [OIII] left/right arms, and a sequence of 17 exposures of 900 sec each with the Hα arm equipped with the narrow Hα filter. No broad R band images were acquired. At the redshift of NGC 628 the observed wavelength of the [OIII] emission is close to 5018 Å. At the location of the [OIII] line for the NGC 628 PNe, the pixel scale in the direction of dispersion is 0.769Å pixel$^{-1}$ **(**or 1.30 pixel Å$^{-1}$**)**. At the same location, the half-power wavelengths for the filter pass- band are 4999 and 5043Å. The wavelength of [OIII] emitters in NGC 628 is very close to the midpoint of the filter passband. Flux calibrations of the PN.S [OIII] and Hα arms were carried out by observing spectro-photometric standard stars at low airmasses during photometric nights. See Appendix C for a detailed description of the zero-point calibration of [OIII] and Hα fluxes. We summarize here the main equations which relate the instrumental magnitude $m_0$ defined as:

$$m_0 = -2.5 \log(\mathrm{counts(ADU)s^{-1}}) \qquad (1)$$

corrected to outside the atmosphere, to the AB magnitude and $m_{5007}$ (Jacoby 1989) systems. The transformations are

$$m_{AB,[OIII]} = m_0 + (18.02 \pm 0.12) \qquad (2)$$

$$m_{5007} = m_0 + 24.66 \qquad (3)$$

for the PN.S [OIII] left/right images, and

$$m_{AB,H\alpha} = m_0 + (22.83 \pm 0.11) \qquad (4)$$

for the Hα images taken with the PN.S Hα camera.

## 2.2. Improved wavelength calibration and geometric distortion corrections for the PN.S

The data reduction pipeline for the PN.S [OIII] left/right arms is a combination of IRAF and FORTRAN scripts and a detailed description is provided in Douglas et al. (2007). The code debiases and flat field corrects the raw images from the PN.S left/right arms (L/R in what follows) using bias frames and sky flats obtained during the observing run. Cosmic rays are removed using a specific developed routine in the pipeline.

The calibration step specific to the PN.S pipeline is the wavelength calibration and distortion correction of the counter-dispersed imaging, or CDI for short. Such a calibration is achieved using a pair of dispersed images of a regular pin-hole mask in the PN.S L/R arms. The pin-hole mask is illuminated with a Cu-Ne-Ar lamp and four arc lines at wavelength 5009, 5017, 5031 and 5038 Å are visible for each pinhole in the mask. An automated task in the PN.S pipeline called *alignspot* identifies and centroids the pin-hole monochromatic arc lamp images and fits a joint polynomial for position and wavelength. The solution represents a map from position on the sky reference frame (X, Y) and wavelength (λ; i.e. the arc lines) to the (x, y) position on the CCDs $\{x, y\} = f(X, Y, \lambda)$. For pairs of detected sources in the L/R CCDs, the mapping can be uniquely inverted to transform $\{x_L, y_L, x_R, y_R\}$ to $\{X, Y, \lambda\}$. These mapping functions allow us to correct spatial distortions introduced by the spectrograph. The velocities of the [OIII] spatially unresolved monochromatic pairs in the L/R arms are evaluated using a rest wavelength of the [OIII] line of 5006.8Å (Douglas et al. 2007). In the standard PN.S pipeline, the geometric transformation corrects for shifts, rotation and linear terms in wavelength, which ensures an accuracy of $\approx 20$ $\mathrm{kms^{-1}}$, as verified by in-depth comparison with independent PN observations in NGC 3379 (Douglas et al. 2007) and in M31 (Merrett et al. 2006). An instrument velocity dispersion of 20 $\mathrm{kms^{-1}}$ is completely adequate to map the kinematics of massive ETGs with LOS velocity dispersions of 100 $\mathrm{kms^{-1}}$ or larger. For the current project, we drastically improved the efficiency of *alignspots* by applying a higher order geometric distortion correction such that the RMS of the velocity calibrations is only a few $\mathrm{kms^{-1}}$ for the high S/N arc lines.

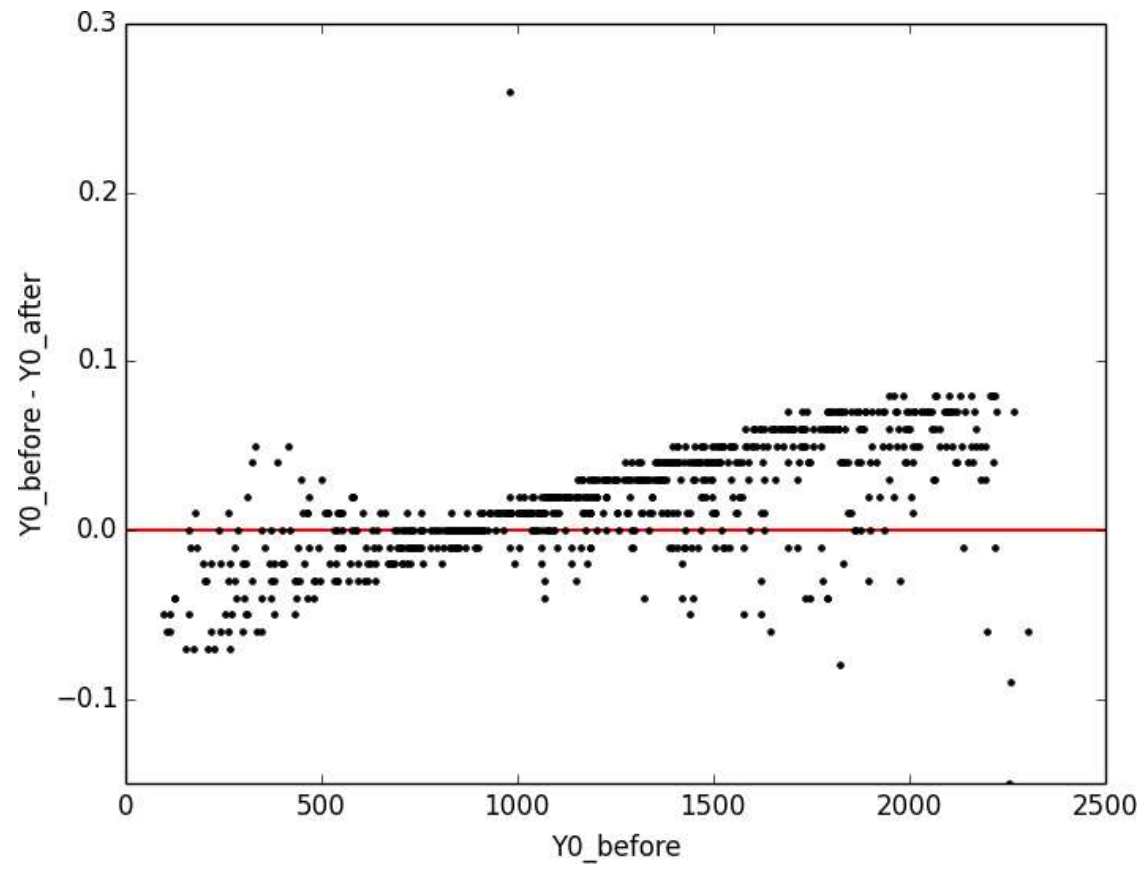

**Figure 2.** RMS plot from the PN.S *alignspots* calibration, before and after the higher order geometric distortion correction. Y0 is the y coordinate of the emission lines in the image of the pin-hole mask before and after the correction of the residual distortion by the larger order polynomial**.** The goal of this step is to correct for the spatial distortions introduced by the spectrograph

Figure 2 shows the RMS of the geometric transformation with the higher order polynomial, up to quadratic terms in (x, y). The L/R images obtained for a given pointing are then stacked to create the deep field images. In the best seeing image, up to 15 – 20 USNO B1 stars are identified, then used to determining the relative offsets and alignment of images. They are further used to transform images to sky coordinates, using the IRAF tasks *geomap* and *geotrans* (Douglas et al. 2007).

*Error determination for the $V_{LOS}$ measurements* – In order to get an empirical estimate of the radial velocity measurement errors associated with PN detections, the 14 science L/R images were divided into two sets and independently identified the unresolved sources in each set. Figure 3 shows the error distributions of the velocity pairs. As most of the PNe candidates are going to have apparent $m_{5007}$ magnitudes fainter than m*= 24.75, the estimated radial velocity errors are typically between 4 and 9 $kms^{-1}$ hence in the range of accuracy required for the measurement of $\sigma_z$ in thin discs; see Figure 3 in Aniyan et al. (2018) for additional details.

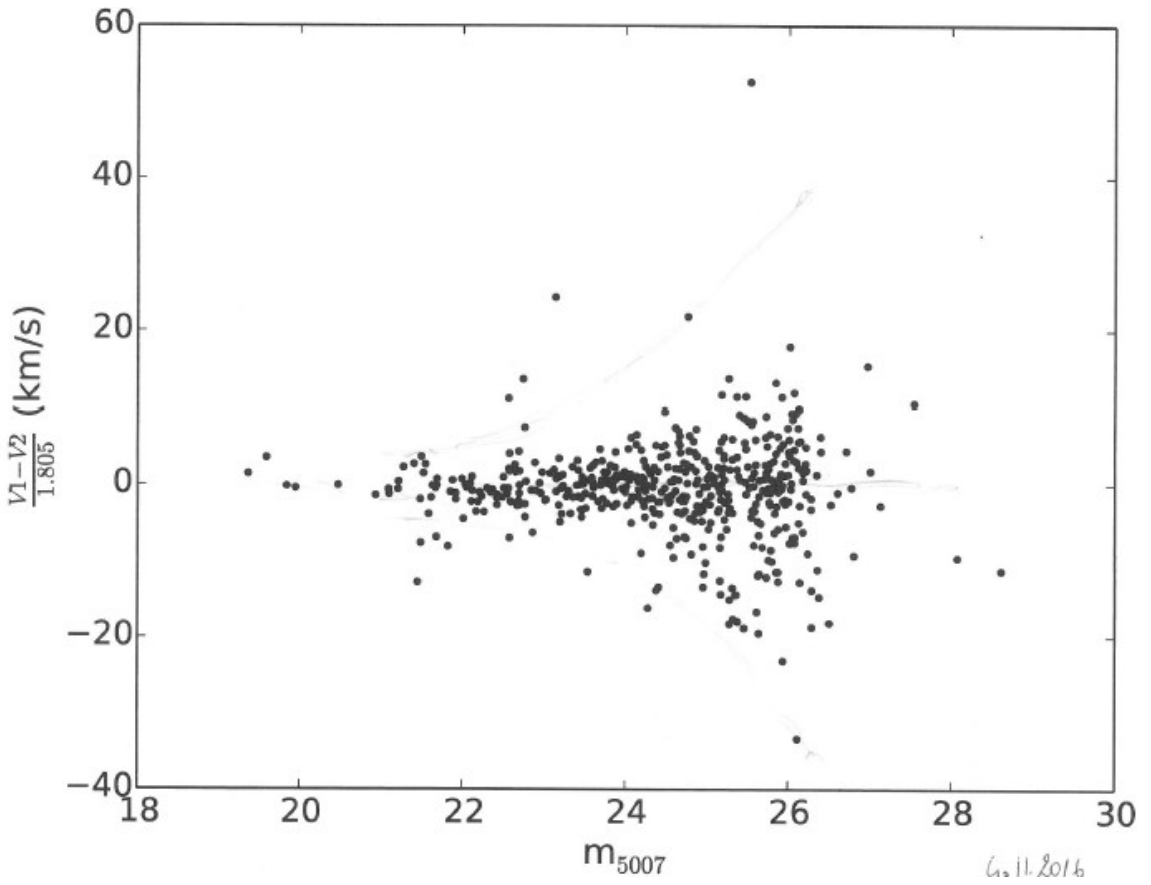

**Figure 3.** Velocity error (pairs) as function of $m_{5007}$. Distribution of the $\Delta V/1.805$ pairs vs $m_{5007}$. The normalization factor is introduced as the two sets of images are registered to the same high image quality reference frame.

*Photometric error and completeness* – The PN.S magnitudes were extensively compared with literature values for PN samples in ETGs and disc galaxies, e.g. M31, giving typical dispersion values of $\sigma_{m5007} = 0.13$ mag[1]. Based on extensive analysis of PN.S samples in ETGs (Douglas et al. 2007; Coccato et al. 2009; Cortesi et al. 2013) detectability reaches ~95% at R > 1$R_e$ radial distances. The scale length in I band of NGC 628 is 73".4 ±3".7 (Möllenhoff 2004), so any PN sample detected at larger radial distances would reach similar completeness.

*Data reduction Hα arm* – As this was the first PN.S project that extensively used the Hα arm, the reduction pipeline for the Hα imaging data was developed from scratch using standard IRAF packages and CCD reduction techniques. The Hα raw images were de-biassed and flat fielded, and then the exposures were aligned and registered with reference to the image with the best seeing; they were then co-added to reach better S/N for the faint sources. Stars were selected in the combined image, and in the DSS image of the same field to compute the astrometric solution for the Hα image. The plate solution was computed with the IRAF package *ccmap* and the world-coordinate system (wcs for short) information was added in the header of the combined Hα image with the package *ccsetwcs*. We now have combined images for the [OIII] 5007Å L/R arms as well as the corresponding combined Hα narrow band image for the same field. In Figure 4 we show the surveyed region in the NGC 628 (M74) disc in the deep PN.S L(eft) dispersed image and the narrow Hα image with few bright HII regions indicated.

## 3. [OIII] 5007Å sources for NGC 628: source selection, catalog extraction and validation

In star-forming late-type discs, HII regions that are ionized by hot OB stars do emit in [OIII] 5007Å, hence additional criteria must be applied to distinguish them from PNe. As the former are nebulae ionized by OB stars, their Hα emissions are the dominant line, e.g. see Peña et al. (2007); Kniazev et al. (2005). Furthermore their apparent [OIII] 5007Å magnitudes may be brighter than that for the bright cut-off of the PNLF at the distance of NGC 628 (Hermann et al. 2008; Kreckel et al. 2017). Ciardullo et al. (2002) and Herrmann & Ciardullo (2009a) showed that the ratio [OIII] 5007Å to Hα can be used as an excellent criterion for discriminating PNe from HII regions, whenever HII regions are spatially unresolved. Once the ratio R = I([OIII]5007Å)/I (Hα + [NII]) is plotted against the $m_{5007}$-m∗, where m∗ is the apparent magnitude of the PNLF bright cut-off at the distance of that PN population, the true PNe occupy a locus corresponding to [OIII] 5007Å line fluxes which are brighter than the Hα fluxes by a factor ≥ 3. Herrmann & Ciardullo (2009a) carried out a survey of PNe in NGC 628 and confirmed them spectroscopically according to the above selection. Our goal is (i) to measure the [OIII] 5007Å and Hα fluxes in the PN.S L/R plus Hα set of images for all [OIII] 5007Å unresolved sources, (ii) match them to the

[1] This value was determined independently from Gaussian fits to residual magnitudes for samples for hundred (NGC 3379; Douglas et al. 2007) and few thousands PNe (M31; Merrett et al. 2006). The error increased to 0.3 at Δm=4 mag below the PNLF bright cut off.

NGC 628 PNe sample from Herrmann & Ciardullo (2009a) and then (iii) infer the selection criteria for the true PNe empirically. We further validate the selection criteria for the PN.S PN sample by cross matching these [OIII] monochromatic spatially un-resolved sources with the SNR candidates selected by Kreckel et al. (2017) from the MUSE observations of NGC 628.

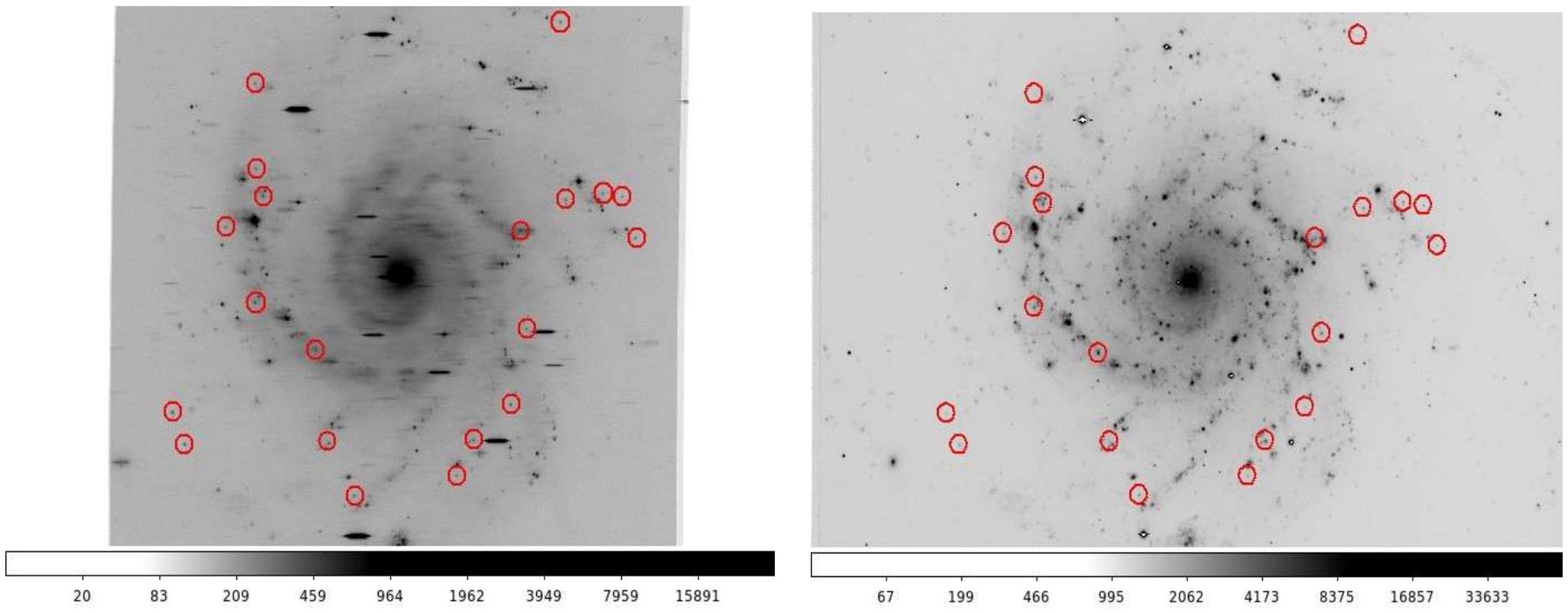

(a) (b)

**Figure 4. a):** Left [OIII] 5007Å Narrow band dispersed image from the PN.S L arm. Monochromatic spatially unresolved sources with red circles are bright HII regions, with no continuum; East is down and North is on the Left. **b)**: Hα image of NGC 628 registered on the [OIII] PN.S L image. This narrow band Hα image is not continuum subtracted.

## 3.1 Selection criteria with PN.S CDI+Hα for PNe in NGC 628 – Matching sources with the Herrmann & Ciardullo (2009a) sample

We proceed as follows. On the NGC 628 PN.S L/R images, pairs of [OIII] 5007Å spatially unresolved monochromatic sources, without continuum, are identified using Sextractor (Bertin & Arnouts 1996). The total number of identified pairs is 693. Then we measured the Hα flux in a matched aperture at the position of each [OIII] 5007Å source on the Hα registered image. Sextractor in dual image mode (e.g. detection of a source on the [OIII] 5007Å PN.S L image, and photometry in matched apertures on the [OIII] PN.S L and Hα images) is utilized to measure positive fluxes for 623 [OIII] 5007Å detected sources. The remaining 70 [OIII] 5007Å L/R pairs do not have corresponding Hα fluxes, either because the PN.S L/R pairs fall in a subregion of the PN.S L/R FoVs that do not overlap spatially with the Hα image, or there is no positive Hα flux. We note here that our color and morphological selection criteria of [OIII] 5007Å sources identify preferentially HII regions which are *isolated* and *spatially unresolved*. We can then use the results from the survey by Leliévre & Roy (2000) in the NGC 628 outer disc as benchmark to verify number density, luminosity function and spatial distribution for the selected HII regions.

As already reported in Aniyan et al. (2018), the number counts of the 693 [OIII] 5007Å emitters show a steep increase at $m_{5007} \approx 25$. The rapid increase in the number counts at this apparent $m_{5007}$ magnitude is consistent with the contribution to the [OIII] 5007Å emitters from the PN population in the NGC 628 disc, as such apparent magnitude is comparable with that of the PN luminosity function (PNLF) bright cut-off at the distance of NGC 628 (D≈9 Mpc, Aniyan et al. 2018). We thus build the color-magnitude diagram (CMD) where the [OIII] 5007 Å - Hα color is computed from the one second normalized instrumental color $m_{0,[OIII]} - m_{0,H\alpha}$ vs. $m_{5007}$. We note that the instrumental magnitude $m_{0,H\alpha}$ may contain a possible contribution from the 120Å wide continuum at the Hα wavelength, in addition to the Hα and NII line fluxes. Differently, the $m_{0,[OIII]}$ of the selected pairs in the PN.S L/R images does not have any contributions from the adjacent continuum, as these PN pairs are selected to be monochromatic emissions with no underlying continuum streaks at the positions of the emissions. The CMDs for the 623 [OIII] 5007Å detected sources with [OIII] and Hα measured fluxes are shown in Figure 5.

We then matched the sky positions of the PN.S [OIII] 5007 Å emitters with Hα flux to the 102 PNe from the Herrmann & Ciardullo (2009a) sample, with a 2'' tolerance aperture and identify them on the CMD. The spectroscopically confirmed PNe from Herrmann & Ciardullo (2009a) do cluster in a locus of the CDM that is delimited by the following selections:

$$m_{H\alpha} \leq 3.2 \times (m_{5007} - m^{*}) + 1.7 \quad ; \quad m_{5007} \geq m^{*} = 24.75 \quad (5)$$

The first criterion specifies a larger [OIII] 5007Å flux in these sources with respect to the flux of the Hα emission: it effectively maps the Herrmann & Ciardullo (2009a) criterion of the line ratios into PN.S magnitudes for the [OIII] CDI and Hα arm. The second one selects the sources with apparent $m_{5007}$ equal or fainter than the apparent magnitude $m^{*} \approx 24.75$ of the PNLF bright cut-off at the distance of NGC 628. Of the 623 PN.S [OIII] 5007 Å emitters with Hα fluxes and hence measured [OIII] - Hα colour, 372 sources are selected as PNe and 251 sources as HII regions, on the basis of the criteria specified in Eq. 5. The selected PNe are indicated by the green triangles in Figure 5. With the addition of the 70 [OIII] 5007 Å emitters with no Hα emission, the entire PN sample in NGC 628 comprises 442 PN candidates. This PN sample extends the known number of classified PNe in NGC 628 by a factor larger than four.

*Possible [OIII] emissions from SNRs in NGC 628* - In a star forming disc galaxy like NGC 628, [OIII] 5007Å emitters may include SNRs as suggested previously in Herrmann & Ciardullo (2009a) and more recently by Kreckel et al. (2017). As in Aniyan et al. (2018), we looked at the IAU Central Bureau for Astronomical Telegrams (CBAT) List of Supernovae website (http: //www.cbat.eps.harvard.edu/lists/Supernovae.html). There are three known historical supernovae in NGC 628; none of these objects made it into the current sample of [OIII] 5007 Å emitters. An [OIII] 5007Å emission line source was found at a distance of 1".6 from SN 2002ap. However, on applying the colour-magnitude cuts in Eq. 5, this object was classified as HII region. We then focus on the list of SNRs candidates provided by Kreckel et al. (2017), from the IFS MUSE study of several fields in NGC 628. We crossmatched the SNR (ra, dec) positions with those of the 693 [OIII] 5007Å pairs. After several tests, we adopted a search radius of 2", as a larger one would lead to double identifications. We matched 22 sources with the SNR candidates listed in Kreckel et al. (2017). Of these sources, 17 fall in the region of the $m_{0,[OIII]} - m_{0,H\alpha}$ vs. $m_{5007}$ CMD excluded by Eq. 5 (see black pentagons in Fig. 5) or are found at radii < 80"; the remaining five SNRs according to Kreckel et al. (2017) have a matching [OIII] 5007 Å pair in our PN sample in the innermost annual radial bin 80'' to 170".

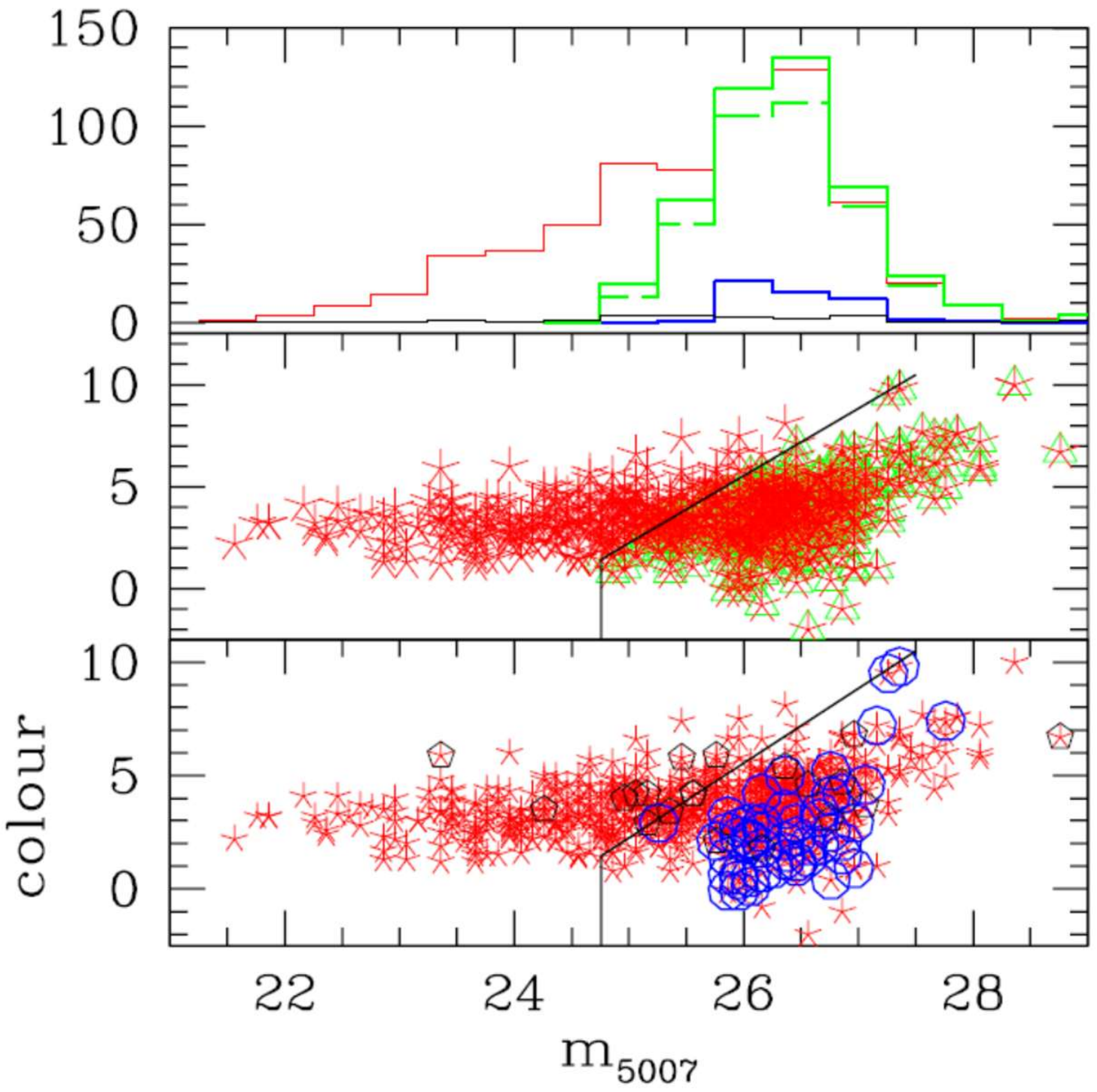

**Figure 5.** Color magnitude diagram ([OIII] - Hα) vs. $m_{5007}$ for the spatially unresolved [OIII] 5007 Å monochromatic emitters with no continuum selected in the PN.S L/R arms + Hα. The continuum black line shows the selection criteria for the PNs vs HII regions expressed in Eq. 5. In the lower panel, blue open circles indicate the PN.S [OIII] emitters matched with the Herrmann & Ciardullo (2009a) spectroscopically confirmed PN sample in NGC 628; black pentagons indicate the PN.S [OIII] emitters matched with SNR candidates from Kreckel et al. (2017). In the middle panel, green triangles indicate the selected PNe in NGC 628 according to Eq. 5. The spatially un-resolved-monochromatic [OIII] emitters with no continuum and no Hα flux in a 2" aperture centered on the [OIII] sources are not shown in these CMDs. The uppermost panel shows the number counts for all the [OIII] 5007Å sources (red histogram), the selected PNe (green dashed histogram, including those with no detected Hα flux with the green continuum line histogram), and the PN matched with Herrmann & Ciardullo (2009a; blue histogram). The number counts per magnitude bin of PN.S [OIII] pairs matched with the Kreckel et al. (2017) SNR candidates are also shown (black histogram).

*Further estimate on possible contaminations by Supernovae (SN) and SNRs* – Possible source of contaminants of any [OIII] 5007 Å selected emission line samples are historical SN. According to Aniyan et al. (2018) there are three historical supernovae in NGC 628, and none of them is included in the NGC 628 selected PN.S sample. Kreckel et al. (2017) identifies SNRs from [OIII] 5007 Å emissions in the MUSE observations of the luminous central regions in NGC 628. As it is discussed in their paper, the [OIII] 5007 Å fluxes from these sources are "bright", and if mis-classified, might affect the estimates of the bright cut-off of the PNLF for a sample PNe selected only on the basis of their [OIII] 5007Å emissions. In Figure 5, the pentagon symbols indicate the matched [OIII] PN.S pairs to the Kreckel et al. (2017) SNR candidates. A sizeable fraction (17/22 or 77%) of this SNR sample is rejected as PN candidates on the basis of our CMD selection, using the [OIII] – Hα color. The remaining five which are classified SNRs from Kreckel et al. (2017) are included in our PN sample in the radial range 80" – 170".

We can then address what the estimate of the fraction of SNRs in extragalactic PN samples would be in other strong star forming galaxies, when the latter are selected from their [OIII] 5007Å emissions only. We proceed from the results of the extended PN survey in the Large Magellanic Cloud by Reid & Parker (2013) who selected 750 PNe. The survey of SNRs in LMC is presented in Maggi et al. (2016); these authors report that there is a total of 59 confirmed SNRs in the LMC, based on their X-ray emissions. These SNRs may contaminate an [OIII] 5007 Å emission only-PN selected sample, if no additional line diagnostic is applied, as discussed in Kreckel et al. (2017).

While the LMC and NGC 628 are galaxies of different morphological types; their optical colour, average oxygen abundance values and radial gradients (in the radial ranges covered by the PN.S survey in NGC 628) are similar though. LMC is one of the very few galaxies for which there is a deep PN survey (with few hundreds PNe) as well as SNR survey, that can be used to simulate a mock to estimate contamination by SNR for M74 OIII/Hα sample.

Any [OIII] 5007Å narrow band survey would include a morphological selection, as the extragalactic PN candidates must be spatially unresolved in natural seeing conditions. The angular size of an unresolved [OIII] 5007Å emitter in the PN.S CDI, under natural seeing (1"), corresponds to 43 pc diameter in NGC 628 (M74). At the distance of the LMC 1''=0.24 pc, hence any unresolved [OIII] emission in M74 would have a diameter of ≤180" at the LMC distance. From the X-ray radius, there are only 13 SNR in LMC whose radius < 100", and about 35 with radius <200". By applying our morphological criteria to the LMC SNRs, only those SNRs with radii < 100" may contaminate a PN sample selected from the [OIII] 5007Å emission and morphological criteria only, at 9 Mpc distance. This is an upper limit, as we are not applying the CMD criterion from the additional Hα emissions to the SNRs in the LMC. On the basis of the LMC PN vs SNR, then about $N_{SNR}/N_{PN} \approx 13/750 = 1.7\%$ of the total PN sample would be misclassified SNRs, at 9 Mpc distance, on the basis of morphological (spatial) criteria applied to the [OIII] 5007 Å emission only.

## 4. Luminosity functions and spatial distributions of PNe and HII regions in NGC 628

Having selected PNe and HII regions on the basis of their instrumental color $m_{0,[OIII]} - m_{0,H\alpha}$ and $m_{5007}$ magnitudes, we now validate their classification by using their LFs and spatial distributions against independent measurements. While the PN.S has been devised to carry out kinematics measurements for discrete sources, it measures consistent magnitudes to PN literature data sets with a dispersion of $\sigma_{m5007} = 0.13$ mag, see Section 2.2 and reference therein. Regarding the spatial completeness: in the case of NGC 628, the dynamical study carried out by Anyan et al. (2018) showed already that the PN sample in the three radial bins investigated in this work is photometrically complete, e.g. PN number density follows the surface brightness profile in V band, as shown by the comparison of their determined dynamical scale length values ($h_{dyn}$, their Figure 17) vs. that computed photometrically by Möllenhoff (2004).

### 4.1. PNLF and HII number counts

Figure 6 shows the HII number counts and PNLF in the three radial bins that have more than 130 candidates in each bin. We describe their properties in turn. In these three radial bins, the HII number counts are nearly flat vs. the apparent $m_{5007}$ in the magnitude range 23 – 26, with a turn-off at $m_{5007} > 26$. We can compare them with the luminosity function (LF) measurements of singularly detected HII regions in the NGC 628 disc by Leliévre & Roy (2000). The HII LF measured by Leliévre & Roy (2000) has a turnover at about log L(Hα )= 37.0 with $\gamma = 0.2 \pm 0.3$, which is consistent with a flat LF. The apparent magnitude of the bright cut-off of the PNLF at the distance of NGC 628, $m_{5007} \approx 25.00$ does correspond to a log L(Hα) value log L(Hα) = 37.5 for a [OIII]/Hα ratio = 1.00[2] (see [OIII]/Hα ratio for the unresolved HII region in the Virgo cluster in Gerhard et al. (2002)). Such a flat LF implies that the number of singularly detected, e.g. isolated, and unresolved HII regions are either constant or decreasing at apparent fluxes $m_{5007} \approx 25.0$. This prediction from the HII LF from Leliévre & Roy (2000) is consistent with the number counts of our isolated spatially unresolved HII regions in the three radial bins in NGC 628.

The spatially unresolved monochromatic [OIII] 5007 Å pairs that are detected in the PN.S L/R arm images at magnitudes $m_{5007} > 25$ show a rapid increase instead. The number counts reach a peak at $m_{5007}$ in the range 26.25-26.5, to be then followed by a decline at fainter magnitudes. Such rise at magnitudes fainter than 25.0 is associated with the population of the PN in the NGC 628. Once we select the *bona fide* PNe using Eq. 5 from the PN.S L/R pairs and Hα sources, their number counts, e.g. PNLFs, in the three radial bins are shown in Figure 6.

[2] For typical HII regions, such ratio is < 1.0, which would lead to [OIII] 5007Å fluxes brighter than corresponding PNLF cut-off $m^*_{5007}$ in NGC 628. Hence the corresponding [OIII] 5007Å sources would not be included in the PN candidate sample for NGC 628.

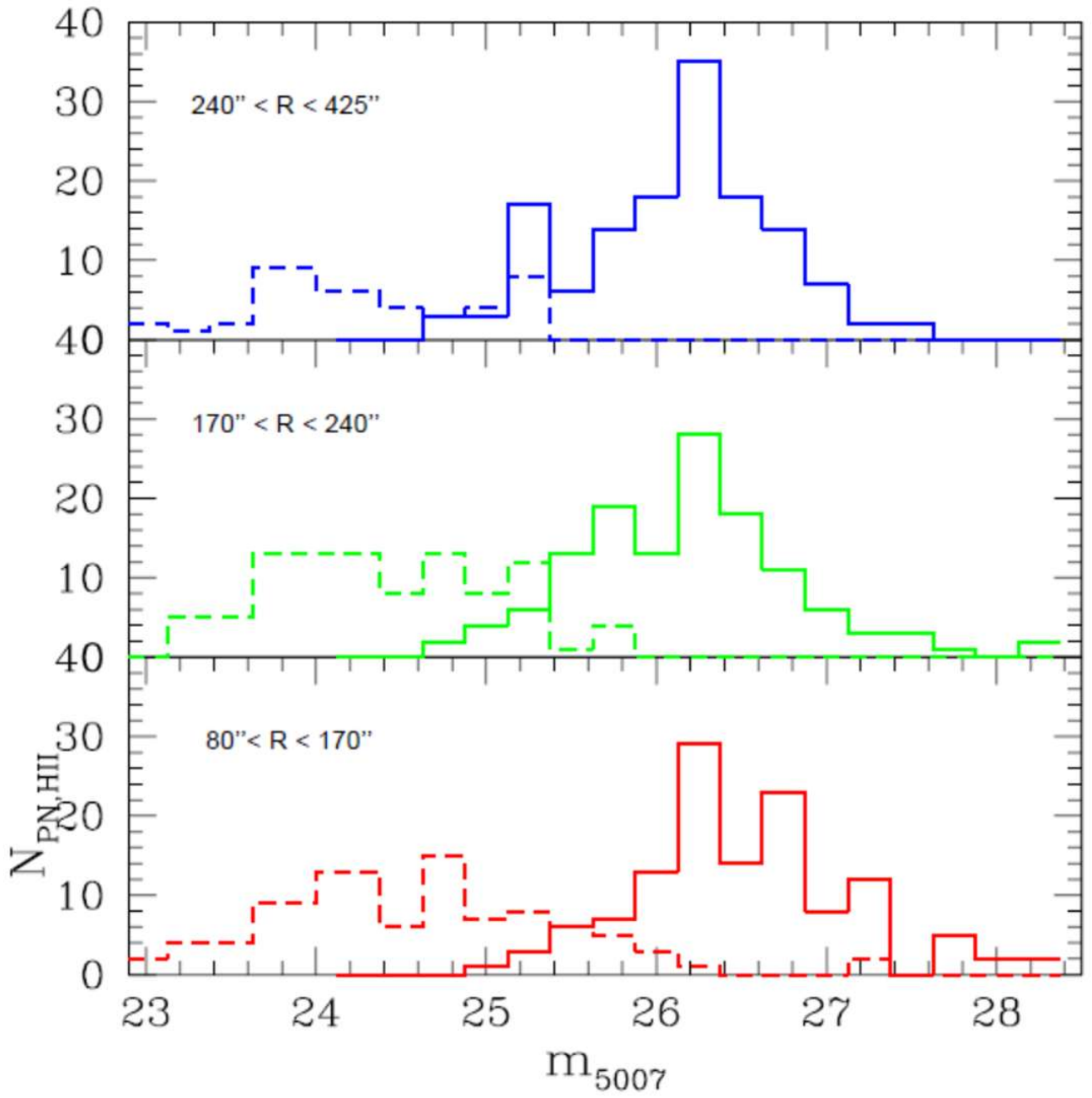

**Figure 6.** Number counts of the isolated, spatially unresolved HII regions (dashed line histograms) and the PNLF of the selected PN candidates in three radial bins (continuum line histogram). The radial bins are selected to have more than 130 PN candidates in each bin to allow for the kinematical decomposition of the LOSVD as in Aniyan et al. (2018). The PNLFs show a clear rise at $m_{5007} \sim 25$ and a decline at magnitude ~ 27. The number counts of isolated, spatially unresolved HII regions are flat in the magnitude range 23 - 26 mag, similarly to what measured by Leliévre & Roy (2000) for isolated HII regions in NGC 628.

## 4.2. Radial number density profiles of PNe and HII regions

We then compare the number density profiles of PNe and HII regions with the surface brightness profiles in the V-band for the stars in the NGC 628 disc. The azimuthally averaged Hα surface brightness as function of galactocentric distance traces the star-formation rate in discs (Kennicutt 1998), while PNe trace light instead. Hence the radial profiles of these discrete tracers may show different behaviours in those discs where there is star formation cut-off at an outer radius, as it occurs in NGC 628. The average Hα surface brightness in NGC 628 shows a cut-off at a galactocentric distance $R \sim R_{25} \approx 312''$ (Figure 3 in Leliévre & Roy 2000), while the light follows an exponential profile at these outer regions. Thus one expects different spatial gradients for PNe/HII regions at these radii, with lower surface number density values for HII regions.

In Figure 7, we show the surface brightness data points in the V-band from Cornett et al. (1994) and compare them with the suitably rescaled surface number density profiles for PNe and HII regions. The offset values are computed independently in the innermost radial bin (80" – 170") for each sample. For the PN sample, the scaling factor in the inner radial bin corresponds to a luminosity specific PN number value of $\alpha_{1.5} = N_{1.5}/L_{bol} = 2.26 \times 10^{-8}$ PN $L^{-1}_{\odot}$ , where $N_{1.5}$ is the number of PNe within 1.5 mag from the PNLF bright cut-off in that radial bin, divided by the bolometric luminosity in the same bin[3] . For the HII regions, the surface number density is scaled to the surface brightness in the same bin; the conversion factor is then kept fixed for the two outer radial bins. The three PN surface number density values follow the surface brightness profile from the exponential disc in NGC 628, as expected from the hypothesis that PN trace the light of their parent stellar population (Buzzoni et al. 2006). At $R \sim R_{25}$ in NGC 628, the outermost PN number density value follows the exponential profile with $\mu_v \sim 21$ mag arcsec$^{-2}$ and $R_h = 73''.4 \pm 3''.7$ (Möllenhoff (2004); see dashed blue line in Fig. 7). Differently, the values for the individual HII regions, selected from their spatially unresolved [OIII] 5007 Å emissions, present a hint of a cut-off in the outermost radial bin. Such a deviation from the exponential profile is in agreement with a sharp decrease of the azimuthally averaged Hα surface brightness Σ(Hα) reported by Leliévre & Roy (2000) at $R \sim R_{25} \approx 312''$ distance.

[3] The luminosity in the bin is estimated from the V band surface brightness profile. The bolometric correction from the optical V-band is applied following the approach illustrated in Buzzoni et al. (2006).

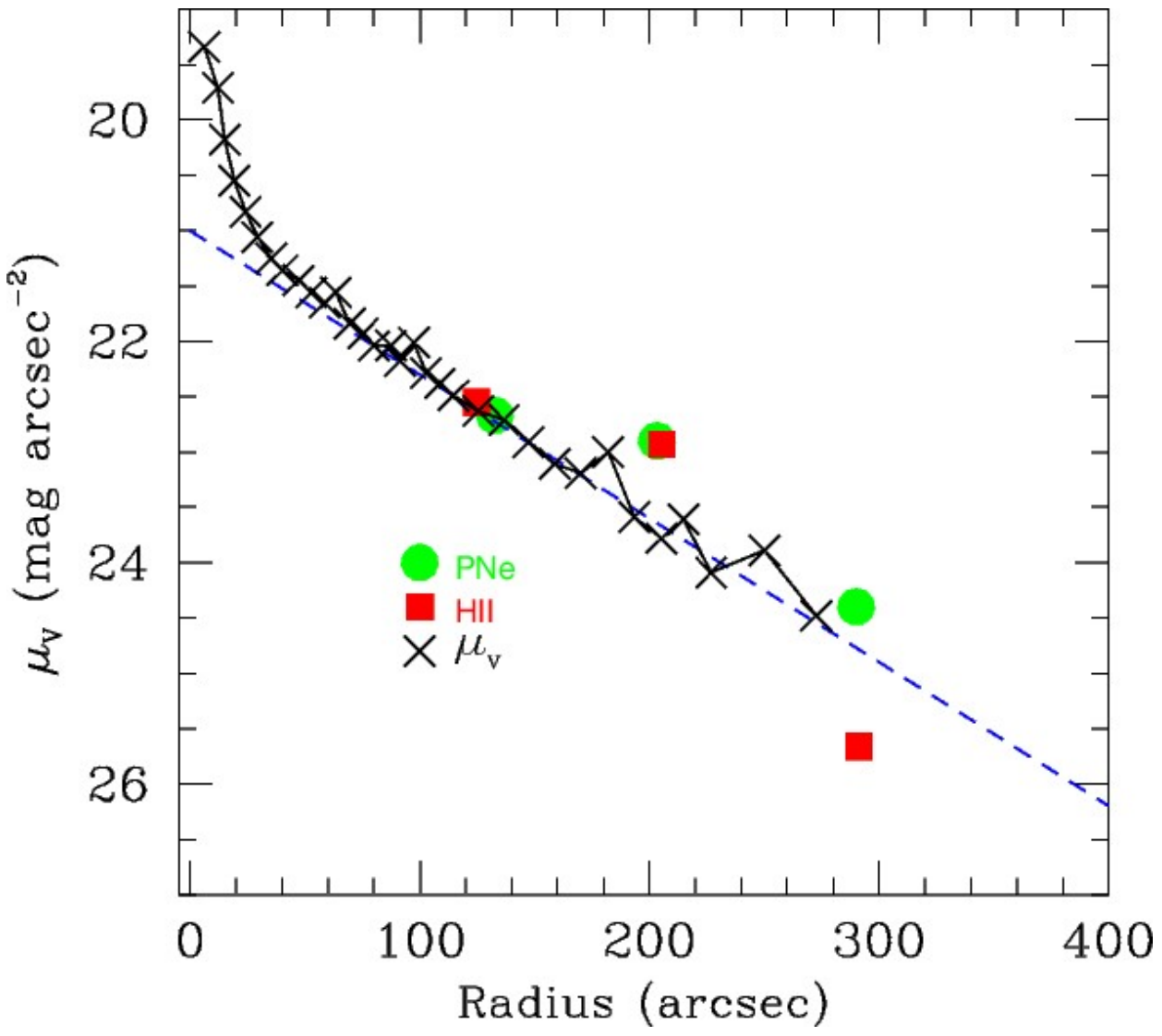

**Figure 7**. Comparison of the NGC 628 surface brightness profile in V band from Cornett et al. (1994) vs. number density values in three radial bins from color selected unresolved HII regions ($\Sigma(N_{H\alpha})$ – full red squares) and PN candidates ($\Sigma(N_{PNe})$ - green full circles). The last point for the HII number density shows a decrease at a radius of ≈ 300" equivalent to $R_{25}$ for this disc galaxy. The blue dashed line shows the exponential profile fit to the NGC 628 surface brightness in V-band (Möllenhoff 2004).

## 5. Distinct velocity distributions for the nebulae populations in the nearly face-on disc galaxy NGC 628

NGC 628 is a nearly face-on disc galaxy whose inclination is estimated to be i = 8°.5 ± 0°.2 (Walter et al. 2008). Such an inclination leads to a measurable overall rotation field in the disc for any tracers as cold as the HI; see the HI velocity field for NGC 628 in Walter et al. (2008) for reference. As the HII regions are tracers of the young massive populations of stars, their LOSVD over the entire disc should display a LOSVD similar to that measured for the integrated 21 cm emission of the neutral atomic HI gas over the same area. Any tracers associated with older stellar populations would be affected by the secular evolution of the disc (Sellwood 2014) or accretions events (Quinn & Goodman 1986, Hopkins et al. 2008). As a consequence, these dynamically older populations would lag behind the rotation of the HI gas and show larger velocity dispersions, see for example the rotational lag of the PNe sample in the disc of Andromeda (Merrett et al. 2006). The effects of secular evolution and/or accretion events in a disc leave imprints like those measured for the age-velocity dispersion relation (AVDR) for the intermediate/older starts in the MW disc (Casagrande et al. 2011; Aniyan et al. 2016), and in the stellar disc of the Andromeda galaxy (Dorman et al. 2015; Bhattacharya et al. 2019). We can thus further corroborate our PN.S selection criteria for PNe and unresolved HII regions by computing their LOSVD over the surveyed region over most of the NGC 628 disc and quantify their residual velocities from that of the HI at the tracers' positions in the disc.

In Figure 8-a, we show the LOSVD for the CMD selected PNe and HII regions respectively and describe them in turn. The LOSVD for the HII regions is characterized by a distribution with a suggestive double peak, which one would associate with the approaching/receding component of maximum $V_{ROT}$ along the line-of-sight. The confirmed presence of a double-peaked distribution would indicate a population of tracers that is dynamically cold and similar to the HI gaseous disc. For the selected PNe, Figure 8-a shows two histograms, one computed for the color selected PNe and a second one for the PNe with no Hα emission. Both LOSVDs are broader than the HII distribution, with wings, and therefore suggestive of populations of tracers which are kinematically "warmer" than the HII regions.

The top flat, possibly double-peaked, LOSVD for the HII regions in NGC 628 is suggestive of a global rotation on top of the distribution of velocities along the line-of-sight. As we are interested in the velocity dispersion component *orthogonal* to the disc plane, then one must account for any residual rotation along the line-of-sight by subtracting it. For each PN/HII region, we then measure the local HI velocity from the THINGS data (Walter et al. 2008) at their position and then compute the residual velocity $\Delta V_{PN\&HII,LOS}=V_{PN\&HII,LOS}-V_{HI}$ for any PN or HII discrete tracer with respect to the HI velocity at the tracer's position in the NGC 628 disc.

*Residual velocities of discrete tracers* – We then compute the residual velocities for each PN (both colour selected and those with no Hα emission alike) and HII region with respect to the HI gas at the tracer's position in the disc. In Figure 8-b we show the histograms of the $\Delta V_{PN\&HII,LOS}=V_{PN\&HII,LOS}-V_{HI}$ values for PNe and HII samples. One can visually assess that the PN $\Delta V_{PN,LOS}$ distribution is broader than that for the HII regions. The distribution of $\Delta V_{PN,LOS}=V_{PN,LOS}-V_{HI}$ is characterized by a mean value $\Delta V_{PN,mean}$ = 0.21 kms$^{-1}$, that is very close to zero within the velocity errors, and a standard deviation of $RMS_{\Delta VPN\&HII,LOS}$ = 25.21 kms$^{-1}$. The distribution of residual velocities for the HII regions $\Delta V_{HII,LOS}=V_{HII,LOS}-V_{HI}$ has a mean value at $\Delta V_{HII,mean}$=-1.092 kms$^{-1}$ and standard deviation $RMS_{\Delta VHII;LOS}$ = 21.63 kms$^{-1}$. Overall the HII line-of-sight $\Delta V_{HII,LOS}$ is somewhat narrower than the corresponding one for PNe, with a possible sharp transition to flat wings, out to ±100 kms$^{-1}$, with respect to the peak of the distribution.

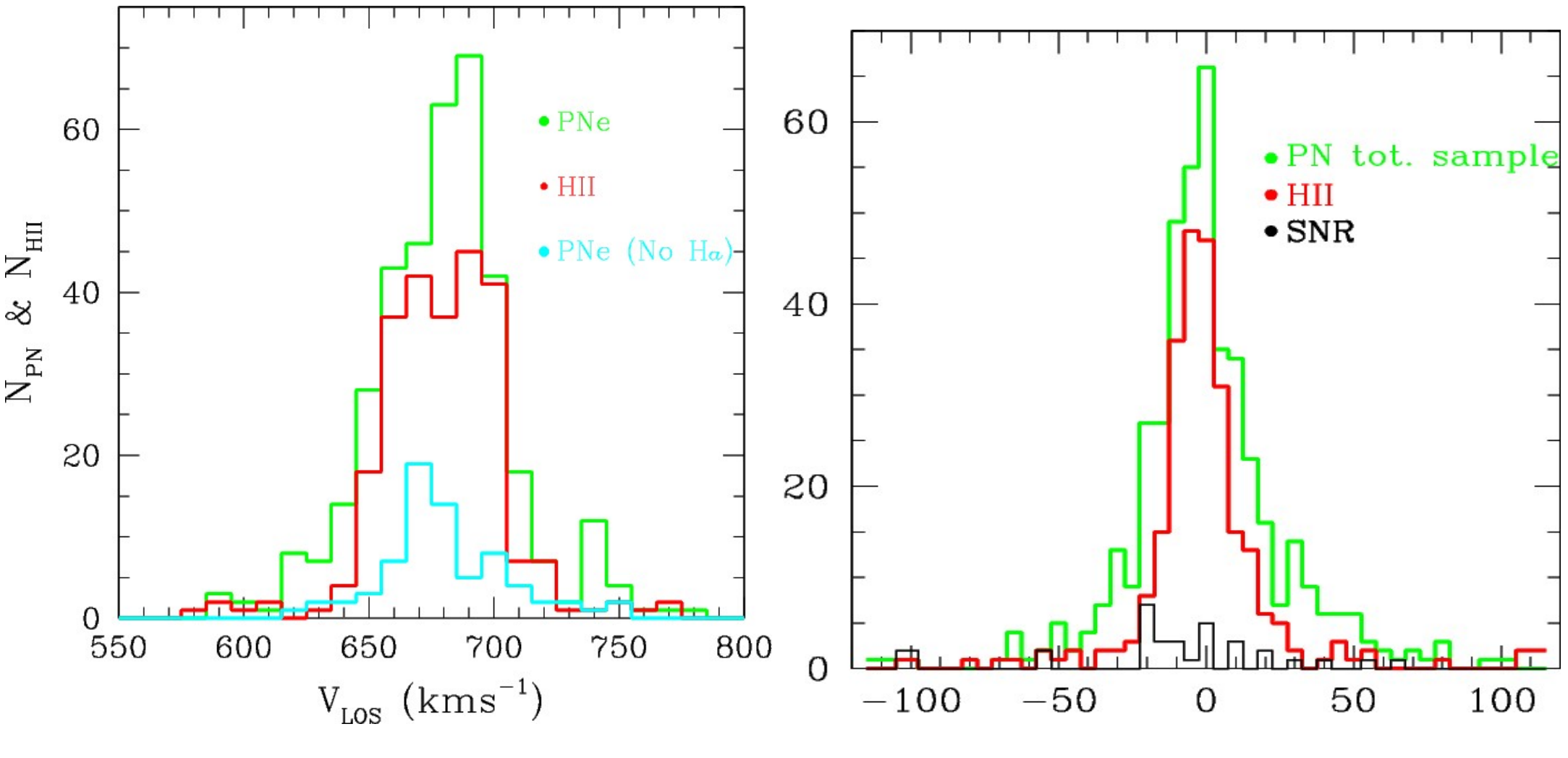

(a) (b)

**Figure 8. (a) Left panel -** Histogram of the line-of-sight velocity $V_{LOS}$ distribution for the color selected PN candidates (green) and the HII regions (red), plus the PN candidates with no Hα emission (cyan). The $V_{LOS}$ is measured from the Doppler shifted $\lambda_{obs}$ with heliocentric correction measured from the unresolved [OIII] emission with the PN.S. **(b) Right panel -** Histogram of the difference between the line-of-sight velocity $V_{LOS}$ of the spatially unresolved, monochromatic [OIII] emitters and the velocity of the HI 21 cm emission at the position of the [OIII] sources, $\Delta V_{PN\&HII}$=$V_{PNe/HII,LOS}$ – $V_{HI}$ . The three histograms show the values for the CMD selected PNe (in green; current histogram includes also the PNe with no Hα emission), for the HII sources (red) and the matched 22 SNR candidates (black) from Kreckel et al. (2017).

*Origin of the flat wings in the* $\Delta V_{HII,LOS}$ –To investigate the origin of the flat wing component in the $\Delta V_{HII,LOS}$ distribution one looks first at the histogram of the $\Delta V_{SNR,LOS}$=$V_{SNR,LOS}$–$V_{HI}$ for the SNRs identified by Kreckel et al. (2017). This histogram is shown in black in Figure 8-b. The histogram is computed from the $V_{LOS}$ velocities measured for the [OIII] pairs whose sky positions are matched with the SNR positions listed in Table 2 from Kreckel et al. (2017), with a tolerance of 2”. The histogram of the SNR $\Delta V_{SNR,LOS}$ is also flat and overlaps in velocities with the flat wings of the histogram of the $\Delta V_{HII,LOS}$ of the HII regions, in the range from -100 to +100 kms$^{-1}$. Clearly such a flat distribution differs from that of a single peaked distribution of velocities expected for an isothermal population of gravitationally bound stars to a massive thin disc (see Appendix A in Aniyan et al. 2018). A flat distribution of velocities can be powered by ISM related phenomena instead, like SN induced shocks or star formation outflows.

Following our CMD selection in Eq. 5, only five SNRs from the Kreckel et al. (2017) are included in the selected PN candidate sample, all in the innermost radial bin (80” – 170”). In Figure 9 we show the histogram of the residual velocities for the PN.S PNe and of the five SNRs in the inner radial bin. In NGC 628, the lower limit to the fraction of misclassified SNRs as PNe is then 5/443 ~ 1.1%, which is about half of the estimated upper limit faction $N_{SNR}/N_{PN}$ ~ 2% from the LMC arguments illustrated in Section 3.

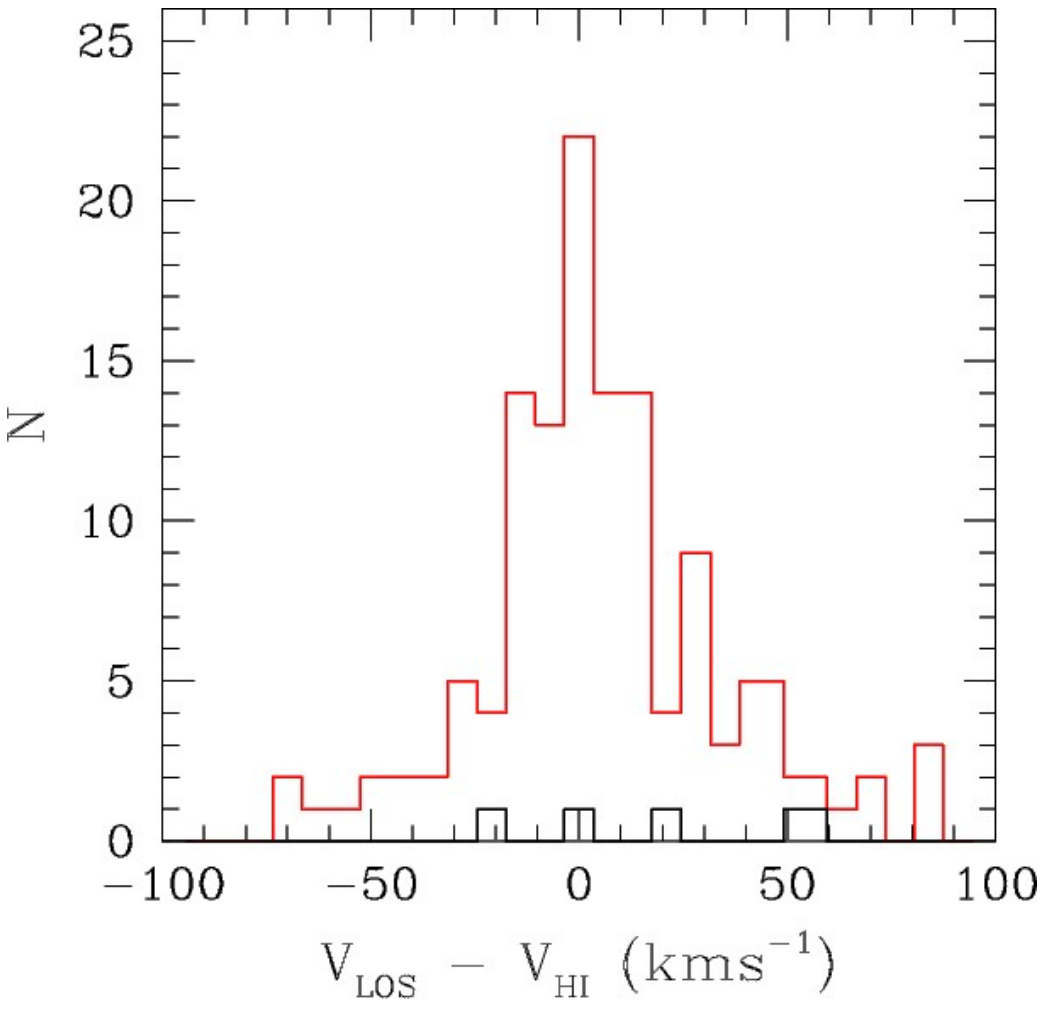

**Fig. 9.** Histogram of the $\Delta V$ = $V_{LOS}$ – $V_{HI}$ for the PN candidates in the inner radial bin 80” < $r$ < 170” (red continuum line) vs the histogram of the $\Delta V$ of the SNR candidates (black continuum line**)**. There are only 5 matched SNRs in this radial interval. The majority of SNRs in Kreckel et al.

(2017) sample matched with PN.S [OIII] 5007 Å pairs are either at smaller radii or are classified as HII regions on the basis of the color criterion (Eq. 5).

*Radial dependance of* $\sigma_{z,PNe}$ *and* $\sigma_{z,HII}$ - The tables in the Appendix list the coordinates of each PN and HII region candidate, their heliocentric LOS velocity and [OIII] 5007 Å PN.S magnitude, including their residual velocity $\Delta V_{PN/HII,LOS} = V_{PN/HII,LOS} - V_{HI}$ vs. the HI 21 cm line at each PN/HII region candidate's sky position; see Appendix A and B. As shown in Figure 9 from Aniyan et al. (2018), in the velocity versus azimuthal angle plots in three radial bins, with 130 PNe each displays a cold population of spatially unresolved emission line objects, plus a hotter population whose velocity dispersion appears to decrease with galactocentric radius.

## 6. PNe as tracers of the kinematically colder and warmer stellar population in the disc of NGC 628

On the basis of the quantitative evidence provided thus far - spatial distributions, number counts and kinematics – we showed that the selection criteria in Eq. 5 identify a PN sample which is distinct from the HII regions in the disc of NGC 628, with an upper limit of 2% contamination by SNRs based on the LMC arguments **(**see Section 3**).** We showed that the PNe have larger velocity dispersion $RMS_{\Delta VPN;LOS}$ than that of the HII regions, with the PN population being progressively dominated by rotation with increasing radius (Aniyan et al. 2018).

As a next step, we wish to explore whether there is any evidence for distinct PN sub-populations in the NGC 628 PN sample. These sub-populations are associated with stellar progenitors of different masses and ages. Peimbert & Torres-Peimbert (1983) classified Galactic PNe in sub-populations, whose chemistry is known to correlate with their velocities and location in the Galactic plane (Stanghellini & Haywood 2018); see also the results of chemically distinct thin and thicker disc PNe in M31 (Bhattacharya et al. 2022; Arnaboldi et al. 2022). Recent theoretical work by Miller Bertolami (2016) on the Post AGB evolution made the case that the lifetimes in the PN phase are very strongly dependent on their progenitor masses, with PN evolving from massive progenitors having very short lives. If this is the case, the kinematics of PN tracers from massive progenitors may not have been modified significantly from that of the HI neutral gaseous disc. In this framework, PNe from massive progenitors may be kinematically cold, with dispersion similar to that of the HI gas, while PNe from less massive progenitors may have larger dispersions, because of the effect of secular evolution on the kinematics of their "older" progenitors; see the results on the AVDR for PN populations of different age progenitors in the M31 disc (Bhattacharya et al. 2019).

PNe at the PNLF bright cut-off are powered by cores that at are least 0.6 $M_{\odot}$ (Ciardullo 2010); given the initial-to-final mass relation, these progenitors are at least 2 $M_{\odot}$ on the Zero-age main sequence and would evolve from a 1 Gyr old or younger stellar population, see Miller Bertolami (2016) and Marigo et al. (2004). As the nebula expand, the measured $m_{5007}$ gets fainter (Henize & Westerlund 1963). At magnitudes fainter than the bright cut-off, there would be a contribution from fading bright PNe, whose disc kinematics is cold, and from PNe evolved from less massive progenitors, whose disc kinematics would be warmer depending on the age of their progenitors[4]. Hence the distribution of the residual $\Delta V_{PN;LOS}$ over a wide radial range in the face-on disc of NGC 628 is expected to be better described by superposition of Gaussians, depending on the fractional contribution from the younger/most massive vs. the older/less massive progenitors to the overall PN population, to be driven by the star formation history. Given that NGC 628 is nearly face-on, the distribution of the $\Delta V_{PN;LOS}$ values as function of the $m_{5007}$ magnitude of the PNe would vary in the sense that "bright, younger progenitor" PNe would display smaller residuals from the HI velocities, i.e smaller values of $\Delta V_{PN;LOS}$ , while "fainter, older progenitor" PNe would have larger residuals, even once one accounts for the larger velocity errors at fainter fluxes.

In Figure 10 we show the distribution of the measured $\Delta V_{PN,LOS}=V_{PN,LOS}-V_{HI}$ for all NGC 628 PNe, with measured Hα emission (green symbols) and those with none (red symbols) separately. Table 1 provides the measurements of the RMS of the $\Delta V_{PN,LOS}$ distribution in magnitude bins for the two samples (second and third column) and of the combined sample (fourth column), the $\sigma_{err}$ for the PN.S velocity measurements (fifth column) and the derived $\sigma_z$ values in NGC 628 measured from PNe subsamples in the different magnitude bins (sixth column); $\sigma_z$ is computed from the quadratic difference $\sigma_z = \sigma_{LOS} - \sigma_{err}$**.** The NGC 628 PNe, with or without Hα emissions, have the same $\Delta V_{PN,LOS}$ as function of magnitude, within the errors. Figure 10 and Table 1 show clearly that the velocity dispersion $\sigma_z$ is a function of the $m_{5007}$ magnitude, with the PNe at the bright cut-off in a 0.5 mag bin being significantly kinematically colder, by a factor ~ 2, than those at Δm ~ 1.0 -1.5 mag fainter than the PNLF bright cut-off magnitude.

---

[4] At fainter magnitudes, there is an effect from larger measurements errors also, that we estimated in Aniyan et al. (2018) and can be accounted for.

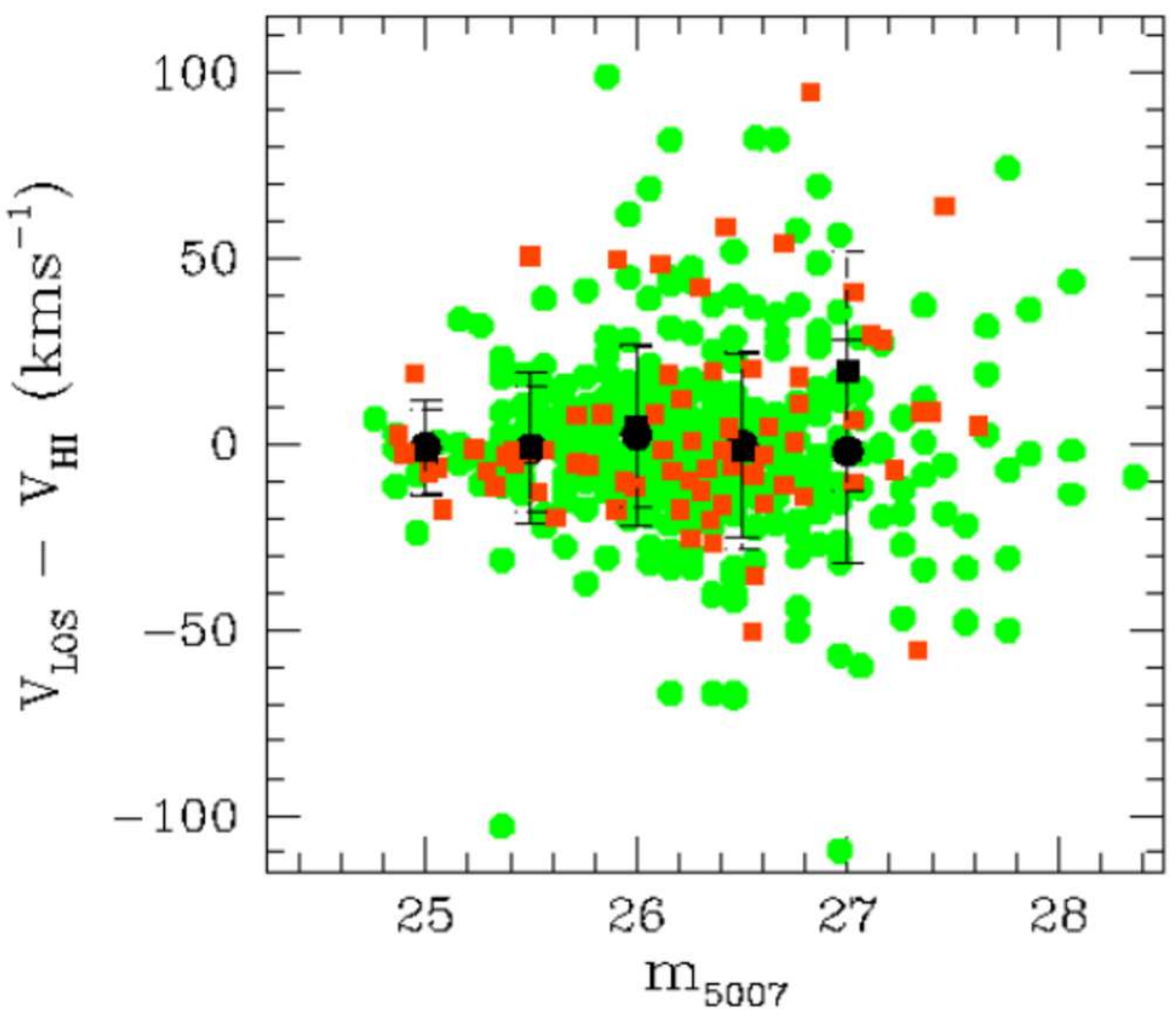

**Fig. 10.** $\Delta V_{PN,LOS}$ vs. $m_{5007}$ distribution for the PN candidates in NGC 628, with detected Hα emission (**green full dots**) and those with none (**red full squares**). The two samples have the same kinematics as function of $m_{5007}$, see Table 1. The full black circle and squared symbols with error-bars in the plot display the average $\Delta V_{PNe,LOS}$ values in magnitude bins of 0.5 mag, the error bars indicate the $\pm 1\sigma$ values, for the PNe with Hα/no Hα emission respectively. The magnitude bin of 0.5 mag is $\approx 3$ x $\sigma_{m5007}$, where $\sigma_{m5007} = 0.13$ mag is the PN.S photometric uncertainty.

**Table 1.** $\sigma_{los}$ vs. $m_{5007}$ in 0.5 mag bins for the PN samples (with and without Hα emission) in NGC 628. The adopted magnitude bin corresponds to ~ 3 times the standard photometric uncertainty of the PN.S (0.13 mag). In column 2 and 3 we report the $\sigma_{LOS}$ values computed as RMS of the $\Delta V_{PN;LOS}$ PN samples (either with or without Hα emission) in magnitude bins. The fourth column reports the $\sigma_{LOS}$ of the $\Delta V_{PN;LOS}$ for the entire PN sample (no distinction applied on the bases of the Hα emission) in each magnitude bin. The fifth column reports the PN.S $\sigma_{err}$ as function of magnitude, and the sixth columns reports the $\sigma_z$ computed from the quadratic difference $\sigma_z = \sigma_{LOS} - \sigma_{err}$.

| $m_{5007}$ | w/ Hα $\sigma_{LOS}$ kms$^{-1}$ | No Hα $\sigma_{LOS}$ kms$^{-1}$ | Total $\sigma_{LOS}$ kms$^{-1}$ | PN.S $\sigma_{err}$ kms$^{-1}$ | NGC 628 $\sigma_z$ kms$^{-1}$ |
|---|---|---|---|---|---|
| 25.0 | 12.90 | 11.25 | 12.06 | 4.0 | 11.4 |
| 25.5 | 20.42 | 16.88 | 19.62 | 4.9 | 19.0 |
| 26.0 | 24.48 | 21.83 | 24.11 | 5.8 | 23.4 |
| 26.5 | 24.54 | 26.46 | 24.78 | 7.3 | 23.7 |
| 27.0 | 30.16 | 32.06 | 31.16 | 8.8 | 29.9 |

## 6.1 Kinematically cold PNe at the PNLF bright cut-off. Are they distinct from the HII population at similar $m_{5007}$ mags ?

Both Figure 10 and 11 show that the kinematics of the PNe within 0.5 mag of the bright cut-off is cold, i.e. the RMS of the $\Delta V_{PN,LOS}$ is 12.06 kms$^{-1}$. It provides the value of the $\sigma_{LOS} = \sigma_z$ for the nearly face-on disk of NGC 628. As we compute the RMS for fainter magnitude bins, it increases progressively to about 24.78 kms$^{-1}$ at 1.5 mag fainter than the bright cut-off of the PNLF, e.g. the population becomes increasingly ``warmer". The RMS values of $\Delta V_{PN,LOS}$ distributions in 0.5 mag bins are listed in Table 1, together with estimated error in that magnitude bin, and the derived $\sigma_z$ value for that population in NGC 628.

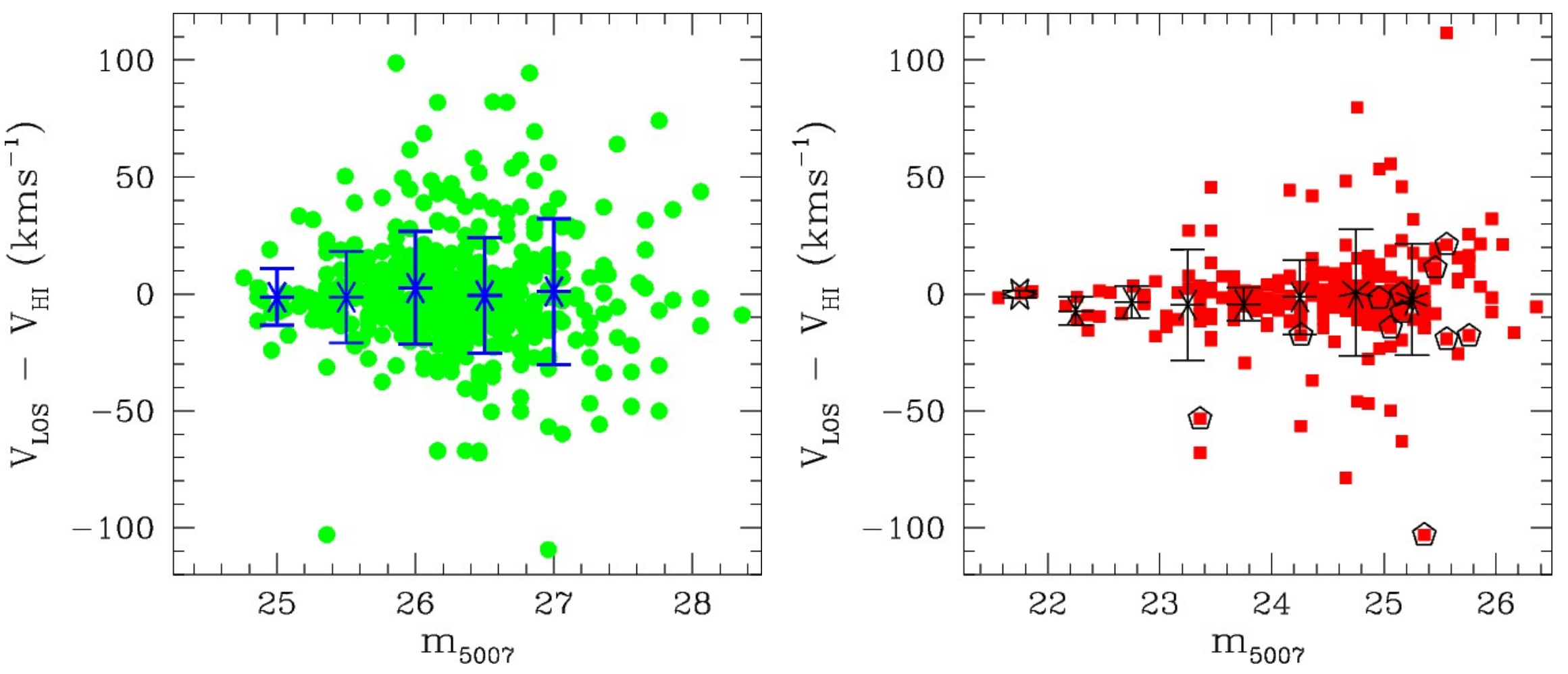

**Figure 11**. a) Distribution of $\Delta V_{PN,LOS}$ values for the PNe as function of $m_{5007}$, green full dots. Blue asterisks indicate average values and $\pm 1\sigma$ blue error bars in magnitude bins . The RMS values of the $\Delta V_{PN,LOS}$ distribution in bins of magnitude increase with $m_{5007}$, consistent with the presence of a kinematically cold PN population at the bright cut-off and a warmer PN population dominating at fainter mags. b) Distribution of the $\Delta V_{HII,LOS}$ value for HII region candidates as function of $m_{5007}$, red full squares. Black asterisks indicate average values and $\pm 1\sigma$ black error bars in magnitude bins. The RMS values of the $\Delta V_{HII,LOS}$ distribution are invariant up to $m_{5007} \sim 24.25$ and then increase in the fainter bins. In the magnitude bins which overlap with the apparent magnitude of the PNLF bright cut-off in NGC 628, the RMS values of the $\Delta V_{HII,LOS}$ distribution are larger than that for $\Delta V_{PN,LOS}$ in the same magnitude range. Black pentagons indicate the matched HII region sources with the catalog of SNRs from Kreckel et al. (2017), in the radial range 80" – 425".

**Table 2.** $\sigma_{los}$ vs. $m_{5007}$ in 0.5 mag bins for the HII candidate sample in NGC 628. The adopted magnitude bin corresponds to ~ 3 times the standard photometric error of the PN.S (0.13 mag). In the second column we report the $\sigma_{LOS}$ values for the HII regions computed as RMS of the $\Delta V_{HII;LOS}$ in each magnitude bin. The third column reports the PN.S $\sigma_{err}$ as function of magnitude.

| $m_{5007}$ | NGC 628 $\sigma_{LOS}$ kms$^{-1}$ | PN.S $\sigma_{err}$ kms$^{-1}$ |
|---|---|---|
| 22.25 | 5.9 | 2.0 |
| 22.75 | 6.8 | 2.0 |
| 23.25 | 23.7 | 3.0 |
| 23.75 | 7.1 | 3.0 |
| 24.25 | 15.8 | 3.0 |
| 24.75 | 27.0 | 4.0 |
| 25.25 | 23.9 | 4.0 |

We wish to address the question whether the PNe within 0.5 mag of the PNLF bright cut-off are misclassified HII regions. We look at the $\Delta V_{HII,LOS}$ vs. $m_{5007}$ distribution shown in Figure 11 and compute the RMS of this distribution in 0.5 magnitude bins. The values are listed in Table 2. At the apparent magnitude corresponding to the PNLF bright cut-off, the RMS value of the $\Delta V_{HII,LOS}$ distribution in 0.5 mag bin is $\approx 25.5$ kms$^{-1}$ which is twice that of the corresponding quantity for the PN population at the same $m_{5007}$. Hence the PNe at the bright cut-off cannot be a sub-sample of the HII regions (randomly extracted at similar apparent $m_{5007}$) because they are significantly kinematically colder. Differently, the HII regions at $m_{5007} = 24.75 – 25.0$, with larger $\Delta V_{HII,LOS}$ , are located preferentially in the inner regions of NGC 628, where the SFR is more intense (Leliévre & Roy 2000). Under such conditions, the related outflows in the ISM are powerful enough to leave imprints in the $\Delta V_{HII,LOS}$. The small $\sigma_z$ of the brightest PNe in our sample also rules out possible contamination from SNRs at these magnitudes, on the basis of the flat and broader $\Delta V_{SNR,LOS}$ of the latter, as shown in Figure 9 and also in Figure 11.

## 6.2. Kinematically cold and warm PN populations in NGC 628: their contribution to the LOSVD and the PNLF

As addressed extensively in the analysis by Aniyan et al. (2018), the PN population in NGC 628 becomes increasingly rotationally supported, i.e. kinematically *cold*, with radius as shown in their Figure 10. In the current investigation we further established that the width of the $\Delta V_{PN,LOS}$ distribution increases towards $m_{5007}$ fainter magnitude, see Fig. 10. In what follows, we would like to quantify the relative contributions of the kinematically *cold* and *warmer* PN populations to i) the global LOSVD orthogonal to the plane of the disc in NGC 628, and ii) the PNLF, in different magnitude bins. The goal is to link the properties of the AVDR for PN populations with their PNLFs in NGC 628. Similar investigations for extragalactic PN populations were carried out by Bhattacharya et al. (2019) in the

M31 disc.

We extract the $\Delta V_{PN,LOS}$ distributions in three radial bins, 80" ≤ $R$ ≤ 170", 170" ≤ $R$ ≤ 240" and 240" ≤ $R$ ≤ 425". We normalize each of them by the ratio $RMS_{bin_i}/RMS_{bin_{middle}}$ and then combine them. The entire distribution over the radial range 80" to 425" over the disc of NGC 628 is shown in Figure 12. A single Gaussian fit to the entire distribution gives $\sigma_{1G,Tot}$ = 21.15 km$^{-1}$. We then model the global RMS homogenized LOSVD as a sum of Gaussian distributions, to evaluate whether it is consistent with a single or multiple components. In order to assign PNe to distinct kinematic (Gaussian) components, we use the Gaussian-mixture method (GMM) as shown in Hartke et al. (2018); Aniyan et al. (2018) and references therein.

We start by fitting a single Gaussian and a double-Gaussian model to the data: the latter is clearly favored, with a lower Bayesian Information Criterion. The combined histogram of the re-scaled RMS homogenized $\Delta V_{PN,LOS}$ over the three radial bins is thus best fitted by two Gaussians. The narrow component has mean value $\mu_1$ = –4.21 kms$^{-1}$ and $\sigma_1$ = 8.8 kms$^{-1}$, and the broader component has $\mu_2$ = 1.95 kms$^{-1}$ and $\sigma_2$ = 26.1 kms$^{-1}$. The *cold* and *warmer* component contribute to the total sample of the PN population in the ratio ≈ 46% (cold) and 54% (warm). These fractions[5] are in line with the estimated contributions to the integrated light spectrum of the cold and warm component in the central IFU spectra for NGC 628 as illustrated in Table 6 of Aniyan et al. (2018). These fractions further prove that the kinematics decomposition in cold vs. warm component is robust towards contamination by SNRs, whose estimated upper limit is 2%, see discussion in Section 3. The rescaled distribution with the Gaussian decomposition is shown in Figure 12. For more information on the Gaussian decomposition in specific radial bins, we refer to the detailed analysis by Aniyan et al. (2018).

*PNLFs of the kinematically cold and warmer population in NGC 628* – In order to quantify the contribution of the two kinematically distinct PN populations in the NGC 628 to the PNLF over the NGC 628 disc, we assign a probability to each PN to be belong to one of the two Gaussians based on its $\Delta V_{PN,LOS}$ velocity values, according to the Gaussian decomposition in radial bins carried out in Aniyan et al. (2018). **In Figure 13**, we then plot the cumulative PNLFs in the three bins for the warm/cold component respectively computed such that each PN contributes for its probability "$p_{cold}$" to the cold component and "1 – $p_{cold}$" to the warmer component. For the outermost bin, the GMM gives a result consistent with a single Gaussian, hence only one cumulative function is shown in green in the plot. Each cumulative curve is normalized by the total number of the respective PNe in each bin.

The cumulative PNLFs are plotted in red for the warm and in blue for the cold component, respectively. In the inner radial bin (range: 80" – 170"; light continuum blue/red lines) the cumulative PNLF for the warmer component presents a steep rise at $m_{5007}$ > 26 mags. The intermediate radial bin (range: 170" – 240") the cumulative PNLF of the warm and cold components (dotted-dashed blue/red lines) show a significant difference at magnitudes near the PNLF bright cut-off m*. The comparison among the cumulative PNLFs for the cold and warmer components in the two inner bins and the single cumulative PNLF in the outermost radial bin (range: 240" - 425"; green continuum line) illustrate that there is a larger number of PNe whose $m_{5007}$ is close the PNLF bright cut-off m* in the outer region (R> 170'').

We find that the PNe whose apparent $m_{5007}$ are closest to the PNLF bright cut-off m* are found at radii > 170", with the bright PNe near the PNLF cut-off being primarily associated with the kinematically cold component in the NGC 628 disc also in the second and innermost radial bins. In the latter, the cumulative PNLF in the innermost bin shows a larger contribution from the *cold* component at mag $m_{5007}$<26.0, while the PN population associated with the kinematically *warmer* component dominates the cumulative PNLF at $m_{5007}$ > 26.5. The presence of PNe at large radii which populate the bright cut-off of the PNLF in NGC 628 is consistent with the inside-out formation of discs (Minchev et al. 2014) and the NGC 628 B-I bluer color in the outermost radial bin (Möllenhoff 2004; see Aniyan et al. 2018, their Table 6).

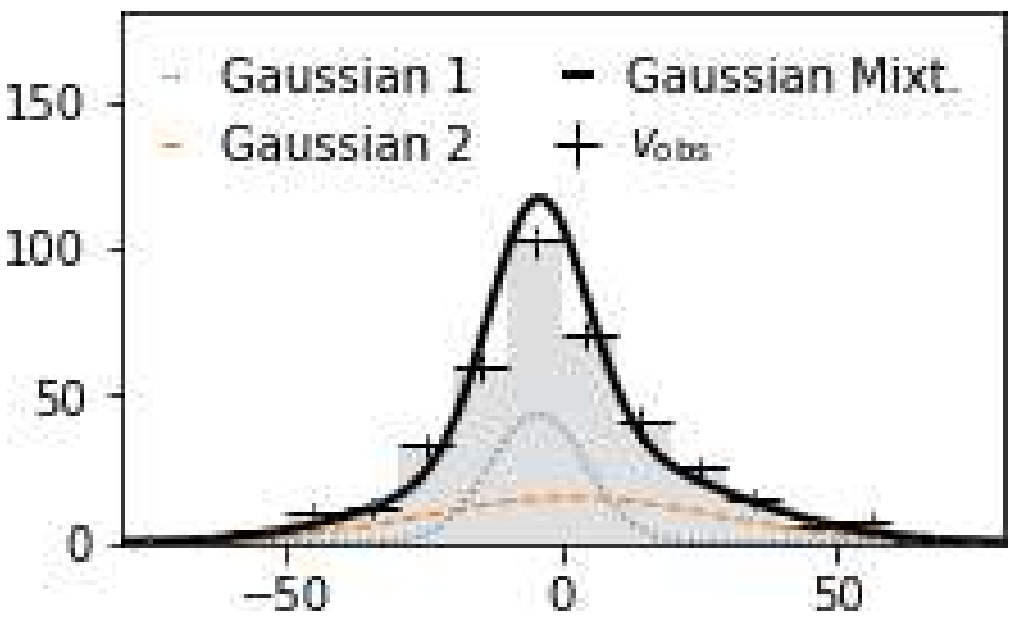

**Figure 12.** Total RMS homogenized $\Delta V_{PN,LOS}$ of the PN populations in NGC 628 covering the radial range 80" ≤ $R$ ≤ 425". This histogram is obtained by merging the PN's ΔV distributions in the three radial bins, once they have been rescaled by the RMS value of $\Delta V_{PN,LOS}$ distribution in the middle radial bin. The RMS of the combined Gaussian is 21.15 kms$^{-1}$, and kurtosis = 0.783. The GMM applied to the merged histogram gives the following parameters for the two Gaussian components: $\mu_1$=–4.21; $\sigma_1$=8.8 kms$^{-1}$ and $\mu_2$=1.95; $\sigma_2$=26.1 kms$^{-1}$.

[5] **Fractions are computed from the integrating the probability of the cold component and then evaluating the warm component as 1-$p_{cold}$**

The results for NGC 628 that show that the colder, dynamically young, component dominates the bright cut-off of the PNLF is also supporting the recent results of Jacoby & Ciardullo (2025) about the critical role of the dust in creating the PNLF bright cut-off, as the most massive and young progenitors would be the most intrinsically dusty (see Bhattacharya et al. 2019 and Ventura et al. 2014). Previous examples of distinct PNLF bright cut-off magnitude values and more in general of deviations of the PNLF shape from the Ciardullo et al. (1989) analytical formula have been measured in the PN populations in the outer disc and halo of M31 (Bhattacharya et al. 2021) and in the outer halos of elliptical galaxies (Longobardi et al. 2015, Hartke et al. 2020); for recent results linking PNLF shapes with the progenitor stellar populations see Valenzuela et al. (2019, 2025).

The results presented in this work on the presence of kinematically distinct PN populations in NGC 628 reflect the composite nature of the stellar populations generating the light in star-forming discs in agreement with an increasing AVDR relation for the older populations also for the face-on galaxy NGC 628.

## 7. Discussion

We wish to compare the vertical velocity dispersions for the PN populations in the NGC 628 with those measured for 1) the outer HI gas disc in NGC 628 and 2) for the stellar populations in the MW and M31 disc. Das et al. (2020) measured the dynamical mass in discs using the HI velocity dispersion at very larger outer radii. In their study which includes NGC 628, they measured the HI vertical dispersion for NGC 628 using the THINGS HI data Walter et al. (2008). From their Figure 2, we estimate the average value $\sigma_{z,HI} \approx 8$ kms$^{-1}$ over the interval of radii sampled by the PN.S PN measurements. This is directly comparable with the vertical velocity dispersion of $\sigma_{PN,cold} \approx 9$ kms$^{-1}$ for the kinematic cold component in the NGC 628 PN sample.

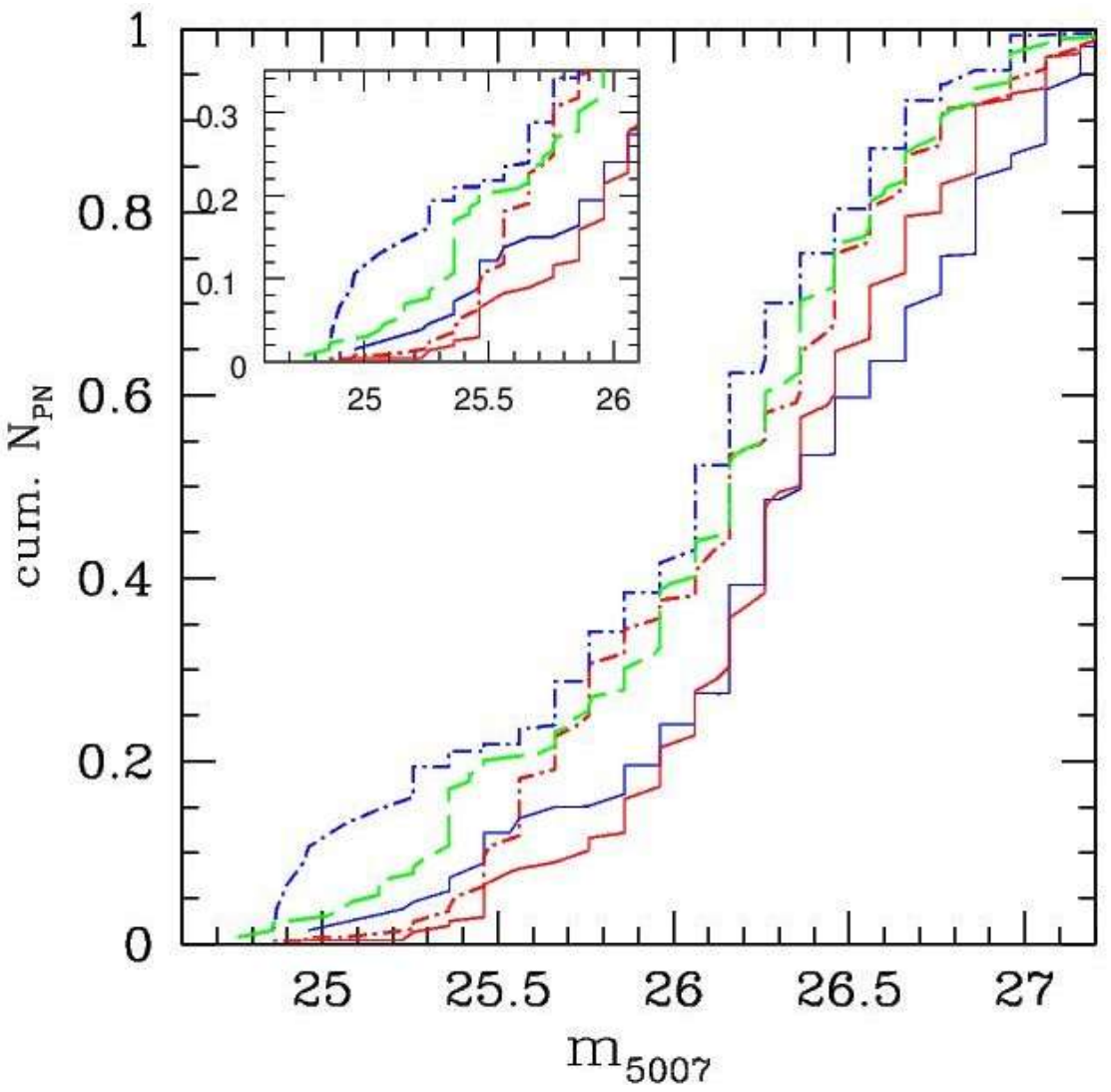

**Figure 13.** Cumulative PNLFs in the three radial bins for the kinematically *cold*-blue lines- and *warmer*-red lines- PN population. Kinematical decomposition from Aniyan et al. (2018). The green line depicts the cumulative PNLF in the outermost radial bin where the GMM decomposition assigns all PNe to the *cold* component in the outermost radial bin (240'' - 425''). In the inner bin (radial range: 80''– 170'', light continuum blue/red lines) the cumulative PNLF for the *warmer* component presents a steep rise at $m_{5007} > 26$ mags. The intermediate radial bin (radial range: 170'' – 240'') the cumulative PNLF of the *warm* and *cold* components (dotted-dashed blue/red lines) show a significant difference at magnitudes near the bright cut -off m*. The comparison among the cumulative PNLFs for the *cold* and *warmer* components in the two inner bins and the single cumulative PNLF in the outermost radial bin illustrate that there is a larger number of PNe whose $m_{5007}$ is close the bright cut off m* in the outer region (R> 170'') consistent with disc stellar population becoming younger at larger radii in NGC 628. In-set in Figure shows enlargement of the bright cut-off magnitudes of the PNLFs.

In the MW, the AVDR for the solar neighborhood[6] is shown in Figure 12 of Aniyan et al. (2018) using the data from the Geneva-Copenhagen survey (Casagrande et al. 2011). The dispersion is of the order 10 kms$^{-1}$ for ages younger than 1 Gyr and then increases to about 20 kms$^{-1}$ at 4 Gyr, with a plateau reached for older ages. The radial range 80'' to 430'' covered by the PN.S PN sample is equivalent to 1 to 5 disc scale length for the thin disc of NGC 628; the disc scale length of NGC 628 is very similar to that of the MW exponential thin disc, see Bland-Hawthorn & Gerhard (2016) for a review. We can then compare the two vertical velocity dispersion measurements for the PNe in NGC 628 to that of the MW stars in the solar neighborhood to infer their stellar ages from the AVDR. We can then associate an age of 1 Gyr to the NGC 628 PNe in the cold component, and an older age of 4 Gyr to the PNe in the warmer component. The ratio of the two samples is consistent with the predictions from a star formation rate that decays with time like $\exp(-t/\tau)$ with $\tau > 3$ Gyr.

[6] This AVDR excludes stars from the thick disc by considering stars with [Fe/H] > –0.3 only.

The AVDR for the M31 disc is much steeper than either that of NGC 628 or the solar neighborhood at the equivalent position in units of disc scale length. As shown by Bhattacharya et al. (2019), the estimated vertical velocity dispersion of PNe in the M31 disc at 2.5 and 4.5 Gyr old progenitors is twice and three times that of the stars in the MW at these reference ages; see additional measurements of the AVDR for the red giant branch stars in the M31 disc by Dorman et al. (2015). The much steeper AVDR for M31 with respect to the MW or NGC 628 (currently understudy) is understood in the framework of a relatively massive merging event with mass ratios 1:4 occurred in the M31 disc about 2 Gyr ago, see simulations by Hammer et al. (2018) and the recent analysis of the projected phase space by Tzakonas et al. (2025). We can then conclude that NGC 628, while forming star at a significant rate, has an AVDR which is quite similar to that of the MW disc and did not experience any recent major merger events.

## 8. Summary and Conclusions

In this study, we showed how the CDI observational technique as implemented with the PN.S to study the kinematics of ETGs can be expanded to study individual nebulae in star forming discs, where HII regions and SNRs contribute to the [OIII] 5007Å sources also. The PN.S equipped with an additional third arm, the Hα arm, allows the identification of *bona-fide* PNe, which are excellent discrete stellar tracers in the outermost region of discs, beyond 1-2 disc scale length and well into the $R_{25}$ outer annuli, where the continuum surface brightness is too faint to allow absorption line spectroscopy against the night sky. The custom built PN.S instrument, equipped with the Hα arm, provides an effective observational parameter space which allows investigations of the kinematics of nebulae as a proxy for discrete stellar tracers in star-forming galaxies, like late-type discs, irregulars and dwarfs.

In this work we described the optical layout, the filter characteristics, and the observational workflow of the PN.S with the Hα arm. We described in detail the improvements of the pipeline data reduction, the results on the velocity errors, the flux calibration of the [OIII] L/R and Hα arms, and the astrometric calibration of the deep image obtained with the Hα arm. For the face-on galaxy NGC 628 (M74), we calibrated the CMD selection criteria for PNe on the sample of 102 spectroscopically confirmed PNe published by Herrmann & Ciardullo (2009a). From these empirically calibrated selection criteria, we extracted an extended sample of 442 PNe and 251 isolated and spatially unresolved HII regions in the NGC 628 disc covering the radial range 80” – 425”, i.e. 1 to 5 disc scale length. The PN.S observations of the NGC 628 lead to a four-fold increase of the number of known PNe in this galaxy. PNe and HII regions selected from the PN.S L/R plus the Hα arms have different number counts as function of apparent $m_{5007}$ and number density profiles. The PN number density profile follows the exponential profile of the light in the NGC 628 disc, while the HII resolved sources display a cut-off where the star formation drops at a 312” radius.

The purpose of the PN.S, which is to measure the kinematics of discrete tracers associated with star light in galaxies, allows one to study the properties of the LOSVD of these populations and validate the selection criteria further. The disc in NGC 628 is seen nearly face-on with $i = 8^{\circ}.5 \pm 0^{\circ}.2$ and the kinematic of its tracers can be compared with the HI 21 cm observations. The LOSVD of the HII region follows the neutral atomic gas 21 cm LOSVD very closely, with the contribution from ISM shocks and outflows in the inner part of the disc. The LOSVD of the entire PN population is wider, e.g. warmer, than that of the HII regions at the same radii. Further analysis of the velocity residuals $\Delta V_{PN,LOS}$ with apparent magnitude $m_{5007}$ provides evidence for distinct PN populations, with kinematically cold PNe dominating the magnitude close to the PNLF bright cut-off, and kinematically warmer PNe dominating at $\Delta m = 1.0 - 1.5$ fainter magnitudes (than the PNLF bright cut-off). The cold and warmer components contribute to the total sample of the PN population in NGC 628 with the ratio 46% (cold) and 54% (warm).

These observational results are important in the current effort to identify distinct components in external discs and measure the disc surface mass density. The use of the PN.S with the additional Hα makes it possible to identify distinct components in the nearby extended face-on discs, in NGC 628 (Aniyan et al. 2018) and NGC 6946 (Aniyan et al. 2021), and address the questions of the Disc-halo degeneracy in these discs. The results from the PN.S observations of NGC 628 clearly indicate the presence of distinct PN sub-populations characterized by different velocities, location, with dominating contributions to the PNLF in different magnitude ranges. These steps set the stage for the investigation and identification of the thin/thick disc components in external galaxies, in addition to M31 and our own MW, using extragalactic PNe. They represent a benchmark to verify the predictions from cosmological motivated simulations of PN populations and their PNLFs (see Valenzuela et al. 2025).

*Acknowledgements.* MAR, OG and KCF wish to acknowledge the support of the ARC Discovery project grant DP150104129 and the Research School of Astronomy and Astrophysics at ANU and Mt. Stromlo Observatory via the Distinguished Visitor fellowship program (in 2018 and 2024). This work was supported by the DAAD under the Australia–Germany joint research program with funds from the Australian Ministry for Science and Education. MAR would like to thank the editors, M. Sarzi, S. Bhattacharya and X. Fang for the invitation to contribute to the Frontieres volume on planetary nebulae populations. The authors would like to thank the Isaac Newton Group staff on La Palma for supporting the PN.S over the years, and the Swiss National Science Foundation, the Kapteyn Institute, the University of Nottingham and INAF for the construction and deployment of the PN.S Hα arm.

## References

Aniyan, S., Freeman, K. C., Gerhard, O. E., Arnaboldi, M., & Flynn, C. 2016, MNRAS, 456, 1484

Aniyan, S., Freeman, K. C., Arnaboldi, M., et al. 2018, MNRAS, 476, 1909

Aniyan, S., Ponomareva, A. A., Freeman, K. C., et al. 2021, MNRAS, 500, 3579

Arnaboldi, M., Aguerri, J. A. L., Napolitano, N. R., et al. 2002, AJ, 123, 760

Arnaboldi, M., Freeman, K. C., Okamura, S., et al. 2003, AJ, 125, 514

Arnaboldi, M., Pulsoni, C., Gerhard, O., & PN. S Consortium. 2017, in IAU Symposium, Vol. 323, Planetary Nebulae: Multi-Wavelength Probes of Stellar and Galactic Evolution, ed. X. Liu, L. Stanghellini, & A. Karakas, 279–283

Arnaboldi, M., Pulsoni, C., Gerhard, O., & ePN. S Team. 2020, in Galactic Dynamics in the Era of Large Surveys, ed. M. Valluri & J. A. Sellwood, Vol. 353, 233–238

Arnaboldi, M., Bhattacharya, S., Gerhard, O. et al. 2022, A&A, 666, 109

Baldwin, J. A., Phillips, M. M., & Terlevich, R. 1981, PASP, 93, 5

Barbosa, C. E., Spiniello, C., Arnaboldi, M., et al. 2021, A&A, 649, A93

Berg, D. A., Skillman, E. D., Croxall, K. V., et al. 2015, ApJ, 806, 16

Bershady, M. A., Verheijen, M. A. W., Swaters, R. A., et al. 2010, ApJ, 716, 198

Bertin, E. & Arnouts, S. 1996, A&AS, 117, 393

Bhattacharya, S., Arnaboldi, M., Caldwell, N., et al. 2019, A&A, 631, A56

Bhattacharya, S., Arnaboldi, M., Gerhard, O. et al. 2021, A&A, 647, 130

Bhattacharya, S., Arnaboldi, M., Caldwell, N. et al. 2022, MNRAS, 517, 2343

Bland-Hawthorn, J. & Gerhard, O. 2016, ARA&A, 54, 529

Buzzoni, A., Arnaboldi, M., & Corradi, R. L. M. 2006, MNRAS, 368, 877

Casagrande, L., Schönrich, R., Asplund, M., et al. 2011, A&A, 530, A138 Ciardullo, R. 2010, PASA, 27, 149

Ciardullo, R., Jacoby, G. H., Ford, H. C. et al. 1989, ApJ, 339, 53

Ciardullo, R., Feldmeier, J. J., Jacoby, G. H., et al. 2002, ApJ, 577, 31

Ciardullo, R. 2010, PASA, 27, 149

Coccato, L., Gerhard, O., Arnaboldi, M., et al. 2009, MNRAS, 394, 1249

Cornett, R. H., O'Connell, R. W., Greason, M. R., et al. 1994, ApJ, 426, 553

Cortesi, A., Arnaboldi, M., Coccato, L., et al. 2013, A&A, 549, A115

Das, M., McGaugh, S. S., Ianjamasimanana, R., Schombert, J., & Dwarakanath,K. S. 2020, ApJ, 889, 10

Dorman, C. E., Guhathakurta, P., Seth, A. C., et al. 2015, ApJ, 803, 24

Douglas, N. G., Arnaboldi, M., Freeman, K. C., et al. 2002, PASP, 114, 1234

Douglas, N. G., Napolitano, N. R., Romanowsky, A. J., et al. 2007, ApJ, 664, 257

Gerhard, O., Arnaboldi, M., Freeman, K. C., & Okamura, S. 2002, ApJ, 580, L121

Greene, J. E., Murphy, J. D., Comerford, J. M., Gebhardt, K., & Adams, J. J. 2012, ApJ, 750, 32

Hammer, F., Yang, Y. B., Wang, J. L., et al. 2018, MNRAS, 475, 2754

Hartke, J., Arnaboldi, M., Gerhard, O., et al. 2018, A&A, 616, 123

Hartke, J. Arnaboldi, M., Gerhard, O. et al. 2020, A&A, 642, 46

Henize, K. G. & Westerlund, B. E. 1963, ApJ, 137, 747

Herrmann, K. A., Ciardullo, R., Feldmeier, J. J., & Vinciguerra, M. 2008, ApJ, 683, 630

Herrmann, K. A. & Ciardullo, R. 2009a, ApJ, 703, 894

Herrmann, K. A. & Ciardullo, R. 2009b, ApJ, 705, 1686

Hopkins, P. F., Hernquist, L., Cox, J.J. et al. 2008, ApJ, 688, 757

Jacoby, G. H. 1989, ApJ, 339, 39

Jacoby, G.H., Ciardullo, R., Roth, M. et al. 2024, ApJS, 271, 40

Jacoby, G.H., Ciardullo, R. 2025, ApJ, 983, 129

Kennicutt, Robert C., J. 1998, ApJ, 498, 541

Kniazev, A. Y., Grebel, E. K., Pustilnik, S. A., Pramskij, A. G., & Zucker, D. B. 2005, AJ, 130, 1558

Kreckel, K., Groves, B., Bigiel, F., et al. 2017, ApJ, 834, 174

Leliévre, M. & Roy, J.-R. 2000, AJ, 120, 1306

Longobardi, A. Arnaboldi, M. Gerhard, O. 2015, A&A, 579, 135

Maggi, P., Haberl, F., Kavanagh, P. J., et al. 2016, A&A, 585, 162

Marigo, P., Girardi, L., Weiss, A., Groenewegen, M. A. T., & Chiosi, C. 2004, A&A, 423, 995

Merrett, H.R., Merrifield, M., Douglas N.R. et al. 2006, MNRAS, 369, 120

Miller Bertolami, M. M. 2016, A&A, 588, A25

Minchev, I., Chiappini, C., & Martig, M. 2014, A&A, 572, A92

Möllenhoff, C. 2004, A&A, 415, 63

Morganti, L., Gerhard, O., Coccato, L. et al. 2013, MNRAS, 431, 3570

Peña, M., Stasińska, G., & Richer, M. G. 2007, A&A, 476, 745

Peimbert, M. & Torres-Peimbert, S. 1983, in IAU Symposium, Vol. 103, Planetary Nebulae, ed. L. H. Aller, 233–242

Pulsoni, C., Gerhard, O., Arnaboldi, M., et al. 2018, A&A, 618, 94

Pulsoni, C., Gerhard, O., Fall, S. M., et al. 2023, A&A, 674, 96

Quinn, P. J. & Goodman, J. 1986, ApJ, 309, 472

Reid, W. A. & Parker, Q. A. 2013, MNRAS, 436, 604

Roth, M. M., Jacoby, G. H., Ciardullo, R., et al. 2021, ApJ, 916, 21

Sellwood, J. A. 2014, Reviews of Modern Physics, 86, 1

Scheuermann, F., Kreckel, K., Gagandeep, S.A. et al. 2022, MNRAS 511, 6087

Soemitro, A.A., Roth, M.M., Weilbacher, P.M. et al. 2023,A&A, 671,142

Spriggs, T. W., Sarzi, M., Galán-de Anta, P. M., et al. 2021, A&A, 653, 167

Stanghellini, L. & Haywood, M. 2018, ApJ, 862, 45

Thomas, D., Maraston, C., Bender, R., & Mendes de Oliveira, C. 2005, ApJ, 621, 673

Tsakonas, C., Arnaboldi, M., Bhattacharya, S. et al. 2025, A&A, 699, 56

van der Kruit, P. C. & Freeman, K. C. 1984, ApJ, 278, 81

Valenzuela, L.M., Mendez R.H., Miller Bertolami, M. 2019, ApJ, 887, 65

Valenzuela, L. M., Miller Bertolami, M. M., Remus, R. et al. 2025, A&A, 699, 371

Ventura, P., Dell'Agli, F., Schneider, R. et al. 2014, MNRAS, 439, 977

Walter, F., Brinks, E., de Blok, W. J. G., et al. 2008, AJ, 136, 2563

## A. Appendix

This appendix provides the PN.S PNe candidate sample for NGC 628 which is utilized for the dynamical model in Aniyan et al. (2018)

**Table A.1.** Table of the PN.S PN candidates in NGC 628. The columns are 1) target ID, 2) right ascension (J2000), 3) declination (J2000), 4) ΔV = $V_{LOS}$ – $V_{HI}$ in kms-1, 5) line-of-sight velocity (heliocentric, $V_{LOS}$), 6) the PN.S [OIII] 5007 Å magnitude and 7) [OIII] – Hα color

| ID | RA (J2000) | DEC (J2000) | ΔV kms-1 | $V_{LOS}$ kms-1 | $m_{5007}$ – 24.66 mag | [OIII] – Hα mag |
|---|---|---|---|---|---|---|
| PN N628 J013623.2+154847.5 | 1:36:23.2 | 15:48:47.5 | 14.5 | 694.2 | 1.6 | 1.3 |
| PN N628 J013623.3+154248.1 | 1:36:23.3 | 15:42:48.1 | 21.3 | 688.1 | 0.6 | 3.3 |
| PN N628 J013623.4+154806.0 | 1:36:23.4 | 15:48:06.0 | -7.0 | 669.4 | 1.7 | 5.0 |
| PN N628 J013623.5+154815.9 | 1:36:23.5 | 15:48:15.9 | -0.4 | 679.2 | 0.4 | 2.5 |
| PN N628 J013623.5+154207.1 | 1:36:23.5 | 15:42:07.1 | 23.1 | 693.7 | 0.6 | 2.3 |
| PN N628 J013623.6+155109.2 | 1:36:23.6 | 15:51:09.2 | -14.2 | 677.3 | 1.3 | 1.9 |
| PN N628 J013623.7+154723.1 | 1:36:23.7 | 15:47:23.1 | -5.4 | 667.9 | 0.4 | 2.2 |
| PN N628 J013623.9+155019.8 | 1:36:23.9 | 15:50:19.8 | 0.5 | 684.7 | 0.3 | 1.6 |
| PN N628 J013623.9+154404.4 | 1:36:23.9 | 15:44:04.4 | 1.0 | 663.1 | 2.0 | 4.2 |
| PN N628 J013624.1+155043.2 | 1:36:24.1 | 15:50:43.2 | -10.2 | 677.2 | 0.7 | 2.0 |
| PN N628 J013624.2+154502.8 | 1:36:24.2 | 15:45:02.8 | -15.0 | 644.1 | 1.2 | 4.7 |
| PN N628 J013624.2+154544.6 | 1:36:24.2 | 15:45:44.6 | 8.4 | 671.4 | 0.6 | 2.9 |
| PN N628 J013624.2+155032.6 | 1:36:24.2 | 15:50:32.6 | 12.9 | 699.2 | 1.6 | 3.0 |
| PN N628 J013624.3+155028.2 | 1:36:24.3 | 15:50:28.2 | -0.1 | 685.0 | 1.9 | 2.0 |
| PN N628 J013624.4+154308.9 | 1:36:24.4 | 15:43:08.9 | 7.2 | 675.9 | 1.1 | 3.4 |
| PN N628 J013624.5+154610.5 | 1:36:24.5 | 15:46:10.5 | -6.8 | 658.5 | 1.5 | 2.7 |
| PN N628 J013624.6+154629.3 | 1:36:24.6 | 15:46:29.3 | 6.9 | 673.1 | 0.0 | 1.4 |
| PN N628 J013624.6+154756.9 | 1:36:24.6 | 15:47:56.9 | 11.3 | 688.0 | 2.0 | 0.4 |
| PN N628 J013624.7+154854.1 | 1:36:24.7 | 15:48:54.1 | 6.6 | 684.4 | 1.9 | 3.7 |
| PN N628 J013624.9+155048.3 | 1:36:24.9 | 15:50:48.3 | -2.2 | 686.5 | 1.3 | 2.4 |
| PN N628 J013624.9+155033.5 | 1:36:24.9 | 15:50:33.5 | 7.1 | 693.1 | 1.4 | 5.1 |
| PN N628 J013625.1+154810.7 | 1:36:25.1 | 15:48:10.7 | 7.6 | 683.1 | 1.8 | 2.9 |
| PN N628 J013625.1+154845.4 | 1:36:25.1 | 15:48:45.4 | 15.1 | 692.6 | 1.6 | 3.1 |
| PN N628 J013625.3+154548.7 | 1:36:25.3 | 15:45:48.7 | 37.5 | 700.0 | 1.6 | 2.0 |
| PN N628 J013625.4+154927.6 | 1:36:25.4 | 15:49:27.6 | -33.3 | 645.2 | 1.5 | 3.0 |
| PN N628 J013625.4+154736.4 | 1:36:25.4 | 15:47:36.4 | -42.3 | 635.4 | 1.7 | 4.3 |
| PN N628 J013625.5+154940.3 | 1:36:25.5 | 15:49:40.3 | 6.5 | 686.6 | 1.1 | 1.7 |
| PN N628 J013625.5+154823.7 | 1:36:25.5 | 15:48:23.7 | 25.2 | 700.4 | 1.9 | 2.5 |
| PN N628 J013625.7+154919.9 | 1:36:25.7 | 15:49:19.9 | -1.9 | 676.9 | 1.6 | 4.8 |
| PN N628 J013625.8+154959.2 | 1:36:25.8 | 15:49:59.2 | -11.6 | 669.8 | 0.1 | 1.6 |
| PN N628 J013626.1+154544.6 | 1:36:26.1 | 15:45:44.6 | -11.9 | 652.0 | 1.2 | 3.3 |
| PN N628 J013626.1+154834.3 | 1:36:26.1 | 15:48:34.3 | -18.5 | 660.4 | 2.1 | 3.9 |
| PN N628 J013626.2+154759.0 | 1:36:26.2 | 15:47:59.0 | -6.3 | 670.2 | 1.1 | 3.4 |
| PN N628 J013626.6+154827.8 | 1:36:26.6 | 15:48:27.8 | 2.7 | 679.3 | 1.9 | 5.6 |
| PN N628 J013626.7+154501.0 | 1:36:26.7 | 15:45:01.0 | 48.5 | 705.3 | 2.1 | 3.3 |
| PN N628 J013626.9+155052.3 | 1:36:26.9 | 15:50:52.3 | -11.7 | 675.7 | 1.6 | 3.2 |
| PN N628 J013627.3+154947.3 | 1:36:27.3 | 15:49:47.3 | -27.7 | 654.5 | 0.9 | 3.0 |
| PN N628 J013627.3+154543.3 | 1:36:27.3 | 15:45:43.3 | -10.0 | 651.7 | 1.6 | 1.6 |
| PN N628 J013627.7+154724.8 | 1:36:27.7 | 15:47:24.8 | 5.8 | 675.5 | 1.0 | 3.5 |
| PN N628 J013627.8+154441.9 | 1:36:27.8 | 15:44:41.9 | -2.4 | 654.4 | 0.5 | 2.9 |
| PN N628 J013628.0+154544.9 | 1:36:28.0 | 15:45:44.9 | -1.8 | 656.7 | 1.4 | 3.7 |
| PN N628 J013628.0+155140.7 | 1:36:28.0 | 15:51:40.7 | -20.4 | 674.8 | 1.2 | 3.0 |
| PN N628 J013628.0+154610.9 | 1:36:28.0 | 15:46:10.9 | 23.1 | 682.0 | 1.7 | 1.2 |
| PN N628 J013628.1+154550.7 | 1:36:28.1 | 15:45:50.7 | 12.8 | 671.4 | 1.7 | 4.5 |
| PN N628 J013628.5+154336.7 | 1:36:28.5 | 15:43:36.7 | -2.0 | 654.5 | 0.6 | 3.3 |
| PN N628 J013628.5+154733.0 | 1:36:28.5 | 15:47:33.0 | 0.4 | 671.9 | 0.3 | 1.8 |
| PN N628 J013628.7+155106.3 | 1:36:28.7 | 15:51:06.3 | 1.3 | 692.4 | 1.7 | 3.4 |
| PN N628 J013628.8+154545.3 | 1:36:28.8 | 15:45:45.3 | 13.6 | 672.2 | 0.8 | 3.2 |
| PN N628 J013629.2+154421.9 | 1:36:29.2 | 15:44:21.9 | -67.2 | 587.0 | 1.7 | 2.9 |
| PN N628 J013629.2+154847.8 | 1:36:29.2 | 15:48:47.8 | 15.4 | 692.6 | 0.8 | 3.5 |
| PN N628 J013629.4+154448.0 | 1:36:29.4 | 15:44:48.0 | -2.3 | 657.8 | 1.2 | 4.4 |
| PN N628 J013629.5+154844.5 | 1:36:29.5 | 15:48:44.5 | -30.6 | 645.6 | 3.0 | 6.0 |
| PN N628 J013629.6+154559.0 | 1:36:29.6 | 15:45:59.0 | 19.1 | 678.5 | 1.1 | -0.1 |
| PN N628 J013629.7+154536.6 | 1:36:29.7 | 15:45:36.6 | 5.9 | 665.8 | 1.2 | 4.3 |
| PN N628 J013629.8+154930.0 | 1:36:29.8 | 15:49:30.0 | 3.4 | 692.6 | 1.5 | 3.7 |
| PN N628 J013630.2+154532.0 | 1:36:30.2 | 15:45:32.0 | 39.0 | 699.3 | 0.8 | 1.1 |
| PN N628 J013630.3+155037.5 | 1:36:30.3 | 15:50:37.5 | 1.6 | 693.8 | 0.6 | 2.2 |
| PN N628 J013630.4+154421.8 | 1:36:30.4 | 15:44:21.8 | 0.4 | 656.3 | 2.6 | 5.0 |
| PN N628 J013630.4+154939.4 | 1:36:30.4 | 15:49:39.4 | -7.9 | 682.1 | 0.6 | 1.3 |
| PN N628 J013630.6+154745.8 | 1:36:30.6 | 15:47:45.8 | 0.9 | 680.0 | 1.8 | 3.8 |
| PN N628 J013630.7+154405.7 | 1:36:30.7 | 15:44:05.7 | 16.8 | 670.1 | 2.2 | 1.0 |
| PN N628 J013630.7+154829.2 | 1:36:30.7 | 15:48:29.2 | 28.8 | 714.3 | 1.7 | 0.3 |
| PN N628 J013630.9+154527.5 | 1:36:30.9 | 15:45:27.5 | -5.9 | 655.8 | 2.0 | 4.1 |
| PN N628 J013630.9+154744.1 | 1:36:30.9 | 15:47:44.1 | -6.8 | 671.7 | 1.4 | 4.8 |
| PN N628 J013630.9+154904.6 | 1:36:30.9 | 15:49:04.6 | -37.6 | 645.7 | 1.0 | 4.4 |
| PN N628 J013630.9+154952.8 | 1:36:30.9 | 15:49:52.8 | -31.9 | 657.1 | 2.2 | 2.2 |
| PN N628 J013631.0+154712.6 | 1:36:31.0 | 15:47:12.6 | 16.5 | 686.9 | 1.2 | 4.3 |
| PN N628 J013631.0+154919.7 | 1:36:31.0 | 15:49:19.7 | -2.0 | 682.8 | 1.4 | 3.7 |
| PN N628 J013631.1+154622.1 | 1:36:31.1 | 15:46:22.1 | 98.8 | 761.8 | 1.1 | 4.4 |
| PN N628 J013631.1+154656.8 | 1:36:31.1 | 15:46:56.8 | 14.2 | 682.7 | 2.3 | 4.3 |
| PN N628 J013631.1+154556.1 | 1:36:31.1 | 15:45:56.1 | 0.2 | 660.5 | 1.9 | 3.4 |
| PN N628 J013631.1+154520.1 | 1:36:31.1 | 15:45:20.1 | 3.9 | 663.8 | 0.8 | 2.4 |
| PN N628 J013631.2+154346.5 | 1:36:31.2 | 15:43:46.5 | -6.7 | 646.1 | 1.3 | 2.5 |
| PN N628 J013631.3+154854.7 | 1:36:31.3 | 15:48:54.7 | 22.4 | 710.3 | 1.7 | 2.9 |
| PN N628 J013631.6+154236.1 | 1:36:31.6 | 15:42:36.1 | -0.3 | 660.4 | 0.6 | 2.4 |
| PN N628 J013631.6+154620.9 | 1:36:31.6 | 15:46:20.9 | 28.6 | 693.3 | 2.3 | 5.5 |
| PN N628 J013632.0+154625.6 | 1:36:32.0 | 15:46:25.6 | -20.5 | 644.4 | 2.0 | 4.1 |
| PN N628 J013632.3+154321.3 | 1:36:32.3 | 15:43:21.3 | 44.8 | 698.4 | 1.2 | 3.0 |
| PN N628 J013632.3+154358.4 | 1:36:32.3 | 15:43:58.4 | -1.1 | 653.2 | 1.4 | 0.7 |
| PN N628 J013632.4+155038.9 | 1:36:32.4 | 15:50:38.9 | -4.8 | 685.6 | 1.9 | 3.9 |
| PN N628 J013632.6+154555.1 | 1:36:32.6 | 15:45:55.1 | -2.0 | 660.0 | 1.6 | 5.3 |
| PN N628 J013632.6+154845.3 | 1:36:32.6 | 15:48:45.3 | -1.4 | 687.6 | 1.4 | 2.4 |
| PN N628 J013632.7+154638.2 | 1:36:32.7 | 15:46:38.2 | 7.8 | 670.9 | 1.9 | 3.1 |
| PN N628 J013632.7+154616.5 | 1:36:32.7 | 15:46:16.5 | 39.0 | 699.7 | 1.3 | 0.1 |
| PN N628 J013632.8+154608.5 | 1:36:32.8 | 15:46:08.5 | 43.5 | 703.5 | 1.5 | 0.9 |
| PN N628 J013632.9+154650.4 | 1:36:32.9 | 15:46:50.4 | 16.1 | 679.2 | 1.2 | 3.8 |
| PN N628 J013633.0+154926.4 | 1:36:33.0 | 15:49:26.4 | -24.5 | 664.5 | 2.0 | 2.7 |
| PN N628 J013633.2+154649.9 | 1:36:33.2 | 15:46:49.9 | 31.5 | 693.1 | 2.9 | 4.4 |
| PN N628 J013633.2+154515.8 | 1:36:33.2 | 15:45:15.8 | -4.9 | 646.6 | 0.7 | 3.1 |
| PN N628 J013633.2+154852.0 | 1:36:33.2 | 15:48:52.0 | 5.6 | 691.2 | 1.4 | 4.9 |
| PN N628 J013633.3+154644.3 | 1:36:33.3 | 15:46:44.3 | 29.6 | 693.1 | 1.9 | 3.1 |
| PN N628 J013633.3+154714.7 | 1:36:33.3 | 15:47:14.7 | 68.7 | 740.8 | 1.3 | 1.8 |
| PN N628 J013633.3+154759.5 | 1:36:33.3 | 15:47:59.5 | 1.7 | 685.7 | 1.8 | 4.7 |
| PN N628 J013633.4+154359.7 | 1:36:33.4 | 15:43:59.7 | -30.9 | 628.0 | 1.4 | 4.7 |
| PN N628 J013633.5+154809.0 | 1:36:33.5 | 15:48:09.0 | 7.0 | 693.8 | 2.5 | 3.8 |
| PN N628 J013633.7+154537.5 | 1:36:33.7 | 15:45:37.5 | 35.5 | 693.7 | 2.2 | 5.7 |
| PN N628 J013633.7+154642.8 | 1:36:33.7 | 15:46:42.8 | 30.8 | 693.7 | 1.4 | 3.4 |
| PN N628 J013633.8+154632.2 | 1:36:33.8 | 15:46:32.2 | 17.4 | 680.8 | 1.5 | 3.5 |
| PN N628 J013633.8+154722.7 | 1:36:33.8 | 15:47:22.7 | 3.1 | 680.2 | 1.0 | 1.9 |
| PN N628 J013633.8+154913.7 | 1:36:33.8 | 15:49:13.7 | -67.0 | 623.2 | 1.4 | 2.3 |
| PN N628 J013633.8+154922.4 | 1:36:33.8 | 15:49:22.4 | -4.7 | 685.5 | 1.1 | 3.2 |
| PN N628 J013633.9+154358.3 | 1:36:33.9 | 15:43:58.3 | -9.4 | 649.4 | 1.0 | 4.2 |
| PN N628 J013634.1+154751.1 | 1:36:34.1 | 15:47:51.1 | 21.2 | 704.1 | 1.3 | 5.6 |
| PN N628 J013634.3+154941.5 | 1:36:34.3 | 15:49:41.5 | -18.7 | 673.0 | 2.7 | 5.7 |
| PN N628 J013634.5+154630.5 | 1:36:34.5 | 15:46:30.5 | 47.2 | 708.3 | 1.5 | 5.5 |
| PN N628 J013634.7+154821.9 | 1:36:34.7 | 15:48:21.9 | -1.9 | 688.7 | 1.7 | 2.3 |
| PN N628 J013634.8+154951.6 | 1:36:34.8 | 15:49:51.6 | -13.7 | 678.6 | 0.7 | 1.7 |
| PN N628 J013634.9+154247.3 | 1:36:34.9 | 15:42:47.3 | -12.4 | 644.3 | 0.9 | 3.4 |
| PN N628 J013635.0+154409.5 | 1:36:35.0 | 15:44:09.5 | -22.9 | 636.8 | 1.8 | 1.4 |
| PN N628 J013635.1+154429.4 | 1:36:35.1 | 15:44:29.4 | 10.0 | 667.5 | 0.9 | 2.2 |
| PN N628 J013635.2+154938.4 | 1:36:35.2 | 15:49:38.4 | -17.5 | 671.8 | 1.0 | 2.5 |
| PN N628 J013635.3+154256.6 | 1:36:35.3 | 15:42:56.6 | -9.0 | 645.8 | 1.7 | 4.2 |
| PN N628 J013635.3+154910.5 | 1:36:35.3 | 15:49:10.5 | -12.9 | 685.7 | 2.1 | 2.1 |
| PN N628 J013635.3+155100.1 | 1:36:35.3 | 15:51:00.1 | -36.0 | 665.3 | 1.7 | 3.4 |
| PN N628 J013635.3+154639.0 | 1:36:35.3 | 15:46:39.0 | -1.8 | 662.6 | 3.3 | 5.8 |
| PN N628 J013635.5+154600.8 | 1:36:35.5 | 15:46:00.8 | 7.5 | 662.8 | 1.7 | 4.2 |
| PN N628 J013635.7+154948.5 | 1:36:35.7 | 15:49:48.5 | -5.6 | 687.4 | 2.7 | 6.2 |
| PN N628 J013635.8+154619.5 | 1:36:35.8 | 15:46:19.5 | -2.5 | 657.5 | 1.2 | 2.7 |
| PN N628 J013636.0+154359.2 | 1:36:36.0 | 15:43:59.2 | 28.7 | 684.0 | 1.1 | 3.9 |
| PN N628 J013636.0+155016.7 | 1:36:36.0 | 15:50:16.7 | -13.4 | 683.1 | 1.7 | 4.1 |
| PN N628 J013636.1+154246.2 | 1:36:36.1 | 15:42:46.2 | 2.6 | 658.7 | 0.1 | 0.8 |
| PN N628 J013636.2+155000.9 | 1:36:36.2 | 15:50:00.9 | -11.4 | 685.6 | 1.9 | 3.7 |
| PN N628 J013636.2+154701.0 | 1:36:36.2 | 15:47:01.0 | 13.6 | 684.3 | 1.7 | 6.5 |
| PN N628 J013636.3+155025.4 | 1:36:36.3 | 15:50:25.4 | -21.8 | 676.4 | 1.4 | 4.8 |
| PN N628 J013636.3+154423.3 | 1:36:36.3 | 15:44:23.3 | 6.0 | 663.5 | 1.7 | 5.3 |
| PN N628 J013636.3+154917.3 | 1:36:36.3 | 15:49:17.3 | 11.8 | 713.9 | 2.2 | 3.4 |
| PN N628 J013636.4+154514.2 | 1:36:36.4 | 15:45:14.2 | 2.6 | 653.8 | 0.6 | 3.1 |
| PN N628 J013636.4+154553.5 | 1:36:36.4 | 15:45:53.5 | 25.1 | 677.6 | 1.6 | 3.7 |
| PN N628 J013636.6+155110.8 | 1:36:36.6 | 15:51:10.8 | 36.5 | 739.3 | 1.8 | 3.5 |
| PN N628 J013636.6+155130.9 | 1:36:36.6 | 15:51:30.9 | 1.1 | 703.3 | 0.6 | 3.6 |
| PN N628 J013636.7+154713.0 | 1:36:36.7 | 15:47:13.0 | -27.6 | 649.1 | 1.3 | 5.1 |
| PN N628 J013636.7+154716.5 | 1:36:36.7 | 15:47:16.5 | 14.8 | 694.5 | 2.1 | 6.8 |
| PN N628 J013636.8+154652.1 | 1:36:36.8 | 15:46:52.1 | 14.7 | 682.5 | 2.3 | 5.3 |
| PN N628 J013637.0+154708.2 | 1:36:37.0 | 15:47:08.2 | -1.1 | 674.3 | 1.5 | 4.3 |
| PN N628 J013637.0+154359.6 | 1:36:37.0 | 15:43:59.6 | -67.2 | 587.7 | 1.4 | 4.5 |
| PN N628 J013637.1+154724.9 | 1:36:37.1 | 15:47:24.9 | 81.9 | 766.2 | 1.4 | 1.5 |
| PN N628 J013637.1+155002.1 | 1:36:37.1 | 15:50:02.1 | 3.7 | 702.8 | 1.0 | 4.4 |
| PN N628 J013637.1+155049.9 | 1:36:37.1 | 15:50:49.9 | -1.2 | 697.4 | 1.9 | 5.4 |
| PN N628 J013637.2+154621.3 | 1:36:37.2 | 15:46:21.3 | -14.3 | 644.4 | 1.9 | 5.5 |
| PN N628 J013637.2+154928.5 | 1:36:37.2 | 15:49:28.5 | 24.0 | 726.7 | 1.1 | 3.8 |
| PN N628 J013637.3+154436.3 | 1:36:37.3 | 15:44:36.3 | 31.8 | 688.6 | 0.5 | 3.3 |
| PN N628 J013637.4+154349.4 | 1:36:37.4 | 15:43:49.4 | 0.6 | 655.4 | 1.3 | 4.2 |
| PN N628 J013637.4+154555.2 | 1:36:37.4 | 15:45:55.2 | 82.1 | 736.0 | 1.8 | 4.9 |
| PN N628 J013637.4+154821.4 | 1:36:37.4 | 15:48:21.4 | -13.6 | 682.1 | 3.3 | 7.2 |
| PN N628 J013637.5+154831.1 | 1:36:37.5 | 15:48:31.1 | -2.6 | 696.5 | 3.1 | 7.6 |
| PN N628 J013637.5+154441.2 | 1:36:37.5 | 15:44:41.2 | -32.1 | 624.3 | 1.3 | 4.4 |
| PN N628 J013637.6+154524.5 | 1:36:37.6 | 15:45:24.5 | -3.1 | 647.2 | 2.1 | 1.5 |
| PN N628 J013637.6+155100.7 | 1:36:37.6 | 15:51:00.7 | 12.6 | 716.1 | 1.4 | 2.9 |
| PN N628 J013637.7+155034.4 | 1:36:37.7 | 15:50:34.4 | -46.9 | 651.7 | 2.5 | 5.3 |
| PN N628 J013637.7+154644.8 | 1:36:37.7 | 15:46:44.8 | -59.9 | 607.4 | 2.3 | 3.7 |
| PN N628 J013637.8+154844.1 | 1:36:37.8 | 15:48:44.1 | 11.5 | 711.6 | 1.6 | 2.1 |
| PN N628 J013637.8+154718.2 | 1:36:37.8 | 15:47:18.2 | 6.7 | 687.0 | 6.1 | 10.3 |
| PN N628 J013637.8+154850.2 | 1:36:37.8 | 15:48:50.2 | -8.3 | 692.4 | 1.5 | 4.5 |
| PN N628 J013637.9+154502.0 | 1:36:37.9 | 15:45:02.0 | -7.1 | 647.8 | 3.0 | 6.8 |
| PN N628 J013637.9+154611.5 | 1:36:37.9 | 15:46:11.5 | -31.3 | 623.1 | 0.6 | 3.3 |
| PN N628 J013638.0+154659.8 | 1:36:38.0 | 15:46:59.8 | -48.2 | 622.9 | 5.0 | 10.4 |
| PN N628 J013638.0+155050.3 | 1:36:38.0 | 15:50:50.3 | -22.4 | 678.6 | 1.5 | 4.6 |
| PN N628 J013638.2+154748.5 | 1:36:38.2 | 15:47:48.5 | -50.1 | 646.7 | 3.0 | 7.4 |
| PN N628 J013638.5+155023.2 | 1:36:38.5 | 15:50:23.2 | -109.4 | 590.1 | 2.2 | 4.2 |
| PN N628 J013638.7+154659.3 | 1:36:38.7 | 15:46:59.3 | -9.2 | 657.4 | 1.4 | 5.6 |
| PN N628 J013638.7+154810.7 | 1:36:38.7 | 15:48:10.7 | 9.1 | 707.4 | 1.4 | 4.2 |
| PN N628 J013638.8+154946.5 | 1:36:38.8 | 15:49:46.5 | 37.3 | 743.1 | 2.0 | 4.1 |
| PN N628 J013638.8+154255.3 | 1:36:38.8 | 15:42:55.3 | -31.7 | 624.4 | 1.8 | 4.0 |
| PN N628 J013638.8+155032.7 | 1:36:38.8 | 15:50:32.7 | 5.3 | 705.0 | 1.9 | 5.6 |
| PN N628 J013638.8+154404.4 | 1:36:38.8 | 15:44:04.4 | 12.6 | 673.0 | 1.1 | 4.7 |
| PN N628 J013639.0+154332.2 | 1:36:39.0 | 15:43:32.2 | -10.1 | 634.0 | 1.0 | 2.4 |
| PN N628 J013639.1+154433.7 | 1:36:39.1 | 15:44:33.7 | -12.9 | 642.8 | 1.1 | 2.8 |
| PN N628 J013639.4+154905.2 | 1:36:39.4 | 15:49:05.2 | -14.3 | 690.7 | 1.5 | 4.7 |
| PN N628 J013639.4+154401.3 | 1:36:39.4 | 15:44:01.3 | -27.3 | 632.2 | 2.1 | 4.2 |
| PN N628 J013639.7+154653.7 | 1:36:39.7 | 15:46:53.7 | 74.1 | 737.7 | 3.0 | 4.9 |
| PN N628 J013639.7+154441.4 | 1:36:39.7 | 15:44:41.4 | -40.0 | 618.6 | 1.7 | 4.8 |
| PN N628 J013639.8+154508.1 | 1:36:39.8 | 15:45:08.1 | 61.7 | 715.6 | 1.2 | 4.1 |
| PN N628 J013639.8+155023.0 | 1:36:39.8 | 15:50:23.0 | -20.8 | 679.5 | 1.9 | 4.2 |
| PN N628 J013640.0+155033.4 | 1:36:40.0 | 15:50:33.4 | -29.7 | 671.3 | 1.5 | 4.9 |
| PN N628 J013640.3+154512.0 | 1:36:40.3 | 15:45:12.0 | 69.4 | 721.4 | 2.1 | 4.1 |
| PN N628 J013640.3+155016.7 | 1:36:40.3 | 15:50:16.7 | -2.7 | 696.1 | 0.5 | 1.9 |
| PN N628 J013640.5+154457.1 | 1:36:40.5 | 15:44:57.1 | 57.3 | 717.3 | 2.0 | 3.1 |
| PN N628 J013640.5+154822.5 | 1:36:40.5 | 15:48:22.5 | -8.5 | 694.0 | 2.6 | 6.6 |
| PN N628 J013640.5+154910.7 | 1:36:40.5 | 15:49:10.7 | -12.2 | 695.2 | 2.5 | 9.5 |
| PN N628 J013640.6+155116.8 | 1:36:40.6 | 15:51:16.8 | -12.3 | 693.2 | 1.2 | 2.6 |
| PN N628 J013640.8+155027.8 | 1:36:40.8 | 15:50:27.8 | -22.1 | 679.8 | 1.6 | 4.6 |
| PN N628 J013640.9+154907.1 | 1:36:40.9 | 15:49:07.1 | 12.3 | 718.4 | 2.6 | 5.2 |
| PN N628 J013640.9+154254.9 | 1:36:40.9 | 15:42:54.9 | 5.2 | 661.4 | 1.6 | 4.3 |
| PN N628 J013641.1+154624.3 | 1:36:41.1 | 15:46:24.3 | -10.7 | 649.8 | 0.9 | 3.5 |
| PN N628 J013641.3+154754.2 | 1:36:41.3 | 15:47:54.2 | 7.6 | 706.0 | 1.6 | 1.7 |
| PN N628 J013641.3+154355.1 | 1:36:41.3 | 15:43:55.1 | -56.9 | 599.5 | 2.2 | 5.0 |
| PN N628 J013641.3+154426.3 | 1:36:41.3 | 15:44:26.3 | 17.7 | 672.2 | 1.4 | 2.0 |
| PN N628 J013641.4+154929.7 | 1:36:41.4 | 15:49:29.7 | -11.4 | 697.1 | 1.9 | 4.5 |
| PN N628 J013641.5+154449.9 | 1:36:41.5 | 15:44:49.9 | 0.8 | 661.8 | 1.4 | 3.8 |
| PN N628 J013641.5+154620.5 | 1:36:41.5 | 15:46:20.5 | -30.7 | 624.0 | 1.1 | 2.5 |
| PN N628 J013641.6+154909.6 | 1:36:41.6 | 15:49:09.6 | 29.6 | 736.3 | 1.5 | 3.7 |
| PN N628 J013641.7+154840.5 | 1:36:41.7 | 15:48:40.5 | 36.1 | 738.1 | 3.1 | 7.6 |
| PN N628 J013641.7+154757.0 | 1:36:41.7 | 15:47:57.0 | 31.4 | 729.6 | 1.4 | 1.8 |
| PN N628 J013641.8+155030.0 | 1:36:41.8 | 15:50:30.0 | -14.4 | 689.3 | 1.8 | 1.9 |
| PN N628 J013641.9+155005.6 | 1:36:41.9 | 15:50:05.6 | -9.0 | 693.3 | 3.6 | 10.0 |
| PN N628 J013641.9+154912.3 | 1:36:41.9 | 15:49:12.3 | 10.2 | 715.0 | 0.7 | 3.0 |
| PN N628 J013642.0+154455.9 | 1:36:42.0 | 15:44:55.9 | 21.1 | 680.5 | 0.8 | 4.2 |
| PN N628 J013642.2+155028.5 | 1:36:42.2 | 15:50:28.5 | -19.6 | 684.8 | 1.3 | 3.1 |
| PN N628 J013642.2+155046.7 | 1:36:42.2 | 15:50:46.7 | -13.3 | 691.1 | 1.8 | -2.0 |
| PN N628 J013642.5+154905.5 | 1:36:42.5 | 15:49:05.5 | -8.9 | 697.7 | 0.7 | 3.3 |
| PN N628 J013642.5+154835.8 | 1:36:42.5 | 15:48:35.8 | 82.0 | 784.9 | 1.9 | 3.3 |
| PN N628 J013642.5+154533.9 | 1:36:42.5 | 15:45:33.9 | -2.8 | 657.6 | 2.3 | 3.7 |
| PN N628 J013642.7+154800.6 | 1:36:42.7 | 15:48:00.6 | -50.2 | 649.6 | 2.0 | 4.1 |
| PN N628 J013642.7+154505.6 | 1:36:42.7 | 15:45:05.6 | -22.3 | 635.1 | 1.8 | 4.6 |
| PN N628 J013643.0+154745.3 | 1:36:43.0 | 15:47:45.3 | -15.3 | 687.5 | 1.4 | 5.2 |
| PN N628 J013643.1+154935.4 | 1:36:43.1 | 15:49:35.4 | 44.6 | 751.9 | 1.4 | 2.7 |
| PN N628 J013643.2+154826.2 | 1:36:43.2 | 15:48:26.2 | 43.8 | 749.5 | 3.3 | 6.0 |
| PN N628 J013643.2+154717.9 | 1:36:43.2 | 15:47:17.9 | 0.0 | 693.1 | 0.4 | 2.9 |
| PN N628 J013643.2+154512.3 | 1:36:43.2 | 15:45:12.3 | -19.7 | 636.3 | 2.4 | 7.2 |
| PN N628 J013643.3+154803.8 | 1:36:43.3 | 15:48:03.8 | 41.3 | 740.5 | 1.0 | 2.1 |
| PN N628 J013643.4+154845.1 | 1:36:43.4 | 15:48:45.1 | -28.9 | 672.8 | 1.4 | 5.2 |
| PN N628 J013643.4+154405.2 | 1:36:43.4 | 15:44:05.2 | -1.2 | 656.9 | 1.3 | 4.5 |
| PN N628 J013643.5+154516.9 | 1:36:43.5 | 15:45:16.9 | 26.9 | 685.7 | 2.4 | 1.0 |
| PN N628 J013643.6+154648.3 | 1:36:43.6 | 15:46:48.3 | 0.2 | 675.6 | 1.8 | 3.8 |
| PN N628 J013643.6+154336.1 | 1:36:43.6 | 15:43:36.1 | 7.1 | 664.9 | 1.5 | 3.2 |
| PN N628 J013643.6+154419.8 | 1:36:43.6 | 15:44:19.8 | -13.1 | 649.2 | 1.8 | 5.0 |
| PN N628 J013643.7+154642.1 | 1:36:43.7 | 15:46:42.1 | -5.7 | 664.4 | 2.1 | 5.2 |
| PN N628 J013643.8+154625.4 | 1:36:43.8 | 15:46:25.4 | 6.3 | 670.4 | 1.7 | 7.1 |
| PN N628 J013643.9+154339.7 | 1:36:43.9 | 15:43:39.7 | -24.1 | 634.4 | 0.2 | 1.0 |
| PN N628 J013644.2+154246.5 | 1:36:44.2 | 15:42:46.5 | 12.3 | 667.6 | 1.2 | 0.6 |
| PN N628 J013644.3+154954.8 | 1:36:44.3 | 15:49:54.8 | 8.2 | 716.0 | 1.1 | 1.4 |
| PN N628 J013644.3+154515.8 | 1:36:44.3 | 15:45:15.8 | 8.9 | 668.4 | 1.4 | 2.5 |
| PN N628 J013644.4+154332.0 | 1:36:44.4 | 15:43:32.0 | -1.4 | 656.5 | 0.1 | 1.6 |
| PN N628 J013644.5+154756.2 | 1:36:44.5 | 15:47:56.2 | -103.0 | 600.3 | 0.6 | 3.4 |
| PN N628 J013644.6+154442.0 | 1:36:44.6 | 15:44:42.0 | -0.5 | 660.0 | 2.4 | 6.4 |
| PN N628 J013644.7+155106.8 | 1:36:44.7 | 15:51:06.8 | -23.5 | 680.5 | 2.0 | 5.4 |
| PN N628 J013644.7+154803.3 | 1:36:44.7 | 15:48:03.3 | -48.1 | 652.6 | 2.8 | 7.7 |
| PN N628 J013644.8+154736.1 | 1:36:44.8 | 15:47:36.1 | -11.8 | 690.4 | 1.6 | 5.4 |
| PN N628 J013644.8+154518.1 | 1:36:44.8 | 15:45:18.1 | 51.9 | 713.2 | 1.7 | 2.8 |
| PN N628 J013644.9+154416.4 | 1:36:44.9 | 15:44:16.4 | 33.4 | 694.8 | 0.4 | 2.5 |
| PN N628 J013644.9+154811.5 | 1:36:44.9 | 15:48:11.5 | 37.1 | 739.9 | 2.6 | 6.7 |
| PN N628 J013645.1+154911.7 | 1:36:45.1 | 15:49:11.7 | -8.3 | 698.2 | 0.2 | 2.2 |
| PN N628 J013645.3+154850.1 | 1:36:45.3 | 15:48:50.1 | -10.6 | 693.0 | 2.1 | 6.7 |
| PN N628 J013645.4+154441.8 | 1:36:45.4 | 15:44:41.8 | 10.3 | 671.9 | 2.6 | 7.0 |
| PN N628 J013645.4+154423.3 | 1:36:45.4 | 15:44:23.3 | -33.5 | 632.3 | 1.7 | 5.1 |
| PN N628 J013645.5+155119.3 | 1:36:45.5 | 15:51:19.3 | -22.0 | 680.1 | 2.8 | 5.2 |
| PN N628 J013645.6+154820.8 | 1:36:45.6 | 15:48:20.8 | -18.8 | 682.7 | 1.7 | 6.0 |
| PN N628 J013645.7+154601.5 | 1:36:45.7 | 15:46:01.5 | 9.2 | 679.9 | 2.0 | 6.0 |
| PN N628 J013645.7+154740.6 | 1:36:45.7 | 15:47:40.6 | -19.2 | 681.5 | 0.8 | 4.2 |
| PN N628 J013645.7+154416.7 | 1:36:45.7 | 15:44:16.7 | -5.1 | 658.7 | 1.0 | 4.5 |
| PN N628 J013645.8+154211.8 | 1:36:45.8 | 15:42:11.8 | -9.3 | 643.7 | 1.5 | 3.2 |
| PN N628 J013645.9+154932.1 | 1:36:45.9 | 15:49:32.1 | -10.1 | 696.0 | 2.0 | 4.5 |
| PN N628 J013645.9+154327.6 | 1:36:45.9 | 15:43:27.6 | 9.5 | 670.9 | 1.3 | 0.5 |
| PN N628 J013645.9+154741.8 | 1:36:45.9 | 15:47:41.8 | 7.7 | 706.9 | 2.1 | 4.7 |
| PN N628 J013646.0+155048.3 | 1:36:46.0 | 15:50:48.3 | -7.2 | 698.6 | 1.2 | 0.4 |
| PN N628 J013646.1+154621.0 | 1:36:46.1 | 15:46:21.0 | -44.3 | 629.7 | 2.0 | 5.0 |
| PN N628 J013646.3+154602.9 | 1:36:46.3 | 15:46:02.9 | 0.3 | 674.6 | 2.2 | 2.7 |
| PN N628 J013646.3+154750.9 | 1:36:46.3 | 15:47:50.9 | -0.2 | 695.3 | 1.4 | 5.3 |
| PN N628 J013646.4+154706.8 | 1:36:46.4 | 15:47:06.8 | -15.7 | 676.0 | 2.2 | 6.8 |
| PN N628 J013646.4+154737.4 | 1:36:46.4 | 15:47:37.4 | -1.8 | 693.7 | 4.0 | 6.7 |
| PN N628 J013646.8+154353.8 | 1:36:46.8 | 15:43:53.8 | -4.9 | 650.5 | 1.3 | 2.3 |
| PN N628 J013646.8+154712.9 | 1:36:46.8 | 15:47:12.9 | 56.2 | 751.9 | 2.2 | 6.3 |
| PN N628 J013646.8+155007.2 | 1:36:46.8 | 15:50:07.2 | -18.1 | 686.1 | 1.5 | 3.6 |
| PN N628 J013646.9+154619.1 | 1:36:46.9 | 15:46:19.1 | 6.5 | 683.1 | 2.0 | 5.7 |
| PN N628 J013647.0+154716.8 | 1:36:47.0 | 15:47:16.8 | 10.9 | 707.1 | 2.1 | 5.7 |
| PN N628 J013647.1+154659.7 | 1:36:47.1 | 15:46:59.7 | 2.5 | 691.1 | 2.9 | 6.2 |
| PN N628 J013647.1+154647.2 | 1:36:47.1 | 15:46:47.2 | 18.8 | 702.1 | 2.9 | 7.4 |
| PN N628 J013647.2+154233.9 | 1:36:47.2 | 15:42:33.9 | -0.7 | 651.7 | 1.7 | 1.0 |
| PN N628 J013647.2+154947.9 | 1:36:47.2 | 15:49:47.9 | -33.8 | 673.3 | 2.6 | 9.8 |
| PN N628 J013647.6+154539.7 | 1:36:47.6 | 15:45:39.7 | 4.6 | 667.7 | 1.3 | 4.7 |
| PN N628 J013647.6+154552.8 | 1:36:47.6 | 15:45:52.8 | 1.8 | 672.2 | 0.7 | 2.5 |
| PN N628 J013647.7+154843.9 | 1:36:47.7 | 15:48:43.9 | -4.4 | 693.5 | 1.5 | 4.5 |
| PN N628 J013647.8+154529.2 | 1:36:47.8 | 15:45:29.2 | -2.6 | 660.0 | 1.5 | 3.7 |
| PN N628 J013648.0+154910.0 | 1:36:48.0 | 15:49:10.0 | -18.8 | 686.7 | 2.5 | 3.3 |
| PN N628 J013648.0+154833.8 | 1:36:48.0 | 15:48:33.8 | 11.4 | 711.3 | 1.2 | 2.8 |
| PN N628 J013648.1+154602.0 | 1:36:48.1 | 15:46:02.0 | 15.6 | 690.9 | 0.9 | 4.3 |
| PN N628 J013648.3+154609.0 | 1:36:48.3 | 15:46:09.0 | -3.0 | 670.6 | 2.3 | 4.6 |
| PN N628 J013648.4+155003.2 | 1:36:48.4 | 15:50:03.2 | -1.7 | 701.4 | 1.1 | 2.6 |
| PN N628 J013648.4+154835.0 | 1:36:48.4 | 15:48:35.0 | 42.9 | 742.7 | 1.4 | 5.2 |
| PN N628 J013648.4+154929.9 | 1:36:48.4 | 15:49:29.9 | 1.9 | 709.4 | 1.8 | 3.8 |
| PN N628 J013648.5+154828.2 | 1:36:48.5 | 15:48:28.2 | -30.4 | 671.3 | 2.0 | 5.2 |
| PN N628 J013648.5+154541.7 | 1:36:48.5 | 15:45:41.7 | 7.0 | 674.7 | 2.3 | 5.1 |
| PN N628 J013648.6+155007.9 | 1:36:48.6 | 15:50:07.9 | -7.3 | 695.5 | 1.2 | 3.7 |
| PN N628 J013648.6+154343.5 | 1:36:48.6 | 15:43:43.5 | -5.2 | 657.4 | 0.6 | 3.6 |
| PN N628 J013648.7+154456.5 | 1:36:48.7 | 15:44:56.5 | -21.5 | 643.1 | 1.9 | 3.1 |
| PN N628 J013648.8+154911.9 | 1:36:48.8 | 15:49:11.9 | -10.1 | 696.4 | 2.1 | 4.2 |
| PN N628 J013648.8+154634.5 | 1:36:48.8 | 15:46:34.5 | 8.7 | 688.2 | 1.6 | 3.5 |
| PN N628 J013648.9+154547.7 | 1:36:48.9 | 15:45:47.7 | -27.2 | 647.5 | 2.5 | 5.3 |
| PN N628 J013649.0+154929.7 | 1:36:49.0 | 15:49:29.7 | -40.6 | 667.2 | 1.6 | 3.8 |
| PN N628 J013649.0+154719.3 | 1:36:49.0 | 15:47:19.3 | -9.6 | 686.8 | 1.3 | 2.9 |
| PN N628 J013649.1+154459.8 | 1:36:49.1 | 15:44:59.8 | 27.9 | 688.9 | 1.2 | -0.1 |
| PN N628 J013649.2+154935.9 | 1:36:49.2 | 15:49:35.9 | -9.0 | 699.2 | 2.1 | 4.0 |
| PN N628 J013649.2+154835.4 | 1:36:49.2 | 15:48:35.4 | -10.9 | 688.2 | 0.5 | 3.3 |
| PN N628 J013649.4+154521.6 | 1:36:49.4 | 15:45:21.6 | -3.6 | 656.8 | 1.3 | 2.1 |
| PN N628 J013649.4+154924.1 | 1:36:49.4 | 15:49:24.1 | 30.2 | 738.0 | 2.1 | 4.5 |
| PN N628 J013649.6+154726.5 | 1:36:49.6 | 15:47:26.5 | 15.1 | 711.8 | 1.5 | 4.9 |
| PN N628 J013649.6+154543.2 | 1:36:49.6 | 15:45:43.2 | -11.9 | 661.7 | 2.3 | 5.6 |
| PN N628 J013649.7+154729.7 | 1:36:49.7 | 15:47:29.7 | 18.8 | 716.2 | 0.7 | 3.4 |
| PN N628 J013649.7+155037.7 | 1:36:49.7 | 15:50:37.7 | -22.3 | 684.6 | 0.8 | 2.9 |
| PN N628 J013649.7+154556.4 | 1:36:49.7 | 15:45:56.4 | -13.9 | 663.6 | 1.6 | 3.7 |
| PN N628 J013649.9+154821.3 | 1:36:49.9 | 15:48:21.3 | -68.2 | 632.3 | 1.7 | 3.8 |
| PN N628 J013650.0+154219.7 | 1:36:50.0 | 15:42:19.7 | -1.5 | 657.6 | 2.2 | 0.9 |
| PN N628 J013650.1+154624.8 | 1:36:50.1 | 15:46:24.8 | -11.5 | 672.3 | 1.9 | 3.2 |
| PN N628 J013650.1+154920.3 | 1:36:50.1 | 15:49:20.3 | -14.2 | 693.9 | 2.1 | 2.9 |
| PN N628 J013650.5+155151.5 | 1:36:50.5 | 15:51:51.5 | -14.5 | 691.0 | 1.4 | 3.1 |
| PN N628 J013650.6+154512.1 | 1:36:50.6 | 15:45:12.1 | 34.7 | 697.7 | 1.9 | 1.3 |
| PN N628 J013651.0+154303.4 | 1:36:51.0 | 15:43:03.4 | 18.6 | 680.5 | 1.3 | 3.3 |
| PN N628 J013651.1+154653.7 | 1:36:51.1 | 15:46:53.7 | -67.1 | 622.4 | 1.6 | 1.7 |
| PN N628 J013651.2+154722.4 | 1:36:51.2 | 15:47:22.4 | -8.4 | 683.1 | 1.0 | 4.2 |
| PN N628 J013651.4+154442.3 | 1:36:51.4 | 15:44:42.3 | -1.9 | 665.0 | 1.4 | 4.5 |
| PN N628 J013651.5+154516.4 | 1:36:51.5 | 15:45:16.4 | -6.5 | 665.7 | 2.0 | 2.5 |
| PN N628 J013651.6+154705.8 | 1:36:51.6 | 15:47:05.8 | -2.5 | 687.7 | 1.4 | 5.1 |
| PN N628 J013651.7+154806.3 | 1:36:51.7 | 15:48:06.3 | 3.5 | 700.0 | 1.6 | 5.8 |
| PN N628 J013651.8+154911.9 | 1:36:51.8 | 15:49:11.9 | -3.7 | 701.6 | 1.2 | 2.3 |
| PN N628 J013652.0+155057.8 | 1:36:52.0 | 15:50:57.8 | 2.0 | 706.4 | 1.4 | 3.3 |
| PN N628 J013652.1+154500.3 | 1:36:52.1 | 15:45:00.3 | -10.5 | 661.2 | 1.4 | 3.9 |
| PN N628 J013652.2+154518.8 | 1:36:52.2 | 15:45:18.8 | 9.0 | 682.0 | 1.8 | 2.6 |
| PN N628 J013652.7+154515.3 | 1:36:52.7 | 15:45:15.3 | -10.7 | 663.0 | 2.2 | 2.9 |
| PN N628 J013653.1+154857.6 | 1:36:53.1 | 15:48:57.6 | -2.2 | 702.7 | 0.7 | 3.6 |
| PN N628 J013653.1+154454.7 | 1:36:53.1 | 15:44:54.7 | -33.3 | 637.9 | 1.4 | -0.8 |
| PN N628 J013653.4+154607.9 | 1:36:53.4 | 15:46:07.9 | 28.9 | 709.9 | 2.1 | -1.0 |
| PN N628 J013653.5+154816.1 | 1:36:53.5 | 15:48:16.1 | 12.7 | 713.5 | 1.1 | 2.9 |
| PN N628 J013653.7+154838.0 | 1:36:53.7 | 15:48:38.0 | -12.1 | 693.0 | 1.8 | 1.4 |
| PN N628 J013653.7+154526.5 | 1:36:53.7 | 15:45:26.5 | -3.9 | 673.6 | 1.6 | 4.1 |
| PN N628 J013653.9+154638.7 | 1:36:53.9 | 15:46:38.7 | 15.4 | 701.0 | 1.5 | 4.1 |
| PN N628 J013654.4+154347.3 | 1:36:54.4 | 15:43:47.3 | -8.2 | 658.8 | 1.1 | 2.8 |
| PN N628 J013654.6+154508.3 | 1:36:54.6 | 15:45:08.3 | -3.0 | 671.3 | 0.9 | 4.1 |
| PN N628 J013654.8+154402.4 | 1:36:54.8 | 15:44:02.4 | 8.1 | 678.3 | 1.4 | 4.1 |
| PN N628 J013654.9+154323.9 | 1:36:54.9 | 15:43:23.9 | -26.3 | 637.0 | 2.2 | 3.4 |
| PN N628 J013655.0+154212.8 | 1:36:55.0 | 15:42:12.8 | 17.3 | 680.2 | 1.5 | 2.4 |
| PN N628 J013655.0+154245.9 | 1:36:55.0 | 15:42:45.9 | 26.1 | 689.6 | 2.1 | 4.6 |
| PN N628 J013655.2+154719.8 | 1:36:55.2 | 15:47:19.8 | 1.9 | 697.7 | 1.1 | 0.6 |
| PN N628 J013655.3+154634.1 | 1:36:55.3 | 15:46:34.1 | -3.7 | 684.3 | 1.6 | 5.0 |
| PN N628 J013655.4+154445.4 | 1:36:55.4 | 15:44:45.4 | -7.0 | 667.3 | 1.4 | 4.1 |
| PN N628 J013655.6+154407.4 | 1:36:55.6 | 15:44:07.4 | 10.4 | 681.3 | 2.0 | 4.8 |
| PN N628 J013655.7+154429.7 | 1:36:55.7 | 15:44:29.7 | 1.1 | 673.3 | 1.4 | 3.9 |
| PN N628 J013655.9+154802.6 | 1:36:55.9 | 15:48:02.6 | -0.3 | 698.6 | 1.7 | 3.1 |
| PN N628 J013656.1+154853.6 | 1:36:56.1 | 15:48:53.6 | 1.3 | 703.8 | 0.9 | 3.9 |
| PN N628 J013656.2+154750.6 | 1:36:56.2 | 15:47:50.6 | -3.7 | 693.5 | 1.8 | 4.7 |
| PN N628 J013656.2+154602.3 | 1:36:56.2 | 15:46:02.3 | -6.3 | 676.9 | 1.3 | 3.5 |
| PN N628 J013656.6+154211.2 | 1:36:56.6 | 15:42:11.2 | 37.0 | 702.5 | 1.8 | 3.1 |
| PN N628 J013656.7+154601.2 | 1:36:56.7 | 15:46:01.2 | 7.7 | 691.4 | 0.9 | 3.0 |
| PN N628 J013656.8+154708.7 | 1:36:56.8 | 15:47:08.7 | 5.4 | 695.8 | 1.6 | 4.4 |
| PN N628 J013657.0+154434.5 | 1:36:57.0 | 15:44:34.5 | -5.2 | 665.7 | 1.5 | 4.6 |
| PN N628 J013657.0+154313.3 | 1:36:57.0 | 15:43:13.3 | 18.2 | 680.9 | 1.0 | 3.3 |
| PN N628 J013657.1+154733.9 | 1:36:57.1 | 15:47:33.9 | -12.6 | 683.0 | 1.6 | 2.3 |
| PN N628 J013657.1+154753.1 | 1:36:57.1 | 15:47:53.1 | 39.7 | 737.5 | 1.7 | 4.5 |
| PN N628 J013657.1+154832.8 | 1:36:57.1 | 15:48:32.8 | -18.1 | 681.0 | 1.5 | 3.3 |
| PN N628 J013657.2+154721.2 | 1:36:57.2 | 15:47:21.2 | -3.6 | 690.3 | 0.5 | 3.1 |
| PN N628 J013657.5+154225.1 | 1:36:57.5 | 15:42:25.1 | -1.6 | 661.9 | 1.0 | 2.9 |
| PN N628 J013657.9+154604.9 | 1:36:57.9 | 15:46:04.9 | -33.4 | 648.6 | 2.8 | 5.9 |
| PN N628 J013657.9+154639.3 | 1:36:57.9 | 15:46:39.3 | 1.8 | 688.2 | 2.2 | 4.5 |
| PN N628 J013658.1+154215.4 | 1:36:58.1 | 15:42:15.4 | -27.8 | 640.7 | 2.2 | 3.3 |
| PN N628 J013658.1+154759.1 | 1:36:58.1 | 15:47:59.1 | 0.5 | 700.1 | 1.5 | 3.1 |
| PN N628 J013658.2+154803.9 | 1:36:58.2 | 15:48:03.9 | 52.3 | 751.9 | 4.3 | 6.5 |
| PN N628 J013658.2+154437.2 | 1:36:58.2 | 15:44:37.2 | -9.2 | 660.3 | 0.7 | 2.8 |
| PN N628 J013658.3+154611.3 | 1:36:58.3 | 15:46:11.3 | -7.9 | 675.3 | 1.7 | 3.6 |
| PN N628 J013658.4+155039.0 | 1:36:58.4 | 15:50:39.0 | -13.0 | 687.7 | 1.4 | 3.7 |
| PN N628 J013658.4+154700.2 | 1:36:58.4 | 15:47:00.2 | -8.9 | 682.3 | 0.9 | 4.1 |
| PN N628 J013658.5+154626.8 | 1:36:58.5 | 15:46:26.8 | 0.1 | 687.4 | 0.7 | 3.3 |
| PN N628 J013658.6+154239.2 | 1:36:58.6 | 15:42:39.2 | 14.2 | 678.2 | 1.5 | 3.4 |
| PN N628 J013658.8+154717.5 | 1:36:58.8 | 15:47:17.5 | -15.4 | 680.3 | 0.9 | 3.4 |
| PN N628 J013658.9+154541.7 | 1:36:58.9 | 15:45:41.7 | -17.9 | 659.3 | 1.6 | 4.5 |
| PN N628 J013659.2+155120.7 | 1:36:59.2 | 15:51:20.7 | 2.6 | 704.0 | 1.6 | 3.8 |
| PN N628 J013659.3+154710.4 | 1:36:59.3 | 15:47:10.4 | 3.4 | 698.1 | 0.7 | 2.5 |
| PN N628 J013659.5+154748.0 | 1:36:59.5 | 15:47:48.0 | -2.0 | 692.6 | 2.4 | 4.9 |
| PN N628 J013659.6+154744.7 | 1:36:59.6 | 15:47:44.7 | 7.4 | 703.0 | 0.8 | 3.4 |
| PN N628 J013659.7+154239.4 | 1:36:59.7 | 15:42:39.4 | 10.9 | 678.6 | 0.9 | 3.4 |
| PN N628 J013659.8+155118.4 | 1:36:59.8 | 15:51:18.4 | -4.9 | 695.7 | 0.4 | 2.8 |
| PN N628 J013659.8+154606.2 | 1:36:59.8 | 15:46:06.2 | 10.5 | 690.4 | 1.0 | 4.1 |
| PN N628 J013659.8+154259.9 | 1:36:59.8 | 15:42:59.9 | 14.1 | 681.2 | 1.3 | 3.4 |
| PN N628 J013659.9+155139.7 | 1:36:59.9 | 15:51:39.7 | -11.3 | 684.4 | 0.6 | 2.5 |
| PN N628 J013700.0+154345.1 | 1:37:00.0 | 15:43:45.1 | -12.5 | 654.5 | 1.2 | 2.9 |
| PN N628 J013700.2+154633.0 | 1:37:00.2 | 15:46:33.0 | -1.9 | 684.5 | 1.4 | 3.6 |
| PN N628 J013700.2+154526.2 | 1:37:00.2 | 15:45:26.2 | -3.0 | 668.9 | 1.3 | 2.2 |
| PN N628 J013700.3+154554.8 | 1:37:00.3 | 15:45:54.8 | -3.8 | 674.6 | 0.5 | 1.9 |
| PN N628 J013700.3+154719.4 | 1:37:00.3 | 15:47:19.4 | -9.7 | 685.5 | 1.9 | 4.3 |
| PN N628 J013700.5+154621.4 | 1:37:00.5 | 15:46:21.4 | -18.6 | 664.5 | 1.6 | 4.6 |
| PN N628 J013700.5+154756.3 | 1:37:00.5 | 15:47:56.3 | -16.6 | 678.0 | 1.2 | 1.4 |
| PN N628 J013700.6+154833.8 | 1:37:00.6 | 15:48:33.8 | 2.4 | 695.1 | 1.4 | 4.7 |
| PN N628 J013700.7+154440.9 | 1:37:00.7 | 15:44:40.9 | -5.8 | 666.3 | 1.7 | 4.1 |
| PN N628 J013701.2+154508.7 | 1:37:01.2 | 15:45:08.7 | 17.7 | 691.5 | 0.6 | 0.8 |
| PN N628 J013655.3+154611.6 | 1:36:55.3 | 15:46:11.6 | 2.3 | 686.8 | 0.1 | (..) |
| PN N628 J013640.5+154418.6 | 1:36:40.5 | 15:44:18.6 | -2.7 | 654.5 | 0.1 | (..) |
| PN N628 J013638.9+154359.6 | 1:36:38.9 | 15:43:59.6 | 18.9 | 676.6 | 0.2 | (..) |
| PN N628 J013701.8+154552.5 | 1:37:01.8 | 15:45:52.5 | -7.5 | 671.7 | 0.2 | (..) |
| PN N628 J013640.8+154701.3 | 1:36:40.8 | 15:47:01.3 | -6.4 | 669.8 | 0.3 | (..) |
| PN N628 J013701.1+154952.2 | 1:37:01.1 | 15:49:52.2 | -17.7 | 680. | 0.3 | (..) |
| PN N628 J013632.4+154631.4 | 1:36:32.4 | 15:46:31.4 | -1.4 | 661.9 | 0.4 | (..) |
| PN N628 J013703.0+154704.8 | 1:37:03.0 | 15:47:04.8 | -7.4 | 672.2 | 0.5 | (..) |
| PN N628 J013623.4+154801.6 | 1:36:23.4 | 15:48:01.6 | -11.4 | 663.9 | 0.5 | (..) |
| PN N628 J013639.6+155027.4 | 1:36:39.6 | 15:50:27.4 | -3.3 | 697.8 | 0.6 | (..) |
| PN N628 J013703.4+154746.3 | 1:37:03.4 | 15:47:46.3 | -1.9 | 682.5 | 0.6 | (..) |
| PN N628 J013703.1+154628.8 | 1:37:03.1 | 15:46:28.8 | -5.1 | 674.9 | 0.6 | (..) |
| PN N628 J013629.7+154925.8 | 1:36:29.7 | 15:49:25.8 | -2.1 | 685.9 | 0.7 | (..) |
| PN N628 J013645.9+154650.1 | 1:36:45.9 | 15:46:50.1 | 50.3 | 732.3 | 0.7 | (..) |
| PN N628 J013638.5+154856.9 | 1:36:38.5 | 15:48:56.9 | -12.8 | 689.8 | 0.7 | (..) |
| PN N628 J013636.9+155046.7 | 1:36:36.9 | 15:50:46.7 | -1.9 | 696.7 | 0.8 | (..) |
| PN N628 J013701.5+155138.5 | 1:37:01.5 | 15:51:38.5 | -19.7 | 676.1 | 0.8 | (..) |
| PN N628 J013625.4+154843.7 | 1:36:25.4 | 15:48:43.7 | -5.5 | 672. | 0.9 | (..) |
| PN N628 J013700.6+155059.7 | 1:37:00.6 | 15:50:59.7 | 7.5 | 706.7 | 0.9 | (..) |
| PN N628 J013643.0+154420.7 | 1:36:43.0 | 15:44:20.7 | -5.2 | 656.2 | 0.9 | (..) |
| PN N628 J013652.7+154315.4 | 1:36:52.7 | 15:43:15.4 | -5.7 | 657.3 | 1.0 | (..) |
| PN N628 J013639.9+154702.3 | 1:36:39.9 | 15:47:02.3 | 7.9 | 681.8 | 1.0 | (..) |
| PN N628 J013703.2+154749.6 | 1:37:03.2 | 15:47:49.6 | -17.7 | 667.7 | 1.1 | (..) |
| PN N628 J013639.5+154653.2 | 1:36:39.5 | 15:46:53.2 | 49.5 | 712.7 | 1.1 | (..) |
| PN N628 J013702.5+154541.3 | 1:37:02.5 | 15:45:41.3 | -10.2 | 667.1 | 1.1 | (..) |
| PN N628 J013702.3+154738.8 | 1:37:02.3 | 15:47:38.8 | -11.3 | 676.7 | 1.2 | (..) |
| PN N628 J013628.9+154724.1 | 1:36:28.9 | 15:47:24.1 | 7.9 | 676.9 | 1.3 | (..) |
| PN N628 J013638.6+154353.7 | 1:36:38.6 | 15:43:53.7 | 48.4 | 701.5 | 1.3 | (..) |
| PN N628 J013640.7+154501.8 | 1:36:40.7 | 15:45:01.8 | -1.5 | 656.4 | 1.3 | (..) |
| PN N628 J013622.4+154315.1 | 1:36:22.4 | 15:43:15.1 | 18.5 | 685.3 | 1.3 | (..) |
| PN N628 J013655.4+154405.6 | 1:36:55.4 | 15:44:05.6 | -7.2 | 663.8 | 1.4 | (..) |
| PN N628 J013635.2+154324.9 | 1:36:35.2 | 15:43:24.9 | -17.9 | 632.3 | 1.4 | (..) |
| PN N628 J013645.1+154438.1 | 1:36:45.1 | 15:44:38.1 | 11.9 | 672. | 1.4 | (..) |
| PN N628 J013628.4+154544.0 | 1:36:28.4 | 15:45:44.0 | -10.0 | 647.8 | 1.4 | (..) |
| PN N628 J013702.7+154450.7 | 1:37:02.7 | 15:44:50.7 | -25.5 | 647.7 | 1.4 | (..) |
| PN N628 J013641.5+154449.9 | 1:36:41.5 | 15:44:49.9 | 0.7 | 661.8 | 1.5 | (..) |
| PN N628 J013645.3+154843.0 | 1:36:45.3 | 15:48:43.0 | 42.1 | 743.9 | 1.5 | (..) |
| PN N628 J013626.6+154925.4 | 1:36:26.6 | 15:49:25.4 | -12.8 | 667.4 | 1.5 | (..) |
| PN N628 J013627.1+154904.2 | 1:36:27.1 | 15:49:04.2 | -6.7 | 672.1 | 1.5 | (..) |
| PN N628 J013633.3+154856.7 | 1:36:33.3 | 15:48:56.7 | -20.6 | 667.2 | 1.5 | (..) |
| PN N628 J013639.5+155112.7 | 1:36:39.5 | 15:51:12.7 | -26.9 | 676.8 | 1.6 | (..) |
| PN N628 J013630.8+154524.1 | 1:36:30.8 | 15:45:24.1 | 19.5 | 681.5 | 1.6 | (..) |
| PN N628 J013629.5+154919.4 | 1:36:29.5 | 15:49:19.4 | -16.3 | 668.3 | 1.6 | (..) |
| PN N628 J013702.3+154457.1 | 1:37:02.3 | 15:44:57.1 | -1.7 | 671.1 | 1.6 | (..) |
| PN N628 J013645.8+154621.6 | 1:36:45.8 | 15:46:21.6 | 58.1 | 728.9 | 1.6 | (..) |
| PN N628 J013648.2+154923.3 | 1:36:48.2 | 15:49:23.3 | 4.0 | 712.9 | 1.6 | (..) |
| PN N628 J013655.5+154428.2 | 1:36:55.5 | 15:44:28.2 | -5.8 | 665.3 | 1.6 | (..) |
| PN N628 J013639.1+154255.6 | 1:36:39.1 | 15:42:55.6 | 20.2 | 676.8 | 1.7 | (..) |
| PN N628 J013700.1+154645.5 | 1:37:00.1 | 15:46:45.5 | -50.5 | 637.5 | 1.7 | (..) |
| PN N628 J013701.1+154806.4 | 1:37:01.1 | 15:48:06.4 | -8.4 | 682.6 | 1.7 | (..) |
| PN N628 J013635.1+154622.0 | 1:36:35.1 | 15:46:22.0 | -35.4 | 623. | 1.7 | (..) |
| PN N628 J013656.0+154357.9 | 1:36:56.0 | 15:43:57.9 | -3.0 | 667.9 | 1.8 | (..) |
| PN N628 J013702.2+154923.7 | 1:37:02.2 | 15:49:23.7 | -16.0 | 677.8 | 1.8 | (..) |
| PN N628 J013639.3+155030.2 | 1:36:39.3 | 15:50:30.2 | 4.5 | 704.9 | 1.8 | (..) |
| PN N628 J013639.1+154632.6 | 1:36:39.1 | 15:46:32.6 | 53.9 | 714.7 | 1.9 | (..) |
| PN N628 J013645.8+154230.2 | 1:36:45.8 | 15:42:30.2 | -11.1 | 640.3 | 1.9 | (..) |
| PN N628 J013627.3+155052.0 | 1:36:27.3 | 15:50:52.0 | 0.3 | 686.9 | 1.9 | (..) |
| PN N628 J013628.3+154632.0 | 1:36:28.3 | 15:46:32.0 | 18.0 | 682.3 | 2.0 | (..) |
| PN N628 J013639.4+154647.2 | 1:36:39.4 | 15:46:47.2 | 10.7 | 675.2 | 2.0 | (..) |
| PN N628 J013701.6+154625.1 | 1:37:01.6 | 15:46:25.1 | -14.3 | 670.1 | 2.0 | (..) |
| PN N628 J013639.4+154630.4 | 1:36:39.4 | 15:46:30.4 | 94.5 | 754.9 | 2.0 | (..) |
| PN N628 J013643.1+154638.6 | 1:36:43.1 | 15:46:38.6 | 6.5 | 673.5 | 2.2 | (..) |
| PN N628 J013639.9+154650.4 | 1:36:39.9 | 15:46:50.4 | 40.8 | 704.5 | 2.2 | (..) |
| PN N628 J013701.5+154601.4 | 1:37:01.5 | 15:46:01.4 | -10.6 | 670.6 | 2.2 | (..) |
| PN N628 J013632.3+155033.0 | 1:36:32.3 | 15:50:33.0 | 29.1 | 721.8 | 2.3 | (..) |
| PN N628 J013638.8+154742.3 | 1:36:38.8 | 15:47:42.3 | 28.0 | 718. | 2.4 | (..) |
| PN N628 J013645.0+155011.2 | 1:36:45.0 | 15:50:11.2 | -7.0 | 697.3 | 2.4 | (..) |
| PN N628 J013653.5+154653.5 | 1:36:53.5 | 15:46:53.5 | -55.7 | 635. | 2.5 | (..) |
| PN N628 J013632.0+155100.1 | 1:36:32.0 | 15:51:00.1 | 8.3 | 699.8 | 2.5 | (..) |
| PN N628 J013643.2+154711.7 | 1:36:43.2 | 15:47:11.7 | 64.0 | 749.8 | 2.6 | (..) |
| PN N628 J013643.7+154621.2 | 1:36:43.7 | 15:46:21.2 | 4.7 | 669.1 | 2.8 | (..) |

## B. Appendix

This section of the appendix provides the isolated, spatially unresolved HII regions candidate sample for NGC 628 selected with the PN.S

**Table B.1.** Table of the PN.S isolated spatially unresolved HII region candidates in NGC 628. The columns are 1) target ID, 2) right ascension (J2000), 3) declination (J2000), 4) $\Delta V = V_{LOS} - V_{HI}$ in kms$^{-1}$, 5) line-of-sight velocity (heliocentric, $V_{LOS}$), 6) the PN.S [OIII] 5007 Å magnitude and 7) the [OIII] – Hα color

| ID | RA (J2000) | DEC (J2000) | ΔV kms$^{-1}$ | $V_{LOS}$ kms$^{-1}$ | $m_{5007} - 24.66$ mag | $[OIII] - H\alpha$ mag |
|---|---|---|---|---|---|---|
| HII N628 J013623.9+154513.7 | 1:36:23.9 | 15:45:13.7 | 2.5 | 664.4 | -0.1 | 2.8 |
| HII N628 J013624.4+154540.1 | 1:36:24.4 | 15:45:40.1 | 2.9 | 665.2 | 0.0 | 2.5 |
| HII N628 J013625.5+154848.9 | 1:36:25.5 | 15:48:48.9 | -7.5 | 669.6 | 0.1 | 3.9 |
| HII N628 J013625.5+154858.0 | 1:36:25.5 | 15:48:58.0 | -3.6 | 673.8 | 0.6 | 3.8 |
| HII N628 J013625.8+154538.8 | 1:36:25.8 | 15:45:38.8 | -8.6 | 652.5 | -2.1 | 2.8 |
| HII N628 J013625.8+155105.6 | 1:36:25.8 | 15:51:05.6 | -11.9 | 677.0 | -0.6 | 2.2 |
| HII N628 J013626.0+154935.2 | 1:36:26.0 | 15:49:35.2 | -8.9 | 672.3 | -0.5 | 2.9 |
| HII N628 J013626.2+154942.3 | 1:36:26.2 | 15:49:42.3 | -9.5 | 669.2 | -1.7 | 1.3 |
| HII N628 J013626.4+154546.2 | 1:36:26.4 | 15:45:46.2 | 1.1 | 666.7 | -1.1 | 1.7 |
| HII N628 J013626.6+154653.3 | 1:36:26.6 | 15:46:53.3 | 0.2 | 669.0 | -0.4 | 2.2 |
| HII N628 J013626.9+155135.3 | 1:36:26.9 | 15:51:35.3 | -4.5 | 689.7 | -1.1 | 1.2 |
| HII N628 J013627.0+154713.8 | 1:36:27.0 | 15:47:13.8 | 8.1 | 677.3 | -0.3 | 3.1 |
| HII N628 J013627.1+154925.3 | 1:36:27.1 | 15:49:25.3 | -13.4 | 667.0 | 0.2 | 2.9 |
| HII N628 J013627.4+154601.9 | 1:36:27.4 | 15:46:01.9 | 12.6 | 672.2 | 0.2 | 3.3 |
| HII N628 J013627.6+154550.0 | 1:36:27.6 | 15:45:50.0 | 7.5 | 667.7 | -1.2 | 3.2 |
| HII N628 J013627.7+154643.1 | 1:36:27.7 | 15:46:43.1 | 1.6 | 669.2 | -0.4 | 2.9 |
| HII N628 J013627.8+154627.6 | 1:36:27.8 | 15:46:27.6 | -7.3 | 654.0 | -0.1 | 3.7 |
| HII N628 J013627.9+154709.9 | 1:36:27.9 | 15:47:09.9 | 7.6 | 675.4 | -1.1 | 2.4 |
| HII N628 J013628.0+154909.3 | 1:36:28.0 | 15:49:09.3 | 5.6 | 687.0 | -0.7 | 2.9 |
| HII N628 J013628.3+154702.0 | 1:36:28.3 | 15:47:02.0 | -25.6 | 640.4 | 0.9 | 5.1 |
| HII N628 J013628.3+154745.8 | 1:36:28.3 | 15:47:45.8 | -7.0 | 669.9 | 0.0 | 2.7 |
| HII N628 J013628.7+154618.2 | 1:36:28.7 | 15:46:18.2 | 8.5 | 669.3 | -0.1 | 3.2 |
| HII N628 J013628.7+154801.7 | 1:36:28.7 | 15:48:01.7 | -1.7 | 677.7 | -1.2 | 2.1 |
| HII N628 J013629.1+154539.5 | 1:36:29.1 | 15:45:39.5 | 19.4 | 679.2 | 0.7 | 4.8 |
| HII N628 J013629.1+154736.3 | 1:36:29.1 | 15:47:36.3 | 3.6 | 674.6 | 0.0 | 2.3 |
| HII N628 J013629.6+154823.0 | 1:36:29.6 | 15:48:23.0 | 11.9 | 688.6 | 0.6 | 4.4 |
| HII N628 J013629.9+154417.5 | 1:36:29.9 | 15:44:17.5 | 1.6 | 655.9 | -0.4 | 3.0 |
| HII N628 J013629.9+154451.4 | 1:36:29.9 | 15:44:51.4 | -0.9 | 660.3 | 0.5 | 3.5 |
| HII N628 J013629.9+154648.7 | 1:36:29.9 | 15:46:48.7 | 4.8 | 673.0 | 0.4 | 3.6 |
| HII N628 J013630.0+154844.7 | 1:36:30.0 | 15:48:44.7 | 17.5 | 694.8 | 0.5 | 4.3 |
| HII N628 J013630.1+154333.5 | 1:36:30.1 | 15:43:33.5 | 0.9 | 656.7 | -1.1 | 3.7 |
| HII N628 J013630.1+154928.5 | 1:36:30.1 | 15:49:28.5 | -11.2 | 678.0 | -2.5 | 3.2 |
| HII N628 J013630.3+154550.7 | 1:36:30.3 | 15:45:50.7 | -7.4 | 653.9 | 0.4 | 3.4 |
| HII N628 J013630.6+154419.0 | 1:36:30.6 | 15:44:19.0 | 3.5 | 660.3 | -1.4 | 1.2 |
| HII N628 J013630.8+154436.0 | 1:36:30.8 | 15:44:36.0 | -68.0 | 592.1 | -1.4 | 3.4 |
| HII N628 J013631.0+154738.4 | 1:36:31.0 | 15:47:38.4 | 4.2 | 680.3 | -0.4 | 2.9 |
| HII N628 J013631.1+154908.7 | 1:36:31.1 | 15:49:08.7 | -1.8 | 682.3 | 0.0 | 4.0 |
| HII N628 J013631.1+154914.4 | 1:36:31.1 | 15:49:14.4 | 1.6 | 686.7 | 0.3 | 3.8 |
| HII N628 J013631.2+154841.9 | 1:36:31.2 | 15:48:41.9 | -23.3 | 661.8 | 0.2 | 4.0 |
| HII N628 J013631.4+154848.0 | 1:36:31.4 | 15:48:48.0 | 5.1 | 691.5 | -0.7 | 3.1 |
| HII N628 J013631.5+154944.6 | 1:36:31.5 | 15:49:44.6 | -10.9 | 672.5 | 0.3 | 3.0 |
| HII N628 J013631.6+154631.6 | 1:36:31.6 | 15:46:31.6 | 6.6 | 671.9 | -0.4 | 3.8 |
| HII N628 J013631.6+154920.5 | 1:36:31.6 | 15:49:20.5 | -1.2 | 684.7 | -2.5 | 2.9 |
| HII N628 J013631.7+154639.2 | 1:36:31.7 | 15:46:39.2 | 14.0 | 682.0 | 1.0 | 5.2 |
| HII N628 J013632.0+154317.1 | 1:36:32.0 | 15:43:17.1 | -2.6 | 653.5 | -1.1 | 2.7 |
| HII N628 J013632.1+154312.0 | 1:36:32.1 | 15:43:12.0 | -1.7 | 655.1 | -3.2 | 2.2 |
| HII N628 J013632.2+154800.3 | 1:36:32.2 | 15:48:00.3 | 15.2 | 698.2 | 0.0 | 3.2 |
| HII N628 J013632.8+154712.7 | 1:36:32.8 | 15:47:12.7 | -56.6 | 614.0 | -0.5 | 4.8 |
| HII N628 J013632.8+154719.5 | 1:36:32.8 | 15:47:19.5 | 79.8 | 753.3 | 0.0 | 2.6 |
| HII N628 J013632.8+154728.6 | 1:36:32.8 | 15:47:28.6 | 4.4 | 676.0 | -0.2 | 3.3 |
| HII N628 J013632.9+154319.8 | 1:36:32.9 | 15:43:19.8 | -1.9 | 651.9 | -1.0 | 2.6 |
| HII N628 J013633.0+154840.4 | 1:36:33.0 | 15:48:40.4 | 2.7 | 689.4 | -1.3 | 3.9 |
| HII N628 J013633.1+154551.3 | 1:36:33.1 | 15:45:51.3 | 10.8 | 673.2 | 0.1 | 4.1 |
| HII N628 J013633.9+154709.7 | 1:36:33.9 | 15:47:09.7 | 46.0 | 716.4 | 0.4 | 6.0 |
| HII N628 J013634.0+155112.0 | 1:36:34.0 | 15:51:12.0 | -2.4 | 694.9 | 0.3 | 3.1 |
| HII N628 J013634.1+154730.5 | 1:36:34.1 | 15:47:30.5 | -6.1 | 673.9 | 0.1 | 4.6 |
| HII N628 J013634.1+154751.1 | 1:36:34.1 | 15:47:51.1 | 21.2 | 704.1 | 1.3 | 5.6 |
| HII N628 J013634.2+154803.1 | 1:36:34.2 | 15:48:03.1 | 15.6 | 703.0 | 0.9 | 4.9 |
| HII N628 J013634.3+154530.7 | 1:36:34.3 | 15:45:30.7 | 20.8 | 679.3 | -0.1 | 4.6 |
| HII N628 J013634.3+154930.1 | 1:36:34.3 | 15:49:30.1 | -5.1 | 685.0 | -0.7 | 2.2 |
| HII N628 J013634.7+154802.7 | 1:36:34.7 | 15:48:02.7 | -37.0 | 648.9 | -0.4 | 4.6 |
| HII N628 J013634.8+155143.7 | 1:36:34.8 | 15:51:43.7 | -13.7 | 681.6 | -0.8 | 2.8 |
| HII N628 J013634.9+154603.4 | 1:36:34.9 | 15:46:03.4 | 25.5 | 681.5 | 1.0 | 5.0 |
| HII N628 J013634.9+154834.9 | 1:36:34.9 | 15:48:34.9 | -2.6 | 682.1 | -0.7 | 2.9 |
| HII N628 J013635.0+154441.4 | 1:36:35.0 | 15:44:41.4 | 1.8 | 659.7 | -0.3 | 2.5 |
| HII N628 J013635.1+154753.1 | 1:36:35.1 | 15:47:53.1 | 8.2 | 693.7 | 0.7 | 4.8 |
| HII N628 J013635.5+154642.9 | 1:36:35.5 | 15:46:42.9 | 44.4 | 710.1 | -0.6 | 4.2 |
| HII N628 J013635.7+154422.5 | 1:36:35.7 | 15:44:22.5 | 7.7 | 667.2 | -1.5 | 3.3 |
| HII N628 J013635.9+154515.0 | 1:36:35.9 | 15:45:15.0 | 1.4 | 652.6 | -1.0 | 3.0 |
| HII N628 J013635.9+154944.5 | 1:36:35.9 | 15:49:44.5 | -8.7 | 684.2 | 0.4 | 4.0 |
| HII N628 J013635.9+154957.7 | 1:36:35.9 | 15:49:57.7 | 2.8 | 698.9 | -1.1 | 2.9 |
| HII N628 J013636.1+154604.2 | 1:36:36.1 | 15:46:04.2 | 122.3 | 774.8 | 0.2 | 4.3 |
| HII N628 J013636.2+154438.0 | 1:36:36.2 | 15:44:38.0 | -46.1 | 607.3 | 0.0 | 4.2 |
| HII N628 J013636.2+154457.0 | 1:36:36.2 | 15:44:57.0 | 53.5 | 709.6 | 0.2 | 4.2 |
| HII N628 J013636.2+154804.1 | 1:36:36.2 | 15:48:04.1 | -47.0 | 644.1 | 0.1 | 5.6 |
| HII N628 J013636.3+155032.4 | 1:36:36.3 | 15:50:32.4 | -11.0 | 687.9 | -1.6 | 3.0 |
| HII N628 J013636.3+155038.9 | 1:36:36.3 | 15:50:38.9 | -2.2 | 696.6 | 0.1 | 2.1 |
| HII N628 J013636.4+155020.0 | 1:36:36.4 | 15:50:20.0 | -10.9 | 685.7 | -0.3 | 3.7 |
| HII N628 J013636.5+155053.0 | 1:36:36.5 | 15:50:53.0 | -10.8 | 688.3 | -1.3 | 2.8 |
| HII N628 J013636.6+155130.9 | 1:36:36.6 | 15:51:30.9 | 1.1 | 703.3 | 0.6 | 3.6 |
| HII N628 J013636.7+154406.8 | 1:36:36.7 | 15:44:06.8 | 25.5 | 682.7 | 1.0 | 5.0 |
| HII N628 J013636.8+154952.8 | 1:36:36.8 | 15:49:52.8 | -11.7 | 686.0 | -1.4 | 3.4 |
| HII N628 J013636.9+155021.0 | 1:36:36.9 | 15:50:21.0 | -5.6 | 690.9 | -0.5 | 4.0 |
| HII N628 J013637.0+154351.9 | 1:36:37.0 | 15:43:51.9 | -2.8 | 652.8 | -0.1 | 1.7 |
| HII N628 J013637.0+154432.5 | 1:36:37.0 | 15:44:32.5 | -9.6 | 645.3 | -2.3 | 3.3 |
| HII N628 J013637.1+154954.0 | 1:36:37.1 | 15:49:54.0 | -5.7 | 692.6 | -0.8 | 2.4 |
| HII N628 J013637.3+154430.8 | 1:36:37.3 | 15:44:30.8 | -4.4 | 650.3 | -1.9 | 3.5 |
| HII N628 J013637.3+154436.3 | 1:36:37.3 | 15:44:36.3 | 31.8 | 688.6 | 0.5 | 3.3 |
| HII N628 J013637.3+154526.4 | 1:36:37.3 | 15:45:26.4 | 8.6 | 659.7 | -0.2 | 4.1 |
| HII N628 J013637.4+155018.9 | 1:36:37.4 | 15:50:18.9 | -19.8 | 674.0 | -1.3 | 2.4 |
| HII N628 J013637.9+154540.6 | 1:36:37.9 | 15:45:40.6 | 1.9 | 657.8 | 0.0 | 3.3 |
| HII N628 J013638.0+154428.4 | 1:36:38.0 | 15:44:28.4 | -62.9 | 591.8 | 0.4 | 3.4 |
| HII N628 J013638.2+154540.8 | 1:36:38.2 | 15:45:40.8 | 111.7 | 766.7 | 0.8 | 5.1 |
| HII N628 J013638.2+154847.5 | 1:36:38.2 | 15:48:47.5 | 17.0 | 717.9 | 0.7 | 7.4 |
| HII N628 J013638.5+154458.5 | 1:36:38.5 | 15:44:58.5 | 4.9 | 658.5 | 0.9 | 5.2 |
| HII N628 J013638.6+154256.5 | 1:36:38.6 | 15:42:56.5 | 3.5 | 659.4 | 0.4 | 3.0 |
| HII N628 J013638.6+154445.9 | 1:36:38.6 | 15:44:45.9 | 15.4 | 674.1 | -0.4 | 4.4 |
| HII N628 J013638.6+154820.4 | 1:36:38.6 | 15:48:20.4 | -5.1 | 695.1 | 0.2 | 4.9 |
| HII N628 J013638.8+154613.4 | 1:36:38.8 | 15:46:13.4 | 6.7 | 669.8 | 0.7 | 4.0 |
| HII N628 J013638.8+154905.6 | 1:36:38.8 | 15:49:05.6 | -18.2 | 687.2 | -1.8 | 3.0 |
| HII N628 J013638.9+155108.3 | 1:36:38.9 | 15:51:08.3 | -14.1 | 690.7 | -1.7 | 2.0 |
| HII N628 J013639.0+155048.3 | 1:36:39.0 | 15:50:48.3 | -0.9 | 699.0 | -0.3 | 3.3 |
| HII N628 J013639.1+154832.8 | 1:36:39.1 | 15:48:32.8 | -16.6 | 682.1 | 1.4 | 6.6 |
| HII N628 J013639.1+155103.4 | 1:36:39.1 | 15:51:03.4 | -6.0 | 699.8 | -0.2 | 1.5 |
| HII N628 J013639.2+154351.4 | 1:36:39.2 | 15:43:51.4 | -18.5 | 634.4 | -1.3 | 3.1 |
| HII N628 J013639.3+154443.3 | 1:36:39.3 | 15:44:43.3 | -0.2 | 659.4 | 0.1 | 4.8 |
| HII N628 J013639.4+154957.9 | 1:36:39.4 | 15:49:57.9 | -0.5 | 706.9 | -0.2 | 1.4 |
| HII N628 J013639.5+154639.5 | 1:36:39.5 | 15:46:39.5 | 3.2 | 666.0 | 1.1 | 6.6 |
| HII N628 J013639.6+154354.4 | 1:36:39.6 | 15:43:54.4 | -8.9 | 644.1 | 0.3 | 3.1 |
| HII N628 J013639.6+154827.1 | 1:36:39.6 | 15:48:27.1 | -7.6 | 695.4 | 1.2 | 5.8 |
| HII N628 J013639.8+154614.6 | 1:36:39.8 | 15:46:14.6 | 32.2 | 693.2 | 1.2 | 6.6 |
| HII N628 J013640.0+154625.5 | 1:36:40.0 | 15:46:25.5 | 2.7 | 661.3 | -0.7 | 3.4 |
| HII N628 J013640.2+155010.5 | 1:36:40.2 | 15:50:10.5 | -11.5 | 689.0 | 0.6 | 4.2 |
| HII N628 J013640.3+154442.2 | 1:36:40.3 | 15:44:42.2 | 16.2 | 676.2 | 0.1 | 2.3 |
| HII N628 J013640.3+154647.5 | 1:36:40.3 | 15:46:47.5 | 7.4 | 669.7 | 0.2 | 4.1 |
| HII N628 J013640.4+154838.2 | 1:36:40.4 | 15:48:38.2 | 48.3 | 751.7 | -0.1 | 5.7 |
| HII N628 J013640.5+154611.9 | 1:36:40.5 | 15:46:11.9 | 13.1 | 671.5 | 0.6 | 4.2 |
| HII N628 J013640.5+154910.7 | 1:36:40.5 | 15:49:10.7 | -12.2 | 695.2 | 2.5 | 9.5 |
| HII N628 J013640.8+154519.7 | 1:36:40.8 | 15:45:19.7 | 3.6 | 657.4 | 0.1 | 3.5 |
| HII N628 J013640.8+154852.7 | 1:36:40.8 | 15:48:52.7 | 1.3 | 704.3 | -2.3 | 4.1 |
| HII N628 J013641.2+154416.2 | 1:36:41.2 | 15:44:16.2 | -12.6 | 643.8 | -0.8 | 4.3 |
| HII N628 J013641.4+154346.9 | 1:36:41.4 | 15:43:46.9 | -3.9 | 650.1 | -0.2 | 3.1 |
| HII N628 J013641.8+155023.5 | 1:36:41.8 | 15:50:23.5 | -5.0 | 698.0 | -1.0 | 3.0 |
| HII N628 J013642.0+154430.7 | 1:36:42.0 | 15:44:30.7 | 4.9 | 662.6 | -0.2 | 3.9 |
| HII N628 J013642.0+154436.3 | 1:36:42.0 | 15:44:36.3 | 0.3 | 658.1 | -3.0 | 3.2 |
| HII N628 J013642.0+154455.9 | 1:36:42.0 | 15:44:55.9 | 21.1 | 680.5 | 0.8 | 4.2 |
| HII N628 J013642.2+154604.4 | 1:36:42.2 | 15:46:04.4 | -78.8 | 578.0 | -0.1 | 5.4 |
| HII N628 J013642.2+154817.0 | 1:36:42.2 | 15:48:17.0 | -53.4 | 647.3 | -1.4 | 5.9 |
| HII N628 J013642.6+154427.6 | 1:36:42.6 | 15:44:27.6 | -4.9 | 655.7 | -0.8 | 2.7 |
| HII N628 J013642.7+154847.6 | 1:36:42.7 | 15:48:47.6 | 6.6 | 710.8 | 0.9 | 4.7 |
| HII N628 J013643.0+154554.7 | 1:36:43.0 | 15:45:54.7 | 11.1 | 669.4 | 0.7 | 5.8 |
| HII N628 J013643.0+155011.8 | 1:36:43.0 | 15:50:11.8 | -6.3 | 697.6 | -0.1 | 3.1 |
| HII N628 J013643.2+154503.3 | 1:36:43.2 | 15:45:03.3 | 8.5 | 664.7 | -0.4 | 5.0 |
| HII N628 J013643.2+154717.9 | 1:36:43.2 | 15:47:17.9 | 0.0 | 693.1 | 0.4 | 2.9 |
| HII N628 J013643.3+154428.6 | 1:36:43.3 | 15:44:28.6 | 0.5 | 661.7 | -2.2 | 3.2 |
| HII N628 J013643.4+154355.2 | 1:36:43.4 | 15:43:55.2 | 1.1 | 658.7 | 0.2 | 2.9 |
| HII N628 J013643.4+154445.8 | 1:36:43.4 | 15:44:45.8 | -0.6 | 656.4 | -0.4 | 4.3 |
| HII N628 J013643.7+154424.6 | 1:36:43.7 | 15:44:24.6 | 5.4 | 666.7 | -1.8 | 3.4 |
| HII N628 J013643.8+154847.4 | 1:36:43.8 | 15:48:47.4 | -8.8 | 694.7 | 0.4 | 5.3 |
| HII N628 J013643.9+154915.7 | 1:36:43.9 | 15:49:15.7 | -4.3 | 695.9 | -1.0 | 1.5 |
| HII N628 J013644.0+154820.0 | 1:36:44.0 | 15:48:20.0 | -5.4 | 696.1 | 0.2 | 3.8 |
| HII N628 J013644.1+154711.1 | 1:36:44.1 | 15:47:11.1 | -4.8 | 688.9 | -0.2 | 5.0 |
| HII N628 J013644.1+154801.0 | 1:36:44.1 | 15:48:01.0 | 16.5 | 717.2 | 1.0 | 5.7 |
| HII N628 J013644.2+154527.3 | 1:36:44.2 | 15:45:27.3 | -14.4 | 648.6 | 0.3 | 4.2 |
| HII N628 J013644.3+154405.3 | 1:36:44.3 | 15:44:05.3 | 0.0 | 660.2 | -0.9 | 3.8 |
| HII N628 J013644.3+154911.7 | 1:36:44.3 | 15:49:11.7 | -8.4 | 694.0 | -0.3 | 3.8 |
| HII N628 J013644.4+154632.4 | 1:36:44.4 | 15:46:32.4 | -1.6 | 670.0 | 1.2 | 7.5 |
| HII N628 J013644.5+154756.2 | 1:36:44.5 | 15:47:56.2 | -103.0 | 600.3 | 0.6 | 3.4 |
| HII N628 J013644.6+154811.8 | 1:36:44.6 | 15:48:11.8 | 55.7 | 758.3 | 0.3 | 3.6 |
| HII N628 J013644.7+154637.1 | 1:36:44.7 | 15:46:37.1 | -5.6 | 670.0 | 1.6 | 8.1 |
| HII N628 J013644.7+154914.8 | 1:36:44.7 | 15:49:14.8 | 13.3 | 718.8 | -1.3 | 3.8 |
| HII N628 J013644.7+154959.3 | 1:36:44.7 | 15:49:59.3 | -11.7 | 693.6 | 0.2 | 4.1 |
| HII N628 J013645.0+154523.6 | 1:36:45.0 | 15:45:23.6 | -7.9 | 658.0 | 0.4 | 4.2 |
| HII N628 J013645.0+154839.1 | 1:36:45.0 | 15:48:39.1 | 27.1 | 727.6 | -1.5 | 4.3 |
| HII N628 J013645.0+154957.5 | 1:36:45.0 | 15:49:57.5 | -4.3 | 700.3 | -1.1 | 2.0 |
| HII N628 J013645.1+154407.1 | 1:36:45.1 | 15:44:07.1 | 18.3 | 679.8 | 0.3 | 4.7 |
| HII N628 J013645.1+154911.7 | 1:36:45.1 | 15:49:11.7 | -8.3 | 698.2 | 0.2 | 2.2 |
| HII N628 J013645.2+154922.4 | 1:36:45.2 | 15:49:22.4 | -11.3 | 695.7 | 0.0 | 2.2 |
| HII N628 J013645.2+155133.4 | 1:36:45.2 | 15:51:33.4 | -14.7 | 689.7 | -0.3 | 2.7 |
| HII N628 J013645.3+154439.5 | 1:36:45.3 | 15:44:39.5 | -12.6 | 648.7 | 0.1 | 2.4 |
| HII N628 J013645.3+154728.3 | 1:36:45.3 | 15:47:28.3 | 0.3 | 701.0 | 0.3 | 5.4 |
| HII N628 J013645.3+155107.2 | 1:36:45.3 | 15:51:07.2 | 3.4 | 704.8 | -2.0 | 3.6 |
| HII N628 J013645.4+154343.3 | 1:36:45.4 | 15:43:43.3 | 4.0 | 661.7 | -0.8 | 2.6 |
| HII N628 J013645.4+154509.7 | 1:36:45.4 | 15:45:09.7 | 27.2 | 687.4 | -1.3 | 4.3 |
| HII N628 J013645.7+154740.6 | 1:36:45.7 | 15:47:40.6 | -19.2 | 681.5 | 0.8 | 4.2 |
| HII N628 J013645.8+154319.2 | 1:36:45.8 | 15:43:19.2 | -10.3 | 647.9 | -1.1 | 2.6 |
| HII N628 J013645.8+154424.7 | 1:36:45.8 | 15:44:24.7 | 7.3 | 672.1 | 0.3 | 4.8 |
| HII N628 J013646.0+154412.1 | 1:36:46.0 | 15:44:12.1 | -11.8 | 647.6 | -0.5 | 2.3 |
| HII N628 J013646.1+154447.7 | 1:36:46.1 | 15:44:47.7 | 11.3 | 678.6 | 0.1 | 2.6 |
| HII N628 J013646.1+154739.2 | 1:36:46.1 | 15:47:39.2 | -1.9 | 695.4 | 0.2 | 4.0 |
| HII N628 J013646.3+154831.0 | 1:36:46.3 | 15:48:31.0 | -8.4 | 691.6 | 0.7 | 4.6 |
| HII N628 J013646.4+154530.6 | 1:36:46.4 | 15:45:30.6 | 0.2 | 661.7 | -1.4 | 4.8 |
| HII N628 J013646.6+154900.2 | 1:36:46.6 | 15:49:00.2 | -8.2 | 695.6 | -2.1 | 3.7 |
| HII N628 J013646.9+154706.4 | 1:36:46.9 | 15:47:06.4 | -17.8 | 672.5 | 1.0 | 5.9 |
| HII N628 J013647.1+154638.1 | 1:36:47.1 | 15:46:38.1 | 21.3 | 704.0 | 1.1 | 5.5 |
| HII N628 J013647.2+154947.9 | 1:36:47.2 | 15:49:47.9 | -33.8 | 673.3 | 2.6 | 9.8 |
| HII N628 J013647.3+154321.6 | 1:36:47.3 | 15:43:21.6 | 3.8 | 664.2 | 0.4 | 3.5 |
| HII N628 J013647.6+154837.9 | 1:36:47.6 | 15:48:37.9 | 6.6 | 704.5 | 0.1 | 4.2 |
| HII N628 J013647.9+154958.0 | 1:36:47.9 | 15:49:58.0 | 1.5 | 707.0 | 0.4 | 4.1 |
| HII N628 J013648.0+154929.9 | 1:36:48.0 | 15:49:29.9 | -4.0 | 704.5 | 0.2 | 3.0 |
| HII N628 J013648.1+154602.0 | 1:36:48.1 | 15:46:02.0 | 15.6 | 690.9 | 0.9 | 4.3 |
| HII N628 J013648.4+154704.7 | 1:36:48.4 | 15:47:04.7 | -17.6 | 675.9 | -0.5 | 3.5 |
| HII N628 J013648.4+154841.4 | 1:36:48.4 | 15:48:41.4 | 9.1 | 709.9 | -0.3 | 4.6 |
| HII N628 J013648.6+154343.5 | 1:36:48.6 | 15:43:43.5 | -5.2 | 657.4 | 0.6 | 3.6 |
| HII N628 J013648.7+154935.4 | 1:36:48.7 | 15:49:35.4 | -11.2 | 697.0 | -0.6 | 3.4 |
| HII N628 J013649.1+154610.5 | 1:36:49.1 | 15:46:10.5 | 3.7 | 679.8 | 0.5 | 3.9 |
| HII N628 J013649.2+154835.4 | 1:36:49.2 | 15:48:35.4 | -10.9 | 688.2 | 0.5 | 3.3 |
| HII N628 J013649.3+154840.4 | 1:36:49.3 | 15:48:40.4 | -11.3 | 689.4 | 0.2 | 3.0 |
| HII N628 J013649.3+154900.0 | 1:36:49.3 | 15:49:00.0 | -3.2 | 701.2 | -1.9 | 1.7 |
| HII N628 J013649.4+154939.8 | 1:36:49.4 | 15:49:39.8 | -13.4 | 693.6 | -0.8 | 3.1 |
| HII N628 J013649.5+154741.6 | 1:36:49.5 | 15:47:41.6 | 11.0 | 706.7 | -0.1 | 4.5 |
| HII N628 J013649.8+154606.9 | 1:36:49.8 | 15:46:06.9 | 41.8 | 719.6 | -0.4 | 2.1 |
| HII N628 J013650.4+154536.7 | 1:36:50.4 | 15:45:36.7 | 6.7 | 677.6 | -0.6 | 3.1 |
| HII N628 J013650.5+154259.4 | 1:36:50.5 | 15:42:59.4 | -0.3 | 660.8 | -1.9 | 1.3 |
| HII N628 J013650.5+154829.0 | 1:36:50.5 | 15:48:29.0 | 45.6 | 744.6 | -1.3 | 4.4 |
| HII N628 J013650.9+154549.6 | 1:36:50.9 | 15:45:49.6 | -5.2 | 672.4 | -2.6 | 4.1 |
| HII N628 J013651.0+154808.0 | 1:36:51.0 | 15:48:08.0 | -14.6 | 678.2 | 0.6 | 3.7 |
| HII N628 J013651.0+155044.8 | 1:36:51.0 | 15:50:44.8 | -12.5 | 692.7 | -0.8 | 3.1 |
| HII N628 J013651.4+154821.2 | 1:36:51.4 | 15:48:21.2 | 2.3 | 703.1 | -1.5 | 3.9 |
| HII N628 J013651.4+154937.4 | 1:36:51.4 | 15:49:37.4 | 7.7 | 715.4 | -0.6 | 2.9 |
| HII N628 J013651.6+154816.7 | 1:36:51.6 | 15:48:16.7 | -5.1 | 695.9 | -1.0 | 3.6 |
| HII N628 J013651.7+154537.1 | 1:36:51.7 | 15:45:37.1 | -3.7 | 672.8 | -0.2 | 4.5 |
| HII N628 J013651.8+154534.7 | 1:36:51.8 | 15:45:34.7 | -27.8 | 648.1 | 0.1 | 4.9 |
| HII N628 J013651.8+154900.5 | 1:36:51.8 | 15:49:00.5 | -15.7 | 687.8 | -2.4 | 2.4 |
| HII N628 J013651.8+154934.7 | 1:36:51.8 | 15:49:34.7 | -49.9 | 656.6 | 0.3 | 4.6 |
| HII N628 J013651.9+154827.6 | 1:36:51.9 | 15:48:27.6 | -19.7 | 686.8 | 0.4 | 4.0 |
| HII N628 J013652.0+154436.7 | 1:36:52.0 | 15:44:36.7 | -1.6 | 667.2 | -0.8 | 3.3 |
| HII N628 J013652.2+154543.6 | 1:36:52.2 | 15:45:43.6 | 1.2 | 682.0 | -0.9 | 4.4 |
| HII N628 J013652.2+154720.8 | 1:36:52.2 | 15:47:20.8 | 1.2 | 693.5 | -2.9 | 3.2 |
| HII N628 J013652.2+154911.4 | 1:36:52.2 | 15:49:11.4 | -13.2 | 691.0 | 0.1 | 2.4 |
| HII N628 J013652.3+154313.2 | 1:36:52.3 | 15:43:13.2 | -1.3 | 662.8 | -1.9 | 2.3 |
| HII N628 J013652.3+154536.4 | 1:36:52.3 | 15:45:36.4 | -2.4 | 675.8 | -0.9 | 4.0 |
| HII N628 J013652.4+154901.2 | 1:36:52.4 | 15:49:01.2 | -4.5 | 700.3 | -1.1 | 2.4 |
| HII N628 J013652.4+154937.5 | 1:36:52.4 | 15:49:37.5 | -11.3 | 693.8 | -1.1 | 1.5 |
| HII N628 J013652.5+154543.0 | 1:36:52.5 | 15:45:43.0 | -9.0 | 671.4 | 0.2 | 3.9 |
| HII N628 J013652.7+154708.4 | 1:36:52.7 | 15:47:08.4 | 3.1 | 696.1 | -0.5 | 5.0 |
| HII N628 J013652.8+154716.4 | 1:36:52.8 | 15:47:16.4 | 22.9 | 716.7 | 0.4 | 3.4 |
| HII N628 J013653.2+154707.1 | 1:36:53.2 | 15:47:07.1 | 0.8 | 691.6 | -1.3 | 3.3 |
| HII N628 J013653.2+154820.0 | 1:36:53.2 | 15:48:20.0 | -3.0 | 697.4 | -0.7 | 2.6 |
| HII N628 J013653.5+154910.2 | 1:36:53.5 | 15:49:10.2 | -22.6 | 681.7 | 0.3 | 3.0 |
| HII N628 J013653.6+154804.2 | 1:36:53.6 | 15:48:04.2 | -8.7 | 689.9 | -1.4 | 2.5 |
| HII N628 J013653.8+154206.1 | 1:36:53.8 | 15:42:06.1 | -2.6 | 657.2 | -0.8 | 6.0 |
| HII N628 J013653.9+154923.8 | 1:36:53.9 | 15:49:23.8 | -4.8 | 700.7 | -1.4 | 1.9 |
| HII N628 J013654.1+154746.3 | 1:36:54.1 | 15:47:46.3 | -4.4 | 690.0 | -0.6 | 2.2 |
| HII N628 J013654.4+154853.7 | 1:36:54.4 | 15:48:53.7 | -3.7 | 700.0 | -0.8 | 2.3 |
| HII N628 J013654.6+154646.1 | 1:36:54.6 | 15:46:46.1 | 5.7 | 691.1 | -0.1 | 3.6 |
| HII N628 J013654.8+154421.1 | 1:36:54.8 | 15:44:21.1 | -0.9 | 666.9 | 0.0 | 2.9 |
| HII N628 J013655.0+154836.6 | 1:36:55.0 | 15:48:36.6 | -7.0 | 695.8 | -0.7 | 3.5 |
| HII N628 J013655.1+154615.2 | 1:36:55.1 | 15:46:15.2 | 0.6 | 685.7 | -1.6 | 3.0 |
| HII N628 J013655.2+154503.5 | 1:36:55.2 | 15:45:03.5 | 12.6 | 687.6 | -0.4 | 2.8 |
| HII N628 J013655.2+154834.8 | 1:36:55.2 | 15:48:34.8 | -6.9 | 695.3 | 0.6 | 4.3 |
| HII N628 J013655.3+154414.5 | 1:36:55.3 | 15:44:14.5 | -5.6 | 663.7 | 0.3 | 3.0 |
| HII N628 J013655.7+154604.8 | 1:36:55.7 | 15:46:04.8 | -3.6 | 681.1 | -0.7 | 1.6 |
| HII N628 J013656.0+154421.0 | 1:36:56.0 | 15:44:21.0 | -2.2 | 670.9 | 0.5 | 3.9 |
| HII N628 J013656.3+154713.4 | 1:36:56.3 | 15:47:13.4 | -29.5 | 664.3 | -1.0 | 3.3 |
| HII N628 J013656.4+154840.8 | 1:36:56.4 | 15:48:40.8 | -3.7 | 697.0 | -0.3 | 1.6 |
| HII N628 J013657.0+154426.5 | 1:36:57.0 | 15:44:26.5 | -7.3 | 663.3 | -0.9 | 1.5 |
| HII N628 J013657.1+154657.4 | 1:36:57.1 | 15:46:57.4 | -6.2 | 683.5 | -0.5 | 3.4 |
| HII N628 J013657.2+154419.8 | 1:36:57.2 | 15:44:19.8 | -14.6 | 654.2 | 0.6 | 3.8 |
| HII N628 J013657.2+154721.2 | 1:36:57.2 | 15:47:21.2 | -3.6 | 690.3 | 0.5 | 3.1 |
| HII N628 J013657.3+154539.9 | 1:36:57.3 | 15:45:39.9 | -8.8 | 675.4 | -1.3 | 3.2 |
| HII N628 J013657.8+154449.2 | 1:36:57.8 | 15:44:49.2 | -0.8 | 671.1 | -0.6 | 3.4 |
| HII N628 J013657.9+154628.7 | 1:36:57.9 | 15:46:28.7 | -7.3 | 678.8 | -0.5 | 3.7 |
| HII N628 J013658.3+154646.3 | 1:36:58.3 | 15:46:46.3 | 9.3 | 697.2 | 1.0 | 5.1 |
| HII N628 J013658.7+154457.5 | 1:36:58.7 | 15:44:57.5 | -5.4 | 668.0 | -0.3 | 3.2 |
| HII N628 J013658.9+154841.4 | 1:36:58.9 | 15:48:41.4 | -12.7 | 684.2 | 0.1 | 3.4 |
| HII N628 J013659.1+154512.3 | 1:36:59.1 | 15:45:12.3 | -9.5 | 664.3 | -1.1 | 3.8 |
| HII N628 J013659.5+154734.4 | 1:36:59.5 | 15:47:34.4 | -0.3 | 698.2 | -0.2 | 3.1 |
| HII N628 J013659.7+154616.3 | 1:36:59.7 | 15:46:16.3 | -9.0 | 671.6 | -2.4 | 2.6 |
| HII N628 J013659.8+155118.4 | 1:36:59.8 | 15:51:18.4 | -4.9 | 695.7 | 0.4 | 2.8 |
| HII N628 J013700.5+154645.7 | 1:37:00.5 | 15:46:45.7 | -4.5 | 683.0 | 0.4 | 4.0 |
| HII N628 J013701.0+154601.3 | 1:37:01.0 | 15:46:01.3 | -20.5 | 661.0 | -0.2 | 1.9 |

# C Appendix - Photometric calibration of the PN.S [OIII] Left/Right and Hα arms

The instrumental magnitude $m_0$ is defined as $m_0 = -2.5 \log(\text{counts(ADU)s}^{-1})$, corrected to outside the atmosphere. From the standard star observations, the calibration of $m_0$ to $m_{AB}$ is

$$m_{AB} = m_0 + (18.01 \pm 0.12) \qquad \text{(C.1)}$$

For $f_\lambda$ (erg cm$^{-2}$ s$^{-1}$ Å$^{-1}$) at the observed wavelength of λ5007Å,

$$m_{AB} = -2.5 \log f_\lambda - 20.90 \qquad \text{(C.2)}$$

The $m_{5007}$ magnitude is defined by

$$-2.5 \log f_\lambda - 2.5 \log T_i(\lambda)d\lambda = m_{5007} + 13.74. \text{ (C.3)}$$

where $T_i$ the transmission curve for filter I. In our case, it is the profile of the [OIII] emission line. To first order it is approximated with a Gaussian with $\sigma_\lambda = 0.6381$Å; the value of this additional term is 0.5. From eqns ( C.1 - C.3), $m_{5007}$ and our instrumental magnitude $m_0$ are related by

$$m_{5007} = m_0 + 24.66 \qquad \text{(C.4)}$$

Using the nominal gain of the CCDs (1.16 $e^-$ ADU$^{-1}$), the mean value of the mean conversion factor, averaged over the two arms of the PN.S, is $2.34 \times 10^{-16}$. Douglas et al. (2007) found a mean value of $2.07 \times 10^{-16}$. We can now estimate our expected instrumental magnitude $m_0$ for the bright end of the PNLF in NGC 628, on which we can base the selection of PNe in NGC 628. The distance modulus of NGC 628 is 29.8 and the bright end of the PNLF is nominally at an absolute magnitude $M^*_{5007} = -4.54 \pm 0.05$ (Ciardullo et al. 2002), hence $m^*_{5007} = 25.3$ or $m_0 = 0.6$.

For the Hα arm we carry out independent calibration using a similar set of equations. From the measurements of the spectrophotometric standard stars observed with the Hα arm, we derive the following relation between the instrument and AB magnitudes:

$$m_{AB,H\alpha} = m_0 + (22.83 \pm 0.11) \qquad \text{(C.5)}$$

For $f_\lambda$ (erg cm$^{-2}$ s$^{-1}$ Å$^{-1}$) at the observed wavelength of λ6563 Å,

$$m_{AB} = -2.5 \log f_\lambda - 21.50 \qquad \text{(C.6)}$$

i.e. for a flat SED spectrum, $I_{H\alpha} = 1.74 I_{[OIII].}$